\documentclass[notitlepage,prb,aps,groupedaddress,onecolumn]{revtex4-1}

\usepackage[utf8]{inputenc}
\usepackage[english]{babel}
\usepackage[T1]{fontenc}
\usepackage{lmodern}
\usepackage{graphicx}
\usepackage{amsmath}
\usepackage{amsfonts}
\usepackage{amssymb}
\usepackage{microtype}
\usepackage{braket}
\usepackage{subfigure}

\usepackage[breaklinks=true,colorlinks, citecolor=blue, linkcolor=black]{hyperref}

\newcommand{\bi}{\bibitem}

\newcommand{\ct}{\cite}

\newcommand{\Tr}{\operatorname{Tr}}

\newcommand{\nn}{\nonumber}
\newcommand{\la}{\langle}
\newcommand{\ra}{\rangle}

\newcommand{\rhoi}{\tilde{\rho}}
\newcommand{\hi}{\tilde{H}}
\nocite{*}
\begin{document}
	
\title{Quantum thermal machines and batteries}%
\author{Sourav Bhattacharjee}%
\email{bsourav@iitk.ac.in}%
\affiliation{Department of Physics, IIT Kanpur, Kanpur-208016, India}%
\author{Amit Dutta}%
\affiliation{Department of Physics, IIT Kanpur, Kanpur-208016, India}%
\date{\today}%
\begin{abstract}
The seminal work by Sadi Carnot in the early nineteenth century provided the blueprint of a reversible heat engine  and the celebrated second law of thermodynamics eventually followed. Almost two centuries later, the quest to formulate a quantum theory of the thermodynamic laws has thus unsurprisingly motivated physicists to visualise what are known as `quantum thermal machines' (QTMs). In this  article, we review the prominent developments achieved in the theoretical construction as well as understanding of QTMs, beginning from the formulation of their earliest prototypes to recent models. We also present a detailed introduction and highlight recent progress in the rapidly developing field of `quantum batteries'.    
\end{abstract}

\maketitle%
\tableofcontents

\section{Introduction}\label{sec_intro}

Thermal machines refer to a broad class of devices whose operation is associated with some form of exchange and/or conversion of heat energy. They usually consist of two or more ‘heat reservoirs’ and a ‘working fluid’ (WF) which facilitates the intended process. Commonly known examples are the classical heat engines and refrigerators \cite{zemansky_mcgraw_book} which form the backbone of almost all mechanical and industrial machines that utilize thermal energy, ranging from the household air conditioners and refrigerators to the fuel based vehicles and the propeller of a spaceship blasting off into space.  Such devices, by construction, operate irreversibly in far from equilibrium settings. The dynamics of the quantities of interest, such as the heat transferred or the work extracted, are governed by the commonly known laws of thermodynamics. However, the tremendous technological advancement achieved in the past few decades has both necessitated as well as facilitated a rapid miniaturization of such thermal machines over the years. The frontiers of such miniaturization  has been pushed down to astonishingly small scales, where quantum effects are prominent and can therefore no longer be ignored. This has resulted in the resurgence of a few age-old questions that have persisted throughout the last century$-$how do thermodynamics, usually associated with macroscopic phenomena, reconcile with quantum mechanics, which describes equations of motion at the microscopic level that are inherently time-reversal symmetric?
 
It is worth noting that quantum mechanics and thermodynamics belong to a set of two immensely successful, albeit independent theoretical frameworks that have withstood the test of rigorous experimental verification. {Yet, it remains an open question as to how the laws of thermodynamics manifest at very small scales, particularly,} where quantum effects are expected to dominate the dynamics of a physical system \ct{kosloff_ent_2100, vinjanampathy_ctp_545, gemmer_berlin_book, campaioli_springer_book, deffner_iop_book, kosloff_jcp_204105, linden_prl_130401, anders_njp_010201, horodecki_natcom_2059, jahnke_epl_50008, tuncer_arxiv_04387}. A priori, one may argue that the well-known laws of thermodynamics (with an exception to the first law) are defined for macroscopic systems described by statistical averages, and hence the question of their validity for microscopic systems consisting of a few particles or qubits may itself appear meaningless. However, in 1959, Scovil et. al. \ct{scovil_prl_262} demonstrated that the working of a quantum three-level maser coupled to two thermal reservoirs resembles that of a heat engine; with  an efficiency  upper bounded by the Carnot limit \ct{carnot}. This work provided the first hint that the laws of thermodynamics, particularly the second law, may have a more fundamental and even quantum origin. In other words, it might be possible to naturally arrive at the thermodynamic laws starting from a microscopic quantum framework. However, the only known model of a `quantum heat engine' at the time, i.e.  the three-level maser, relied on a quasi-static description based on the equilibrium population of the energy-levels and hence could not provide any further insight into the dynamical processes involved. 

It was not until the 1990's  when researchers, motivated by the developing field of open quantum systems, began to look for new toy models of `quantum thermal machines' (QTMs). The aim was to design simplistic models that have the same functional behavior as classical thermal machines, i.e. conversion of heat energy into useful work and vice-versa, yet at the same time that could be analyzed within the dynamical framework of open quantum systems. 
The advent of the Lindbladian framework \cite{lindblad_cmp_119, gorini_jmp_821} made plausible the construction of physically meaningful models of QTMs that could  operate at far from equilibrium settings. The challenge was however to demarcate the dynamical energy exchanges into parts that could be associated with quantum analogues of `heat' and `work' as well as to formulate a second law in terms of the entropic changes involved in a cycle of operation. The formulation of minimal working models of reciprocating or stroke engines soon followed; these works paved the way for an explosion of research that extensively analyzed the performance of such models in diverse scenarios, from exploring the role of entanglement and coherences to the consequences of using non-thermal baths and many more. However,  to this day, the debate regarding a unique definition of quantities such as quantum work and heat as well as a universal formulation of the second law is far from being fully settled.   

Over the years, a plethora of such simple models of QTMs \cite{alicki_osid_1440002, he_pre_036145, geva_jmo_635, geva_jcp_7681, segal_pre_026109, allahverdyan_prl_050601, agarwal_pre_012130, alicki_jpa_015002, brunner_pre_032115, chen_epl_40003, correa_pre_042131, correa_scirep_3949, gallego_njp_125009, goswami_pra_013842, levy_prl_070604,lin_pre_046105, fialko_prl_085303, zagoskin_prb_014501, linden_prl_130401,scully_prl_220601,segal_prl_260601, quan_prl_180402, venturelli_prl_256801, klimovsky_pre_012140, bender_jpa_4427,erdman_prb_245432, chattopadhyay_epj_302, chattopadhyay_sr_16967, ono_arxiv_10181,usui_arXiv_03832, manzano_arxiv_03830,dann_arxiv_02801, huangfu_arxiv_12486,oliveira_arxiv_11694,kaur_arxiv_10258,fang_arxiv_04856,bouton_arxiv_10946,oliveira_pre_014149,khandelwal_arxiv_11553,chatterjee_pre_062109,harunari_prr_023194,ghoshal_pra_042208,zheng_pre_012110}  have been proposed  
to probe the thermodynamic laws at the quantum domain, a few of which have also been realized experimentally \cite{abah_prl_203006, robnagel_sc_325, blickle_natphys_12, lindenfels_prl_080602, peterson_prl_240601, klatzow_prl_110601, lin_pre_046105,passos_pra_022113}. Extensive analysis of these quantum models have strongly pointed to the presence of an upper bound to the efficiency and performance of such quantum heat engines and refrigerators. The existence of the Carnot bound, which is a manifestation of the celebrated second law of thermodynamics, at such small scales re-established a strong case for the validity of thermodynamics principles down to microscopic scales, thereby necessitating further scrutiny of the emergence of the thermodynamic laws at the fundamental level. 

The exhaustive analysis of QTMs, nevertheless, has led to an unprecedented understanding of how simple few-level quantum systems exchange energy as well as information with other such systems or with an external environment. Such understanding has in turn opened up the possibility of engineering microscopic devices in a way that may revolutionize nano-scale engineering. As for example, the application of `quantum probes', which are essentially simple quantum systems such as a qubit or a harmonic oscillator, to quantum metrology  have only been recently realized. \ct{giovannetti_np_222, degen_rmp_035002, kurizki_tech_1, pezze_rmp_035005} Particularly, they have been shown to be suitable candidates for high precision measurements in thermometry \ct{hofer_prl_090603, mukherjee_cp_162, mondal_arxiv_08509} (i.e., temperature measurements of nanoscale devices) as well as  magnetometry \ct{bhattacharjee_njp_013024} (magnetic field measurements). 

As much as  the focus has been on thermal energy conversion in the quantum regime, the subtleties of quantum phenomena affecting the process of energy storage and its subsequent extraction had not received much attention until recently. Alicki and Fannes, in their pioneering work \cite{alicki_pre_042123}, showed that a \textit{quantum battery} composed of many identical copies of a single quantum system can, in principle, facilitate a higher energy extraction per cell through cyclic unitary process when compared to a single cell. In addition, they concluded that the maximal work extraction is possible only if the battery is driven through intermediate entangled states while \textit{discharging} the battery. However, it was later proved \cite{hovhannisyan_prl_240401} that although ~entangling' or non-local operations are required for maximal work extraction, it is not necessary to  generate entanglement, per se, in the battery during the discharging process. Further, the use of non-local operations was also shown \cite{campaioli_prl_150601} to result in a faster scaling of the speed of discharging or charging (depositing energy) the battery with the battery size, as compared to local driving protocols. The  above results have also been verified in a number of models. In spite of rapid developments, a robust mechanism to identify and utilize quantum affects for optimizing the usage of quantum batteries has not been established yet.  
   
In this article, we present a brief review of the basic design of some of the broad class of QTMs that are widely studied in literature.  As already  mentioned, the concepts of quantum heat and quantum work as such are not yet uniquely defined and several definitions of these can be found in literature (see Ref.~[\onlinecite{vinjanampathy_ctp_545}] for a review). However, in this review article, we limit ourselves to a handful of these definitions relevant for understanding the working of the thermal machines discussed. Two parameters of paramount importance which characterize the performance of QTMs and which we will repeatedly encounter in the course of discussion, are the efficiency ($\eta$) and the coefficient of performance (COP). The efficiency is defined in as the ratio of work output to the heat supplied from the hot bath, when the QTM operates as an engine. Similarly, the COP is defined as the ratio of heat extracted from the cold bath to the work performed on the working fluid, when the QTM acts as a refrigerator. In this regard, it is useful to recall the second law which states that the maximum efficiency $\eta_c=1-T_c/T_h$ and $\mathrm{COP}_c=T_c/(T_h-T_c)$ is attained in a Carnot cycle which is a reversible cycle operating between baths with temperatures $T_h$ and $T_c$.  We also review a couple of applications of QTMs in the field of quantum metrology \cite{paris_ijqi_125, ma_pr_89, correa_prl_220405, zhou_natcom_78, giovannetti_np_222, degen_rmp_035002, kurizki_tech_1, pezze_rmp_035005}, particularly in thermometry \cite{correa_prl_220405, pasquale_natcom_12782, hofer_prl_090603, hovhannisyan_prb_045101, shevchenko_prap_014013, palma_pra_052115, mehboudi_jpa_303001, potts_q_161,  mukherjee_cp_162} and magnetometry \cite{muessel_prl_103004, brask_prx_031010, albarelli_njp_123011, polino_aqs_024703, bhattacharjee_njp_013024}. Finally, we also outline recent developments in theoretical modeling of quantum batteries, with majority of the discussion focused on cyclic unitary protocols.

We would like to mention here that this review article is in no way exhaustive. In particular, given the vast amount of literature available as far as QTMs are concerned, this review aims for a brisk introduction to the basics of QTMs outlining the essential underlying principles. On the other hand, given that the theoretical concept of quantum batteries is relatively new and still in its nascent stages, we have thus taken care to provide a more in-depth discussions on its fundamentals as well as recent developments. {Finally, we have tried to maintain a consistent notational convention throughout the article as best as possible. However, in some sections, we had to resort to use  particular notational conventions used in previous works in order to maintain consistency with the relevant figures that we have reused from those works. Hence, the reader may come across a few instances where different symbols may have been used to represent the same quantity in different sections, although we have taken care to clearly spell out all such redefinitions.}


\section{Preliminary models}\label{sec_prelim}
In this section, we outline the working of two very simple yet insightful models of QTMs, which operate quasi-statically and are capable of working as quantum heat engines or quantum refrigerators. The first model we introduce is the three-level maser \ct{scovil_prl_262}  which, as already mentioned, is the earliest prototype of a  quantum heat engine. The other model we discuss, is the
realisation of a four-stroke Carnot engine where the working substance is comprised of the text-book system of a single particle in a one-dimensional box potential as working fluid \cite{bender_jpa_4427, bender_prsla_1519}. This model is unique in the way that it first identifies the analogue of classical `force' and uses the same to calculate the quantum work performed. Although numerous other models \cite{geusic_pr_343, alicki_jpa_L103, geva_jcp_3054, feldmann_ajp_485, kosloff_jap_8093, kosloff_jcp_1625, geva_pre_3903,palao_pre_056130, feldmann_pre_4774, feldmann_pre_046110} were also proposed in the early days of QTMs, we however, begin by discussing the two  models mentioned above. We highlight these particular models to make the reader appreciate the fact that thermodynamic signatures, as these toy models demonstrate, can manifest in the working of QTMs even when one does not explicitly resort to open system dynamics to analyze them.

\begin{figure*}
	\subfigure[]{
	\includegraphics[width=0.4\textwidth]{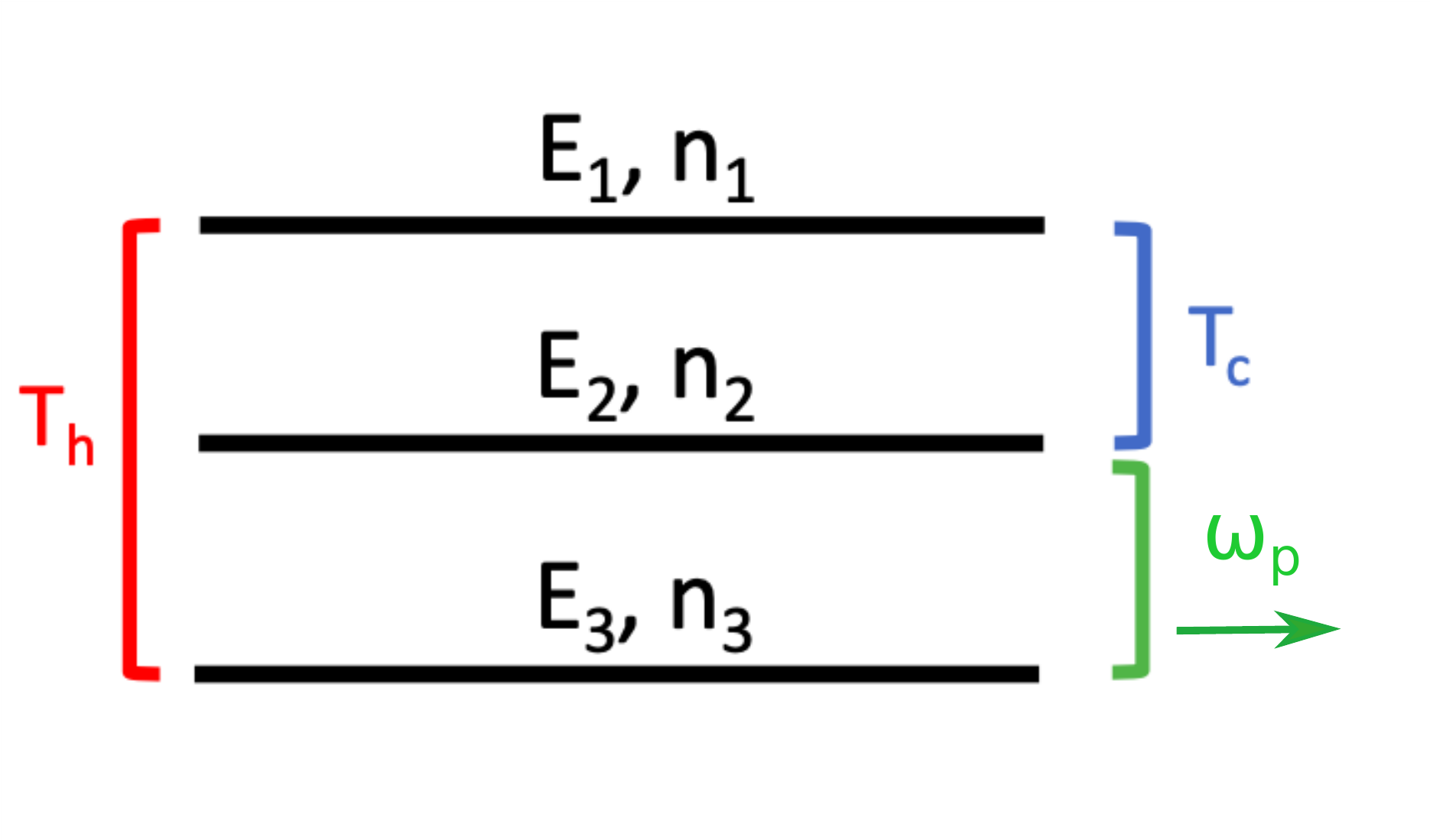}
	\label{fig_3lvl}}
	\subfigure[]{
	\includegraphics[width=0.4\textwidth]{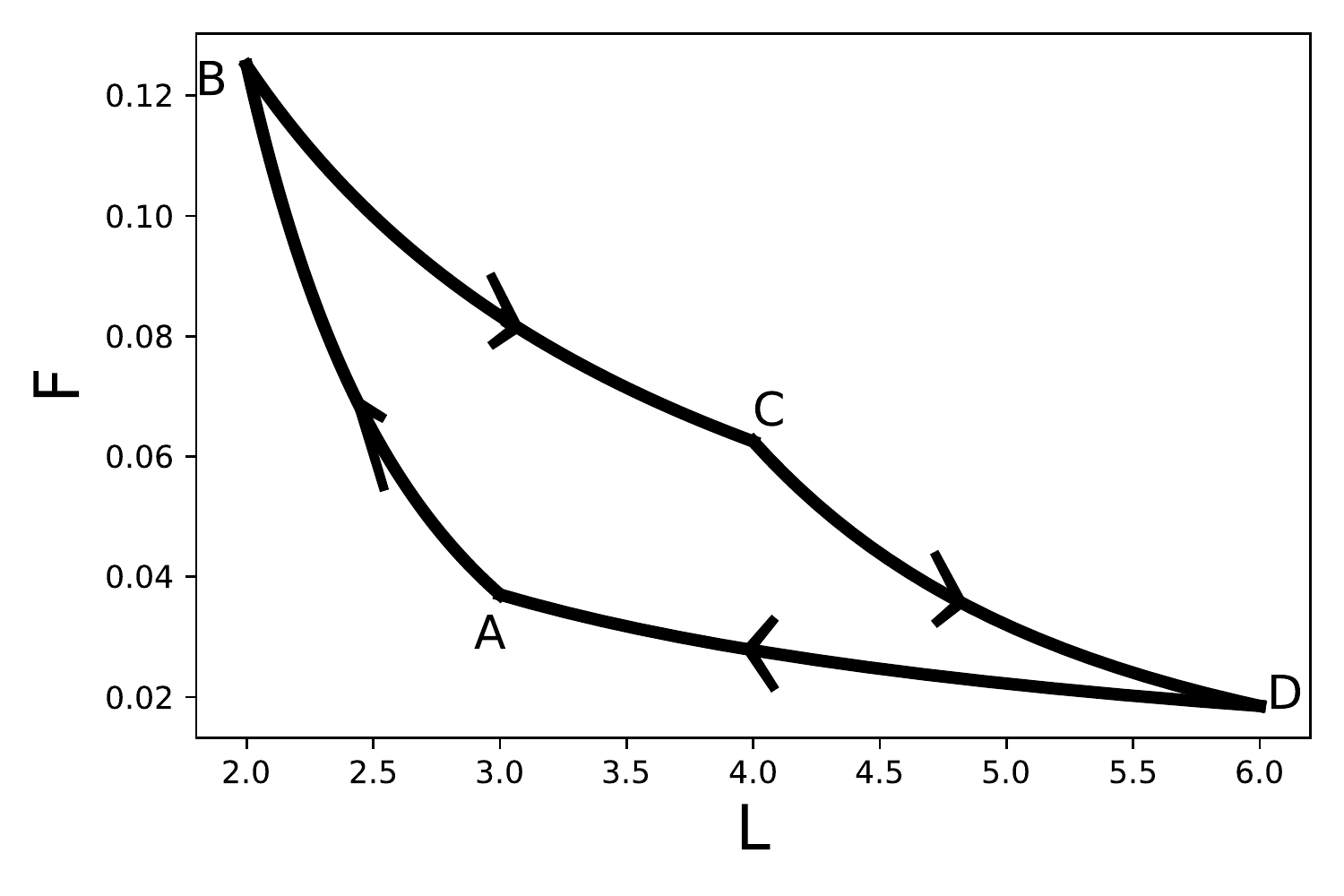}
	\label{fig_carnot}}

	\caption{(a)Schematic of the three-level maser which can act as a quantum heat engine. A hot bath with temperature $T_h$ induces excitations between energy levels $E_1$ and $E_3$, while a cold bath with temperature $T_c$ induces excitations between $E_1$ and $E_2$. Work is extracted by an external field resonant with the energy gap $E_2-E_3$. (b) $F-L$ diagram of the quantum Carnot cycle discussed in Sec.~\ref{subsec_box}, depicting the four strokes of the cycle. We have assumed $m=1$ while calculating the force $F(L)$ and further scaled it as $F(L)/\pi^2\to F(L)$. Note the striking similarity of the quantum Carnot cycle with the classical version in which the working medium is composed of an ideal gas enclosed in a three-dimensional volume; a similar cycle is then manifested in the $P-V$ diagram where $P$ and $V$ are the pressure and volume of the gas enclosed, respectively.}
\end{figure*}
\subsection{Three level maser}\label{subsec_maser}
In 1959, Scovil \textit{et.al.}  \ct{scovil_prl_262} realized that the steady-state operation of a three-level maser (see Fig.~\ref{fig_3lvl}) connected to two thermal baths through appropriately chosen frequency filters resemble that of a heat engine or refrigerator. The `quantum' working fluid is a three-level system with energy levels $E_1$, $E_2$ and $E_3$ ($E_1>E_2>E_3$) and populations $n_1$, $n_2$ and $n_3$, respectively. {Herein and throughout the rest of the review, we shall set the Boltzman and Planck's constants to unity, $k_B=\hbar=1$, unless explicitly mentioned otherwise.} A \textit{hot} bath with temperature $T_h$ is coupled to the system through a frequency filter so that it can induce excitations only between $E_1$ and $E_3$ with energy $\omega_h\sim |E_1-E_3|$. Similarly, a cold bath with temperature $T_c$ is allowed to induce transitions between $E_2$ and $E_3$ with energy $\omega_c= |E_1-E_2|$ in the system. Further, the system is also coupled resonantly with a radiation field with frequency $\omega_p= |E_2-E_3|$.

After the system attains (dynamical) equilibrium, the populations of the energy-levels satisfy,
\begin{subequations}
\begin{equation}
\frac{n_1}{n_3}=e^{-\frac{\omega_h}{T_h}},
\end{equation}
\begin{equation}
\frac{n_1}{n_2}=e^{-\frac{\omega_c}{T_c}}.
\end{equation}
\end{subequations}
Within this equilibrium regime, for each quanta of excitation $\omega_h$ induced by the hot bath, the system loses energy $\omega_c$ to the cold bath and $\omega_p$ to the radiation field, so that the populations are held steady. The energy exchanged with the baths can be thought of as `heat' transferred while the energy supplied to the radiation field is identified as the \textit{work} extracted from the system with the radiation field playing the role of the classical `piston'. Note that the preceding characterization of heat and work trivially satisfies the first law,
\begin{equation}\label{eq_maser_firstlaw}
\omega_h=\omega_c+\omega_p.
\end{equation}

The most remarkable result however appears when the efficiency of the system  is considered. An engine-like operation is possible when a `population inversion' is achieved between the levels $E_2$ and $E_3$, i.e. $n_2\geq n_3$, which leads to the condition,
\begin{equation}
\frac{n_2}{n_3}=\frac{n_2}{n_1}\frac{n_1}{n_3}=\exp\left(\frac{\omega_c}{T_c}-\frac{\omega_h}{T_h}\right)\geq 1
\end{equation}
or, 
\begin{equation}
\frac{\omega_c}{\omega_h}\geq\frac{T_c}{T_h}.
\end{equation}
The efficiency, i..e., the ratio of the work extracted to the energy supplied by the hot reservoir, is obtained as,
\begin{equation}
\eta=\frac{\omega_p}{\omega_h}=1-\frac{\omega_c}{\omega_h}\leq 1-\frac{T_c}{T_h},
\end{equation}
where we made use of the first law (see Eq.~\eqref{eq_maser_firstlaw}) for obtaining the second equality. 

Note that the dynamical equilibrium assumed at all instants of time means that the maser operation is reversible. In other words, the engine-like operation discussed above can be reversed to obtain refrigerator-like operation, in which a quanta of excitation $\omega_p$ induced by the radiation field leads to the extraction of an energy quanta $\omega_c$ from the cold bath and the simultaneous loss of an energy quanta $\omega_h$ to the hotter bath. The coefficient of performance is easily calculated as,
\begin{equation}
\mathrm{COP}=\frac{\omega_c}{\omega_p}=\frac{\omega_c}{\omega_h-\omega_c}\leq\frac{T_c}{T_h-T_c}.
\end{equation}
The efficiency as well as the coefficient of performance of the three-level maser when working as a heat engine and refrigerator, respectively, therefore appears to be identical to those known for the classical Carnot cycle. This observation hence provided the astonishing result that the second law can   hold true even for few-level quantum systems.
\subsection{Particle in a one-dimensional box}\label{subsec_box}
Another simple model of a quantum Carnot engine was provided by Bender \textit{et.al.} \ct{bender_jpa_4427, bender_prsla_1519} which, much like its classical counter-part, is operated in a cycle comprising of discrete strokes. The working fluid is made up of a particle confined in a one-dimensional hard-walled box, in which one of the two walls is movable. The length of the box $L$ is therefore an continuous variable which can be controlled externally to perform work on the system. To elucidate, let us consider the particle wave function $\ket{\psi}=\sum_na_n\ket{\phi_n}$, where $\ket{\phi_n}$ are the energy eigen states of the system. The energy expectation value of the system is
\begin{equation}\label{eq_carnot_energy}
\la E\ra=\sum_n|a_n|^2E_n(L),
\end{equation}
with $E_n(L)=n^2\pi^2/2mL^2$, where $m$ is the mass of the particle.  Invoking upon the notion of classical mechanics, the instantaneous `force'  which performs the work is defined as \cite{bender_prsla_1519,quan_pre_041129}
\begin{align}
F(L)&=-d\la E\ra/dL=-\sum_n|a_n|^2\frac{dE_n(L)}{dL}\nonumber\\
&=\sum_n|a_n|^2\frac{n^2\pi^2}{mL^3}.
\end{align}
It is important to note that in defining the force we have used
$ \la E\ra$. In general, the derivative with respect to the energy expectation value in the above expression should also contain terms such as $E(L)d|a_n|^2/dL$. However, we recall that any change in the populations $|a_n|^2$ is necessarily associated with changes in the Von-Neumann entropy of the system. For  quasi-static processes, which are the only processess involved in the working of the quantum Carnot cycle, an entropy change can result only from heat transfer with external baths or environment and not from any external force acting on the system. Hence, it is justified to neglect derivatives with respect to $|a_n|^2$ while defining the force.
 
Having defined the force, the quantum analogues of adiabats and isotherms are identified as follows. The \textit{quantum adiabats} correspond to the processes {in which the system is kept isolated from the environment and the length $L$ of the box is slowly changed. In the ideal situation, quantum mechanical adiabatic theorem then implies that the populations $a_n$ are held constant.} On the other hand, the \textit{quantum isotherms} are defined as operations in which the coefficients $a_n$ change along with $E_n$ on tuning $L$, but in such a way that the energy expectation value defined in Eq.~\eqref{eq_carnot_energy} remain constant. {Strictly speaking, `isotherms' demand a constant temperature which is itself not well-defined for an isolated single particle quantum system. Nevertheless, the quantum isotherms are defined here in analogy with classical isothermal processes of ideal systems (e.g. ideal gas) in which the temperature is equivalent to the internal energy and remains constant during a classical isothermal process.}

For purpose of simplicity, consider the situation in which only the two lowest lying energy eigen states contribute to the wave function, so that $a_1+a_2=1$. Let us assume that the system is initialized in the ground state,  with the box length $L=L_A$, so that $a_1(L_A)=1$. The energy expectation $\la E_A\ra$ is,
\begin{equation}\label{eq_carnot_en1}
\la E_A\ra=E_1(L_A)=\frac{\pi^2}{2mL_A^2},
\end{equation} 
where the subscript $A$ marks the initial point of the cycle. The start of the subsequent strokes will likewise be labeled as $B,C$ and $D$.
The quantum Carnot cycle is then constructed as follows (see Fig.~\ref{fig_carnot}):
\begin{enumerate}
\item \textit{Adiabatic compression ($A\to B$)} -- Length of the box is quasi-statically compressed to $L=L_B$ with $L_B<L_A$ so that the system remains in the instantaneous ground state. The force acting on the system during this stroke ($A\to B$) is given by
\begin{equation}
F_{AB}(L)=\frac{\pi^2}{mL^3},
\end{equation}
which performs work and increases the energy of the system to	
\begin{equation}
\la E_B\ra=\frac{\pi^2}{2mL_B^2},
\end{equation}
where $\la E_B\ra$ is the energy expectation value at the end of this stroke. The populations do not change and hence we have $a_1(L_B)=a_1(L_A)=1$. 
\item \textit{Isothermal expansion ($B\to C$)} -- Length of the box is expanded to $L=L_C=2L_B$ such that the energy expectation value remains constant throughout, i.e.
\begin{equation}\label{eq_iso2}
|a_1(L)|^2\frac{\pi^2}{2mL^2}+(1-|a_1(L)|^2)\frac{2\pi^2}{mL^2}=\frac{\pi^2}{2mL_B^2}.
\end{equation}
Note that the populations depend on $L$ during the process $B\to C$ due to the constraint on the  energy expectation. At the end of the expansion,  the system therefore reaches the first excited state with $a_1(L_C=2L_B)=0$. Hence, we have,
\begin{equation}
\la E_C\ra=\la E_B\ra=\frac{\pi^2}{2mL_B^2}
\end{equation}

The force acting on the system during this step is,
\begin{equation}
F_{BC}(L)=|a_1(L)|^2\frac{\pi^2}{mL^3}+(1-|a_1(L)|^2)\frac{4\pi^2}{mL^3},
\end{equation}
where $a_1(L)$ is constrained by Eq.~\eqref{eq_iso2}.
\item \textit{Adiabatic expansion ($C\to D$)} -- As in step~1, the length is quasi-statically changed to $L=L_D=2L_A$ with $L_D>L_C$ such that the system remains in the first excited state, i.e. $a_1(L_D)=a_1(L_C)=0$. The energy expectation at the end of this stroke is therefore,
\begin{equation}
\la E_D\ra=\frac{4\pi^2}{2mL_D^2}=\frac{\pi^2}{2mL_A^2},
\end{equation}
and the instantaneous force acting on the system during the expansion is
\begin{equation}
F_{CD}(L)=\frac{4\pi^2}{mL^3}
\end{equation}

\item \textit{Isothermal compression ($D\to A$)} -- Finally, the length is tuned back to $L=L_A$ in a way that the energy expectation value remains constant, 
\begin{equation}
|a'_1(L)|^2\frac{\pi^2}{2mL^2}+(1-|a'_1(L)|^2)\frac{2\pi^2}{mL^2}=\frac{\pi^2}{2mL_A^2},
\end{equation}
where the population functions are primed to distinguish them from the functions in the isothermal stroke $B\to C$. This also ensures that the system returns to the instantaneous ground state, $a_1'(L_A)=a_1(L_A)=1$, hence closing the cycle. The force acting on the system during this final stroke is given by,
\begin{equation}
F_{DA}(L)=|a'_1(L)|^2\frac{\pi^2}{mL^3}+(1-|a'_1(L)|^2)\frac{4\pi^2}{mL^3}.
\end{equation}

\end{enumerate}

The total work performed during the cycle is now calculated as
\begin{gather}
W=\int_{L_A}^{L_B}F_{AB}(L)dL+\int_{L_B}^{L_C=2L_B}F_{BC}(L)dL\nn\\+\int_{L_C}^{L_D=2L_A}F_{CD}(L)dL+\int_{L_D}^{L_A}F_{DA}(L)dL\nn\\
=\frac{\pi^2}{m}\left(\frac{1}{L_B^2}-\frac{1}{L_A^2}\right)\log 2
\end{gather}
Heat is absorbed by the system only during the isothermal expansion and equals the work done during the expansion, i.e.,
\begin{equation}
Q_H=\int_{L_B}^{2L_B}F_{BC}(L)dL=\frac{\pi^2}{mL_B^2}\log 2.
\end{equation}
The efficiency of this quantum engine cycle is therefore
\begin{equation}
\eta=\frac{W}{Q_H}=1-\frac{L_B^2}{L_A^2}=1-\frac{\la E_A\ra}{\la E_B \ra}
\end{equation}
Note that, by construction, $\la E_A\ra$ and $\la E_B\ra$ play the role of constant temperatures during the isothermal compression and expansion steps, respectively. The efficiency therefore turns out to be identical to that of the classical Carnot cycle.

{Despite its elegance, the setup discussed above is however a highly idealistic one that neglects the effect of the `actual temperatures' of the external environment, which shall inevitably effect the population of the energy levels. Nevertheless, the fact that a Carnot like bound emerged even within such idealistic setups was an intriguing result that invited further scrutiny of the second law.}

As already mentioned, the preliminary models discussed in this section are based on quasi-static processes which can be readily analyzed without any need for  explicit equations of motion that govern the dynamics of the working fluid. On the down side, such simplistic description, {besides being highly idealistic}, do not provide any fundamental insights regarding the dynamical processes involved. In the following sections, we will take a look at models that operate at far from equilibrium settings and importantly whose operation bear the hallmarks of classical thermodynamic principles such as a maximum efficiency bounded by the Carnot efficiency.

\section{Open Quantum Systems}\label{sec_open}
An open quantum system refers to a quantum system $S$ which obeys the quantum mechanical laws of motion and is coupled to an external environment $E$. The environment, in principle, can either be quantum in nature with a discreet energy spectrum, or classical in the continuum limit of vanishing energy gaps. 
Although, the dynamics of the composite system (i.e., the system $S$ and the environment $E$ taken together) can be described by the Schrodinger or the Von-Neumann equation, the exponentially large degrees of freedom of the environment renders it practically impossible to analytically solve these equations. One of the ways to resolve this issue is to formulate equations in terms of the reduced state of the system by tracing out the relevant degrees of freedom of the environment from the dynamics of the composite system. In this section, we discuss one such equation, namely the Gorini–Kossakowski–Sudarshan–Lindblad (GKSL) equation \cite{lindblad_cmp_119, gorini_jmp_821}, which is relevant for the rest of this review.

\subsection{Dynamical maps}
To begin with, let us assume that the system and the environment are initially decoupled, $\rho(0)=\rho_s(0)\otimes\rho_E(0)$, where $\rho_s(0)$, $\rho_E(0)$ and $\rho(0)$ represent the initial states of the system, environment and the composite system, respectively. The temporal evolution of the system is described by a dynamical map $V(t,0)$, which satisfies,
\begin{equation}\label{eq_dyn_map}
\rho_s(t)=\Tr_E\left[U(t,0)\rho_s(0)\otimes\rho_E(0)U^{\dagger}(t,0)\right]=V(t,0)\rho_s(0),
\end{equation}
where $\rho_s(t)=\Tr_E\rho(t)$ is the reduced state of the system obtained by tracing over the environmental degrees of freedom and $U(t,0)$ is the unitary evolution operator acting on the composite system.

A few remarks about the dynamical map $V(t,0)$ are in order. Firstly, $V(t,0)$ is a self-mapping on the density matrix space which implies that it must be completely positive (maps only to positive eigen-values) and trace preserving (so that all density matrices have unit trace), as can be  verified from Eq.~\eqref{eq_dyn_map}. Importantly, one can also check that $V(t,0)$  only comprises of operators defined on the Hilbert space of  $S$. \\

Secondly, if the dynamics is Markovian (memory-less) in nature, the family of maps $V(t,0)$, $\forall t>0$ constitutes what is known as a \textit{quantum dynamical semi-group}, which satisfies  $V(t,0)=V(t,t')V(t',0)$. In the absence of explicit time dependence of the system Hamiltonian and environmental couplings, the semi-group property requires the dynamical map $V(t,0)$ to be of the form $V(t,0)=e^{\mathcal{L}t}$, where $\mathcal{L}$ is referred to as the Linbladian super-operator. Substituting in Eq.~\eqref{eq_dyn_map}, we therefore find that the \textit{master equation} governing the evolution of the system is of the general form \cite{lindblad_cmp_119, gorini_jmp_821},
\begin{equation}\label{eq_gen_lindblad}
\frac{d\rho_s(t)}{dt}=\mathcal{L}\rho_s(t).
\end{equation}
Note that in the case of isolated system, the evolution of the system is governed by the Von-Neumann equation,
\begin{equation}
\frac{d\rho_s(t)}{dt}=-i\left[H_s,\rho_s(t)\right],
\end{equation}
from which one can immediately identify the super-operator $\mathcal{L}_{U}$ for unitary evolutions as,
\begin{equation}
\mathcal{L}_{U}=-i\left[H_s,\rho^{iso}_s(t)\right].
\end{equation}

In what follows, we will explicitly see these emergence of the features discussed above as we outline a short derivation of the GKSL master equation. We shall keep ourselves restricted to the cases where the Hamiltonian of the system is either independent \cite{lindblad_cmp_119, gorini_jmp_821, breuer_oxford_book, kryszewski_arxiv_1757} or periodically modulated \cite{kosloff_ent_2100, alicki_arxiv_4552,  levy_pre_061126, mukherjee_pre_062109, kosloff_arpc_365, klimovsky_atmop_329, kolar_prl_090601} in time, as these are the ones relevant in the context of QTMs. Although periodically modulated Hamiltonians are strictly not time-independent, the invariance of the Floquet \cite{floquet_sens_47, bukov_ap_139, eckardt_rmp_011004} Hamiltonian at stroboscopic instants, as we shall elaborate below, permits a dynamical equation of the form in Eq.~\eqref{eq_gen_lindblad}.

\subsection{The GKSL equation for static Hamiltonians}\label{subsec_open_static}

Consider a system-environment composite represented by a time-independent Hamiltonian,
\begin{equation}
H=H_s+H_E+H_I,
\end{equation}
where $H_s$ and $H_E$ correspond to the Hamiltonians of the system and the environment, respectively, while $H_I$ encapsulates the coupling between them and is of the form,
\begin{equation}
H_I=\sum_iS_i\otimes B_i.
\end{equation}
The operators $S_i$ and $B_i$ are local Hermitian operators pertaining to the Hilbert spaces of the system and the environment, respectively. As an illustration, the interaction between a two-level system and a photonic cavity (bath) can be of the form, $H_I = \sigma_x\otimes(a^\dagger+a)$, where $\sigma_x$ is a Pauli matrix and $a$ ($a^\dagger$) is the photonic annihilation (creation) operator. We start with the von-Neumann equation governing the evolution of the composite system in the interaction picture,
\begin{equation}\label{eq_lind_1}
\frac{d\rhoi(t)}{dt}=-i\left[\hi_I(t),\rhoi(0)\right]\\-\int_0^t\left[\hi_I(t),\left[\hi_I(t'),\rhoi(t')\right]\right]dt'.
\end{equation}
In the above equation, for any observable $O$,  $\tilde{O}$ represents the same in the interaction picture.\\

We now make an important approximation, namely the \textit{Born} or the \textit{weak-coupling} approximation, which assumes that (i) the system does not influence the environment so that $\rhoi_E(t)=\rhoi_E$ and (ii) the composite system exists in a tensor-product state at all times, $\rhoi(t)=\rhoi_s(t)\otimes\rhoi_E$. The Born approximation is valid for fast decaying environmental correlations and we shall return to it later. Under the above approximations, the equation of motion for the reduced state of the system can be obtained as,
\begin{equation}\label{eq_lind_2}
\frac{d\rhoi_s(t)}{dt}=-i\Tr_E\left[\hi_I(t),\rhoi_s(0)\otimes\rhoi_E\right]\\-\int_0^t\Tr_E\left[\hi_I(t),\left[\hi_I(t'),\rhoi_s(t')\otimes\rhoi_E\right]\right]dt'
\end{equation}

Next, we decompose the system operators $S_i$ into projections on the eigen-space of the Hamiltonian $H_s$ as,
\begin{equation}
S_i=\sum_{\varepsilon,\varepsilon'}\bra{\varepsilon}S_i\ket{\varepsilon'}\ket{\varepsilon}\bra{\varepsilon'}=\sum_{\omega=\varepsilon-\varepsilon'}S_i(\omega)\ket{\varepsilon}\bra{\varepsilon'},
\end{equation}
where $\ket{\varepsilon}$ are the energy eigen states of the system and  $\omega=\varepsilon-\varepsilon'$ are the possible excitation energies. The above decomposition leads to a particularly simple form of the Hamiltonian $H_I$ in the interaction picture,
\begin{align}\label{eq_hi_int}
\hi_I(t)&=\sum_ie^{iH_st}S_ie^{-iH_st}\otimes\sum_ie^{iH_Et}B_ie^{-iH_Et}\nn\\
		&=\sum_{i,\omega}e^{-i\omega t}S_i(\omega)\otimes B_i(t),
\end{align}
where,
\begin{equation}\label{eq_bi}
B_i(t)=\sum_ie^{iH_Et}B_ie^{-iH_Et} 
\end{equation} 
One can now derive the following relations,
\begin{subequations}

\begin{equation}
\Tr_E\left[\hi_I(t),\rhoi_s(0)\otimes\rhoi_E\right]=\sum_{i,\omega}\left[e^{-i\omega t}S_i(\omega),\rhoi_s(0)\right]\la B_i(t)\ra_E,
\end{equation}
\begin{widetext}
\begin{equation}
\Tr_E\left[\hi_I(t),\left[\hi_I(t'),\rhoi_s(t')\otimes\rhoi_E\right]\right]=\sum_{i,j,\omega,\omega'}e^{i(\omega' t-\omega t') }\left(S_j^\dagger(\omega')S_i(\omega)\rhoi_s(t')-S_i(\omega)\rhoi_s(t')S_j^\dagger(\omega')\right)\la B_j^\dagger(t)B_i(t')\ra_E+h.c.,
\end{equation}
\end{widetext}	
\end{subequations}
where $\la \cdot\ra_E$ denotes averaging over the state $\rhoi_E$. In most cases, the average of the environmental operators $\la B_i(t)\ra_E$ vanish. However, even if they are finite, one can re-scale the system Hamiltonian appropriately to set $\la B_i(t)\ra_E=0$ \cite{kryszewski_arxiv_1757} and thus we shall ignore these averages in all the scenarios which we will consider in the rest of the paper.  Substituting the above relations in Eq.~\eqref{eq_lind_2} and shifting the time-coordinate $t'=t-t'$, we obtain,
\begin{equation}\label{eq_lind_3}
\frac{d\rhoi_s(t)}{dt}=\sum_{i,j,\omega,\omega'}\int_0^tdt'e^{i(\omega'-\omega)t}
\Big(S_i(\omega)\rhoi_s(t-t')S_j^\dagger(\omega')\\-S_j^\dagger(\omega')S_i(\omega)\rhoi_s(t-t')\Big)e^{i\omega t'}\la B_j^\dagger(t)B_i(t-t')\ra_E+h.c.
\end{equation}

The next step is to invoke upon  the Markovian approximation, which assumes that the two-time environment correlation functions decay rapidly within increasing time separation $t-t'$, so much so that the system does not change appreciably within their decay time. In other words, if the environment correlations decay within a time $\tau_B$ and the system relaxes over a time-scale $\tau_R$, then $\tau_R\gg\tau_B$. Thus, if we consider a \textit{coarse-grained} evolution of the system with a time-scale $\sim \tau_R$, we can replace $\rhoi_s(t-t')$ with $\rho(t)$ in Eq.~\eqref{eq_lind_3} and extend the integral to infinity, as all contributions from time $t'>\tau_B$ can be neglected. We therefore arrive at,
\begin{equation}\label{eq_lind_4}
\frac{d\rhoi_s(t)}{dt}=\sum_{i,j,\omega,\omega'}e^{i(\omega'-\omega)t}
\Big(S_i(\omega)\rhoi_s(t)S_j^\dagger(\omega')\\-S_j^\dagger(\omega')S_i(\omega)\rhoi_s(t)\Big)\Gamma_{i,j}(\omega)+h.c.
\end{equation}
where
\begin{align}
\Gamma_{i,j}(\omega)&=\int_0^\infty dt' e^{i\omega t'}\la B_j^\dagger(t)B_i(t-t')\ra_E\nn\\
 &=\int_0^\infty dt' e^{i\omega t'}\la B_j^\dagger(t')B_i(0)\ra_E.
\end{align}
We have used the fact that $\rhoi_E$ is stationary, $\left[\rhoi_E,H_E\right]=0$ in deriving the second equality. Note that in Eq.~\eqref{eq_lind_4}, the evolution at a particular time $t$ is only determined by the state of the system at time $t$; hence there are no `memory' effects.

Finally, we make the rotating wave or secular approximation which allows us to ignore all fast oscillating terms, i.e. we retain only terms with $\omega'=\omega$. Note that this approximation, like to Markovian approximation, is also valid under a coarse-grained picture of the time evolution. The secular approximation allows to cast Eq.~\eqref{eq_lind_4} in the Linbladian form (see Eq.~\eqref{eq_gen_lindblad}),
\begin{subequations}\label{eq_gksl}
\begin{equation}\label{eq_static_gksl}
\frac{d\rhoi_s(t)}{dt}=\tilde{\mathcal{L}}\rhoi_s(t),
\end{equation}
where,
\begin{equation}\label{eq_lind_static}
\tilde{\mathcal{L}}\rhoi_s(t)=-i\left[H_{LS},\rhoi_s(t)\right]\\+\sum_{\omega,i,j}\gamma_{i,j}(\omega)\left(S_i(\omega)\rhoi_s(t)S_j^\dagger(\omega)-\frac{1}{2}\{S_j^\dagger(\omega)S_i(\omega),\rhoi_s(t)\}\right).
\end{equation}  
The renormalized or \textit{Lamb shifted} Hamiltonian $H_{LS}$ commutes with $H_s$ and is given by,
\begin{equation}
H_{LS}=\sum_{\omega,i,j}\eta_{i,j}(\omega)S_j^\dagger(\omega)S_i(\omega),
\end{equation}
\end{subequations}
where $\gamma_{i,j}(\omega)=2\mathrm{Re}\left[\Gamma_{i,j}(\omega)\right]$ and $\eta_{i,j}(\omega)=\mathrm{Im}\left[\Gamma_{i,j}(\omega)\right]$. he equation derived in Eq.~\eqref{eq_gksl} is famously known as the GKSL master equation. 
Note that the first term in Eq.~\eqref{eq_lind_static} captures the unitary part of the evolution while the second term encapsulates the dissipative part of the evolution.

Let us now return to the Born approximation. The assumption $\rho_E(t)=\rho_E$ is strictly not valid because there always exist a finite relaxation time which the environment requires to equilibriate, even if this relaxation time is very small in comparison with the relaxation time of the system. However, if one considers a  `coarse-graining' of the time-scale as we have done above in the case of the Markovian and secular approximations, the above assumption is perfectly valid as . Similarly, let us consider that a finite correlation $\chi$ is built up between the system and the environment, so that at a given time $t$, $\rho(t)=\rho_s(t)\otimes\rho_E+\chi(t)$. If one explicitly calculates the contribution of these correlation in the evolution after a small time $\Delta t$ by integrating Eq.~\eqref{eq_lind_1}, one gets \cite{kryszewski_arxiv_1757},
\begin{align}
\Delta\rhoi_s^{cor}(t)&=-\int_t^{t+\Delta t}dt\int_0^t\Tr_E\left[\hi_I(t),\left[\hi_I(t'),\chi(t')\right]\right]dt'\nn\\
&\propto \sum_{i,j}\int_t^{t+\Delta t}dt\int_0^t\la B_j^\dagger(t)B_i(t') \ra_E dt',
\end{align}
where $\Delta\rhoi_s^{cor}$ quantifies the extra contribution arising from the correlations $\chi(t)$.
One can see that a finite contribution will only arise for $\Delta t<\tau_B$ as the environmental correlations $\la B_j^\dagger(t)B_i(t') \ra_E$ are negligible for $|t-t'|> \tau_B$. Hence, once again, the coarse-grained picture of the time evolution permits us to neglect this extra contribution arising only for a very short duration.

To obtain the asymptotic steady-state attained by the system, we revert back to the Schrodinger picture. In the static case, the GKSL equation in Eq.~\eqref{eq_static_gksl} assumes the form,
\begin{equation}
\frac{d\rho_s(t)}{dt}=\mathcal{L}\rho_s(t) =-i[H_s,\rho_s(t)] +\tilde{\mathcal{L}}\rho_s(t),
\end{equation}
in the Schrodinger picture. The steady state $\rho_{ss}$ is therefore determined by solving the characteristic equation $\mathcal{L}\rho_{ss}=0$.

We note in passing that the set of reasoning in the derivation above can also be generalized for slowly varying Hamiltonians $H_s(t)$. The rate of change should be slow enough so that the \textit{quantum adiabaticity} is maintained. In other words, the time-scale over which $H_s(t)$ changes appreciably is much greater than the time-scale of relaxation of the system as well as the baths. Under such conditions, the GKSL equation assumes the form,
\begin{equation}
\frac{d\rhoi_s(t)}{dt}=\mathcal{L}(t)\rhoi_s(t),
\end{equation} 
where all the quantities in the time-dependent super-operator is derived in terms of the instantaneous Hamiltonian of the system. Likewise, the steady state $\rho_ss(t)$ is also slowly-varying and is dependent on the instantaneous energy eigen-values of the system.

\subsection{The GKSL equation for periodic Hamiltonians}\label{subsec_open_periodic}
We now consider the case of a periodically driven system coupled to an external environment \cite{kosloff_ent_2100, alicki_arxiv_4552,  levy_pre_061126, kosloff_arpc_365, mukherjee_pre_062109, klimovsky_atmop_329, kolar_prl_090601, brandner_pre_062134, hofer_njp_123037},
\begin{equation}
H(t)=H_s(t)+H_E+H_I,
\end{equation}
where $H_s(t+T)=H_s(t)$. A dynamical equation in Linbladian form can be derived in a way similar to the case of static Hamiltonian discussed above, albeit with some alterations. The essential ingredient in the case of periodic Hamiltonians is the so called Floquet \cite{floquet_sens_47, bukov_ap_139, eckardt_rmp_011004} Hamiltonian $H_F$, {which is defined as follows. Consider the unitary evolution operator over a single time period $T$, in the absence of any coupling to the environment,
\begin{equation}
U_s(T,0)=\mathcal{T}e^{-i\int_0^TH_s(t)dt}=e^{-iH_FT},
\end{equation}
where $\mathcal{T}$ is the time-ordering operator and the Floquet Hamiltonian $H_F$ is defined as,
\begin{equation}
H_F=\frac{i}{T}\ln U_s(T,0).
\end{equation}
The Floquet Hamiltonian $H_F$ thus acts as an effective static Hamiltonian which drives the evolution of the system when observed at stroboscopic instants of time, i.e., $U_s(mT,0)=\exp\left(-imH_FT\right)$. It possesses a set of time-independent \textit{quasi-energy} eigen-states $\ket{\phi_n}$, which satisfy $H_F\ket{\phi_n}=\varepsilon_n\ket{\phi_n}$, where $\varepsilon_n$ are refereed to as the quasi-energies.}

Now, let us consider the time evolution operator generating the evolution up to an arbitrary time,
\begin{equation}
U_s(t,0)=U_s(t,0)e^{iH_Ft}e^{-iH_Ft}=R(t,0)e^{-iH_Ft},
\end{equation} 
where $R(t)=U_s(t,0)e^{iH_Ft}$. Using the relations $U_s(mT,0)=\exp(-imH_FT)$ and $U_s(mT+t',mT)=U_s(t',0)$ where $0<t'<T$, one can easily verify that $R(t+T)=R(t)$. {Thus, $R(t)$ has a discrete time-translational invariance}, which permits its Fourier decomposition as,
\begin{equation}
U_s(t)=\sum_qR(q)e^{-iq\Omega t}e^{-iH_Ft}, 
\end{equation}
where $\Omega=2\pi/T$ is the frequency of the periodic modulation and
\begin{equation}
R(q) = \frac{1}{T}\int_0^TR(t)e^{iq\Omega t}dt.
\end{equation}
The operators $S_i$ are transformed in the interaction picture as,
\begin{align}
\tilde{S}_i&=U^\dagger_s(t,0)S_iU_s(t,0)\nn\\
		   &=e^{iH_Ft}\sum_{q',q}\left(R^\dagger(q')S_iR(q)e^{-i(q-q')\Omega t}\right)e^{-iH_Ft}\nn\\
		   &=\sum_me^{iH_Ft}S(m)e^{-im\Omega t}e^{-iH_Ft}\nn\\
		   &=\sum_{m}\sum_{\omega=\varepsilon_{n'}-\varepsilon_{n}}\bra{\phi_n}S_i(m)\ket{\phi_{n'}}e^{-im\Omega t}e^{i(\varepsilon_n-\varepsilon_{n'})t}\nn\\
		   &=\sum_{m,\omega}S_i(m,\omega)e^{-i(\omega+m\Omega)t}
\end{align} 
where $\omega=\varepsilon_{n'}-\varepsilon_n$ now denotes the difference in quasi-energies or eigen-values of the Floquet Hamiltonian $H_F$. This is unlike the static case where $\omega$ refereed to the difference in energy eigen-values of the time independent $H_s$.
The Hamiltonian $H_I$ hence assumes a similar form in the interaction picture as in Eq.~\eqref{eq_hi_int},
\begin{equation}
\hi_I=\sum_{i,\omega,m}e^{-i(\omega +m\Omega)t}S_i(m,\omega)\otimes B_i(t),
\end{equation}
where $B_i(t)$ is given by Eq.~\eqref{eq_bi}. Assuming the same set of approximations as in Sec.~\ref{subsec_open_static}, one arrives at the following equation of the super-operator $\mathcal{L}$ ,
\begin{equation}\label{eq_gksl_periodic}
\tilde{\mathcal{L}}\rhoi_s(t)=-i\left[H_{LS},\rhoi_s(t)\right]+\\\sum_{m,\omega,i,j}\gamma_{i,j}(\omega+m\Omega)\Big(S_i(m,\omega)\rhoi_s(t)S_j^\dagger(m,\omega)\\-\frac{1}{2}\{S_j^\dagger(m,\omega)S_i(m,\omega),\rhoi_s(t)\}\Big),
\end{equation}
where $H_{LS}$ commutes with the Floquet Hamiltonian $H_F$ and is given by,
\begin{equation}
H_{LS}=\sum_{m,\omega,i,j}\eta_{i,j}(\omega+m\Omega)S_j^\dagger(m,\omega)S_i(m,\omega).
\end{equation}

Unlike the static case, transforming the GKSL equation to the Schrodinger picture is not trivial in general. Instead, we first obtain the steady state in the interaction picture itself by solving $\tilde{\mathcal{L}}\rhoi_{ss}=0$. The corresponding state in the Schrodinger picture is given by,
\begin{equation}
\rho_{ss}(t)=U_s(t,0)\rhoi_{ss}U_s^\dagger(t,0), 
\end{equation} 
and satisfies $\rho_{ss}(t+T)=\rho_{ss}(t)$ as $U_s(t+T,0)=U_s(t,0)$. Hence, the steady state is periodic in nature and is in sync with the modulating Hamiltonian.

\section{Continuous thermal machines}\label{sec_conti}

The three-level maser discussed in Sec.~\ref{subsec_maser}, apart from being the earliest prototype of quantum thermal machines, can also be identified as the simplest form of \textit{continuous thermal machines} \ct{geusic_pr_343, geva_pre_3903,kosloff_arpc_365, mukherjee_pre_062109} (CTMs). This class of thermal machines are characterized by their perpetual (continuous) coupling with both the heat source as well as the sink, unlike the reciprocating class of machines discussed in the next section. CTMs have a greater experimental relevance as they, unlike their reciprocating counter-parts, do not require intermittent couplings and decouplings between the working fluid and the baths that are particularly difficult to implement at microscopic scales. Moreover, such intermittent coupling-decoupling mechanisms are bound to generate some finite energetic-costs on the performance of reciprocating machines which are mostly ignored in theoretical calculations. 

Work extraction or refrigeration in CTMs is usually enforced by a periodic modulation of the system Hamiltonian, which in general, drives the system to a periodic steady state. The dynamical exchanges of energy between the system and the baths as well as the \textit{work reservoir} (energy source of the external agent which modulates the system) in the steady state are, in general, out-of-equilibrium processes. In this regard, the three-level maser can be considered as a special case of CTMs in which a dynamical equilibrium {(populations of energy levels held steady) is assumed throughout the energy conversion process of each quanta of excitation}, thereby rendering the operation perfectly reversible. On the other hand, for a generic out-of-equilibrium process, the dynamics is irreversible in nature and the performance is found to be worse than the reversible Carnot engine or refrigerator. In what follows, we shall adopt the master equation approach, detailed in Sec.~\ref{subsec_open_periodic}, to analyze the performance of a simple CTM operating in the steady state. 

For purpose of simplicity, let us consider a periodically modulated two-level system (TLS) in contact with two thermal baths having temperatures $T_h$ and $T_c$ ($T_h>T_c$) \ct{klimovsky_pre_012140}. The Hamiltonian of the composite system reads,
\begin{equation}
H(t)=H_s(t)+H_h+H_c+H_I,
\end{equation} 
where $H_s(t)$ and $H_h(H_c)$ are the Hamiltonians of the system and the hot (cold) bath, respectively. The modulation is performed on the energy-gap of the TLS so that $H_s(t)$ is of the form,
\begin{equation}\label{eq_cont_hs}
H_s(t)=\frac{1}{2}\omega_s(t)\sigma_z,
\end{equation}
with $\omega_s(t+T)=\omega_s(t)$ and $\sigma_z$ being a Pauli matrix. The interaction between the system and the baths is chosen to be,
\begin{equation}\label{eq_cont_hi}
H_I=\sigma_x\otimes(B_h\otimes\mathcal{I}_c+\mathcal{I}_h\otimes B_c),
\end{equation}
where $\sigma_x$ is a Pauli matrix, while $B_h$ and $B_c$ are Hermitian operators which act locally on the Hilbert spaces of the hot and cold baths, respectively. Note that the modulation imposed is such that $\left[H_s(t), H_s(t')\right]=0$, which ensures that the external driving only modulates the energy levels and do not generate any excitations in the system. On the other hand, the interaction between the system and the baths is chosen such that  $\left[H_s(t),H_I\right]\neq 0$ and thus the baths can induce excitations and affect the population of the energy levels. 

On plugging in Eqs.~\eqref{eq_cont_hs} and~\eqref{eq_cont_hi} in the GKSL equation derived Eq.~\eqref{eq_gksl_periodic} and simplifying, we obtain,
\begin{subequations}\label{eq_cont}
\begin{equation}
\frac{\partial\rho(t)}{\partial t}=\mathcal{L}\rho(t)=\sum_{j,m}\mathcal{L}_{m}^j\rho(t),
\end{equation} 
where,
\begin{equation}
\mathcal{L}_{m}^j\rho=P_m\Big(\gamma^j(\omega_0+m\Omega)(\sigma^-\rho\sigma^+-\frac{1}{2}\{\sigma^+\sigma^-,\rho\})\\+\gamma^j(-\omega_0-m\Omega)(\sigma^+\rho\sigma^--\frac{1}{2}\{\sigma^-\sigma^+,\rho\})\Big),
\end{equation}
\end{subequations}
where we have ignored the Lamb-shift corrections to the energy levels. The superscript $j=h,c$ label the operators or correlation functions defined on the hot and cold baths, respectively. Likewise, $\omega_0$ denotes the mean gap of the two-level system averaged over $T$, $\Omega=2\pi/T$ is the frequency of modulation and $m=0,\pm 1, \pm 2\dots$ corresponds to the different photon sectors or side-bands created as a result of the modulation. The coefficient $P_m$ assigns a weight to the contribution from the $m^{th}$ side-band and is given by, 
\begin{equation}
P_m = \left|\frac{1}{T}\int_0^Te^{-i\int_0^t\left(\omega_s(t')-\omega_0\right)dt'}e^{-im\Omega t}dt\right|^2
\end{equation}
Note that, unlike in the previous subsections, we have not used $\tilde{O}$ to denote an operator $O$ in the interaction picture for simplicity in notation. 

A closer look at Eq.~\eqref{eq_cont} suggests that the effective action of the periodic modulation in conjugation with the coupling to thermal baths can be interpreted as a dissipative evolution driven by infinite copies of each of the thermal baths. Each of the copies, which we henceforth refer to as sub-baths, of a particular thermal bath couples to different side-bands of the Floquet spectrum.  The super-operator $\mathcal{L}^j_m$ encodes the dissipative action  arising due to the coupling of one of the copies of the $j^{th}$ bath to the $m^{th}$ side-band. The sub-baths therefore induce excitations in the system having energies equal to the energy gaps of the different side-bands. As discussed in Sec.~\ref{subsec_open_periodic}, the steady state $\rho_{ss}$ in the interaction picture is obtained by solving the eigenvalue equation $\mathcal{L}\rho_{ss}=0$, which transforms to a periodic steady state in the Schrodinger picture. 

A crucial assumption in determing the steady state is that the thermal baths  satisfy the Kubo-Martin-Schwinger condition $\gamma^j(-\omega)=e^{-\beta_j\omega}\gamma^j(\omega)$, where $\beta_j=1/T_j$ is the inverse temperature of $j^{th}$ bath. The steady state is then found to be,
\begin{subequations}\label{eq_ctm_steady}
	\begin{equation}
	\rho_{ss}=\frac{1}{1+r}\begin{pmatrix}
	r & 0\\0 & 1\end{pmatrix},
	\end{equation}
where,
	\begin{equation}\label{eq_r_ori}
	r=\frac{\sum_{m,j}P_m \gamma^j\left(\omega_0+m\Omega\right)e^{-\frac{\omega_0+m\Omega}{T_j}}}{\sum_{m,j}P_m \gamma^j\left(\omega_0+m\Omega\right)},
	\end{equation} 
\end{subequations}
To determine the heat currents and work, we first note that each sub-bath, when acting independently on the system, can in principle drive the system to a Gibbs-like steady state determined by the eigenvalue equation $\mathcal{L}_{m}^j\rho_{m,ss}^j=0$. These steady states are of the form \cite{alicki_arxiv_4552, kosloff_arpc_365, klimovsky_atmop_329},
\begin{equation}\label{eq_cont_sub_steady}
\rho_{m,ss}^j=\frac{1}{\mathcal{Z}}\exp{\left(\frac{\omega_0+m\Omega}{\omega_0}\beta_j H_{F}\right)},
\end{equation}
where $H_F=\omega_0\sigma_z/2$ and $\mathcal{Z}=\Tr\left(\exp{\left(\frac{\omega_0+m\Omega}{\omega_0}\beta_j H_{F}\right)}\right)$. Next, we calculate the rate of change of von-Neumann entropy $S(t)=-\Tr\left(\rho(t)\ln\rho(t)\right)$,
\begin{equation}
\frac{d S(t)}{dt}=-\Tr\left(\dot{\rho}(t)\ln\rho(t)\right)=-\sum_{j,m}\Tr\left(\mathcal{L}_{m}^j\rho(t)\ln\rho(t)\right).
\end{equation}
Further, it follows from from Spohn's inequality \ct{spohn_jmp_1227} for Markovian dynamics that $\Tr\left(\mathcal{L}_{m}^j\rho(\ln\rho-\ln\rho_{m,ss}^j\right)\leq 0$. Substituting this inequality in the above equation yields,
\begin{equation}
\frac{d S(t)}{dt}\geq-\sum_{j,m}\Tr\left(\mathcal{L}_{m}^j\rho(t)\ln\rho^j_{m,ss}\right)=\sum_j\frac{J_j(t)}{T_j}.
\end{equation}
The above inequality can be considered to be a dynamical version of the second law. In the steady state, the Von-Neumann entropy maximizes, and we obtain,
\begin{equation}
\sum_j\frac{J_j}{T_j}=-\sum_{j,m}\Tr\left(\mathcal{L}^j_{m}\rho_{ss}\ln \rho^j_{m,ss}\right)\leq 0.
\end{equation}
Substituting Eq.\eqref{eq_cont_sub_steady} in the first equality above, the steady state heat currents can be identified as,
\begin{equation}
J_j=\sum_m\left(\frac{\omega_0+m\Omega}{\omega_0}\right)\Tr\Big(\mathcal{L}_{m}^j\rho_{ss}H_F\Big),
\end{equation}
while the power is calculated from the principle of energy conservation (first law),
\begin{equation}
P=\dot{W}=-\sum_jJ_j.
\end{equation}

The working and the mode of operation of the thermal machine discussed above depend crucially on the form of modulation as well as the bath spectral functions $\gamma^j(\omega)$. To act as a heat engine or refrigerator, the two baths need to be spectrally separated \ct{klimovsky_pre_012140, klimovsky_atmop_329}. As for example, consider the case of a sinusoidal modulation of the system with the bath spectra separated as $\gamma^h(\omega)=0~\forall\omega<\omega_0$, $\gamma^c(\omega)=0~\forall\omega>\omega_0$. One can then show that the QTM operates as a refrigerator, i.e. $J_h<0$ and $J_c$, $P>0$, if the modulation frequency $\Omega>\Omega_{cr}$, where the critical frequency $\Omega_{cr}$ given by \ct{klimovsky_pre_012140},
\begin{equation}
\Omega_{cr}=\omega_0\frac{T_h-T_c}{T_h+T_c}.
\end{equation}
For $\Omega<\Omega_{cr}$, the QTM operates as a heat engine, characterized by reversal of the signs of the heat currents and power, i.e. $J_h>0$ and $J_c$, $P<0$. In this regime, the efficiency is found to be
\begin{equation}
\eta=\frac{2\Omega}{\omega_0+\Omega},
\end{equation}
The maximum efficiency is achieved as $\Omega\to\Omega_{cr}$; at $\Omega=\Omega_{cr}$ the machine achieves the Carnot efficiency $\eta=1-T_c/T_h$, but the power as well as the heat current vanishes. Similarly, in the refrigerator regime, the COP is also found to be limited by the Carnot bound. Note that, although the efficiency and COP in general depends upon the choice of bath spectral function, they are nevertheless, always restricted by the Carnot bounds. Recently, it has also been demonstrated that using an asymmetric pulse \cite{mondal_arxiv_08509,ono_prl_207703}, the switching between different modes of operation can also be achieved by tuning the up (or equivalently down) time duration of the pulse. Additionally, when modulated at resonance $\Omega=\omega_0$, tuning the up time duration also allows to QTM to function as a heater. In this mode of operation, the power supplied to the system $P>0$ is used to heat up both the thermal baths, i.e. $J_h<0$, $J_c<0$.

The above framework of continuous thermal machines with TLSs as working fluid has also been extended to the case of multi-level systems with degenerate excited states \cite{klimovsky_scirep_14413,niedenzu_pre_042123}. When compared with the performance of TLSs, it is found that the presence of degeneracy in the case of multi-level systems can boost the heat currents and power of the thermal machine; however, the efficiency or the coefficient of performance remains identical to that of the case of TLSs. Similarly, it has been shown that using $N$ two-level atoms as working fluid in place of the TLS enhances the power output (as well as cooling capacity in refrigerator mode) of the thermal machine when compared with the net power output from $N$ independent machines \cite{niedenzu_njp_113038}. Further, it has also been demonstrated that the hot bath can be cooled to very low temperatures \cite{klimovsky_pra_023431} if one considers a  modified version of the continuous thermal machine discussed above, with the system and interaction Hamiltonians chosen as,
\begin{subequations}
	\begin{equation}
	H_s(t)=\frac{1}{2}\omega_0(t)\sigma_z+\frac{1}{2}g\left(\sigma_+e^{-i\nu t}+\sigma_-e^{i\nu t}\right)
	\end{equation}
	and
	\begin{equation}
	H_I=\sigma_z\otimes B_h\otimes\mathcal{I}_c+\sigma_x\otimes\mathcal{I}_h\otimes B_c,
	\end{equation}
\end{subequations}
respectively. The first equation models a laser driven TLS \cite{szczygielski_pre_012120} with the coupling strength between the TLS and the laser as $g$, while the second equation describes the coupling of the TLS to a dephasing (does not induce transitions between energy levels of $\sigma_z$) hot bath and a cold bath. The particular advantage of this model is that, unlike the model governed by Eqs.~\eqref{eq_cont_hs} and~\eqref{eq_cont_hi}, no spectral separation of the two baths is required for the cooling operation. Finally, we note that the working of CTMs in the regime of non-Markovian dynamics has aslo been explored \cite{mukherjee_cp_8}.

\section{Reciprocating thermal machines}\label{sec_recip}
A reciprocating thermal machine operates in a cycle which is composed of discrete strokes. Notably, the working fluid is coupled to the baths only in some of the strokes of the cycle. This class of thermal machines are much simpler to analyze in comparison to continuous ones, although they are in general trickier to implement experimentally. The particle in box system, discussed in Sec.~\ref{subsec_box} is an example of reciprocating thermal machine. In spite of it's tantalizing similarity to the classical Carnot cycle \cite{carnot}, it does not offer much insight into the dynamical exchange of energy between the working fluid and the environment. In particular, the heat exchanged with the baths is indirectly found from the work performed in the isothermal strokes where the net energy change is zero. In this section, we will first discuss the common notions of quantum heat and work widely used in literature, which will be subsequently used to review some of the commonly studied reciprocating thermal machines.  

Although there are no universally accepted definitions of `quantum work' or `quantum heat' till date, we briefly underline here a set of definitions that are particularly useful in the context of reciprocating devices, especially, in the limit of weak system-bath coupling. Consider a system $S$ with reduced density matrix $\rho_s(t)$ evolving under a time-dependent Hamiltonian $H_s(t)$ and coupled to an external environment. A change in energy expectation value of $S$ after duration $\tau$ can be expressed as \ct{alicki_jpa_L103, vinjanampathy_ctp_545},

\begin{align}\label{eq_heatwork}
\Delta E&=\int_0^\tau\frac{d}{dt}\left[{\rm Tr}\left(\rho_s(t)H_s(t)\right)\right]dt\nn\\&=\int_{0}^{\tau}{\rm Tr}\left(\frac{d\rho_s(t)}{d t}H_s(t)\right)dt+\int_{0}^{\tau}{\rm Tr}\left (\rho_s(t)\frac{d H_s(t)}{d t}\right)dt\nn\\&=W+Q,
\end{align}
where the work is identified as,
\begin{subequations}
\begin{equation}\label{eq_qwork}
W=\int_{0}^{\tau}{\rm Tr}\left(\rho_s(t)\frac{d H_s(t)}{d t}\right)dt,
\end{equation}
and the heat exchanged with the environment as,
\begin{equation}\label{eq_qheat}
Q=\int_{0}^{\tau}{\rm Tr}\left(\frac{d\rho_s(t)}{d t}H_s(t)\right)dt.
\end{equation}
\end{subequations}
The above definitions of quantum work and heat can be intuitively justified as follows: If the system $S$ was isolated, any change in energy expectation can only be associated with a work performed, as there is no environment with which $S$ can exchange heat. This is consistent with the definition of heat in Eq.~\eqref{eq_qheat} as for an isolated system, one can easily check that,
\begin{align}
Q_{iso}&=\int_{0}^{\tau}{\rm Tr}\left(\frac{d\rho_s(t)}{d t}H_s(t)\right)dt\nn\\&=-i\int_{0}^{\tau}{\rm Tr}\left(\left[H_s(t),\rho_s(t)\right]H_s(t)\right)dt=0,
\end{align} 
where we have used the Liouville's equation to arrive at the second equality. Further, both the quantum work and heat, as defined above, depend on the evolution process and are therefore not state functions, similar to their classical counterparts. One can therefore, consider Eq.~\eqref{eq_heatwork} as the quantum equivalent of the first law. 

Further, the definitions in Eq.~\eqref{eq_heatwork} also permits a natural formulation of a dynamical version of the second law. Within the framework of GKSL equations derived for time-independent and slowly varying system Hamiltonians $H_s(t)$ in Sec.~\ref{sec_open}, the rate of heat exchanged can be calculated using  Eq.~\eqref{eq_qheat} as,
\begin{equation}
J(t)=\frac{dQ}{dt}={\rm Tr}\left(\frac{d\rho_s(t)}{d t}H_s(t)\right)={\rm Tr}\left(\mathcal{L}(t)\rho_s(t)H_s(t)\right)
\end{equation}
Next, we invoke the Spohn's inequality \cite{spohn_jmp_1227},
\begin{equation}
\Tr\left[\mathcal{L}(t)\rho_s(t)\left(\ln\rho_s(t)-\ln\rho_{ss}\right)\right]\leq 0,
\end{equation}
where $\rho_{ss}$ is the steady state which satisfies $\mathcal{L}\rho_{ss}=0$. For baths obeying the Kubo-Martin-Schwinger (KMS) condition \cite{kubo_jpsj_570, martin_pr_1342, haag_cmp_215}, $\rho_{ss}(t)$ corresponds to the thermal Gibbs state $\rho_{ss}(t)=e^{-\beta H_s(t)}/\Tr\left(e^{-\beta H_s(t)}\right)$, where $\beta=1/T$ is the inverse bath temperature. Substituting in the above equation, we arrive at,
\begin{equation}
\frac{dS(t)}{dt}-\frac{J(t)}{T}\geq 0,
\end{equation}
where $S(t)=-\Tr\left(\rho_s(t)\ln\rho_s(t)\right)$ is the von Neumann entropy of the system. The above equation is the dynamical version of the second law which we had also obtained in Sec.~\ref{sec_conti} for time-periodic Hamiltonians.
Finally, we would like to mention here that the above definitions of work and heat need to be modified in some cases, such as in the case of autonomous or self-contained quantum thermal machines \ct{tonner_pre_066118,niedenzu_q_195, brask_njp_113029, klimovsky_atmop_329, klimovsky_pre_022102, klimovsky_scirep_7809, strasberg_prl_180605, hammam_njp_043024}.
\subsection{Four stroke devices}\label{subsec_4stroke}
Carnot, Otto and Stirling engines are prime examples of four stroke thermal machines as their operation are based on cycles that are made up of four sequential strokes. The quantum analogues of these four stroke machines have been extensively studied, particularly the Otto engine has received the most attention followed by the Carnot engine. This is because the Otto cycle is the simplest to analyze owing to the fact that the heat and work exchanges occur separately in different strokes, unlike the Carnot and Stirling cycles, in which both the processes occur simultaneously in the isothermal strokes. In particular, note that the isothermal stroke requires a dissipative evolution with a time varying Hamiltonian. However, exact formulations in terms of the Lindbladian framework exist only for the cases of static and infinitesimally slowly varying Hamiltonians. Thus, the Lindbladian framework becomes inadequate when one is interested in studying finite-time performances of the thermal machines. On the contrary, the Otto cycle provides no such hindrance, as the natural separation of the work producing and heat exchanging strokes allows the former to be rephrased in terms of unitary evolution in isolated conditions and the latter in terms of dissipative dynamics with time-independent Hamiltonians.  

In the following, we therefore restrict ourselves to reviewing the working principles of the quantum Otto cycle \cite{vinjanampathy_ctp_545, kieu_prl_140403, quan_pre_031105, wang_pre_041113, he_scpma_1317, he_scpma_1751, zhang_prl_150602, campo_sr_6208, campisi_natcom_11895, kosloff_ent_136, huang_pre_051105, robnagel_prl_030602, manzano_pre_052120, thomas_pre_062108, pezzutto_qst_025002, halpern_prb_024203, hartmann_arxiv_08689, deng_pre_062122, henrich_epj_157, rezek_njp_83, thomas_pre_031135} using the definitions of quantum heat and work discussed previously. We note at the outset that a vast amount of research have gone into analyzing numerous aspects of the quantum Otto cycle and as such, a full-fledged discussion of all such work is not feasible. We therefore, resort to highlighting only fundamental aspects of its working and outline some of the interesting results that have been reported over the past few years.

\begin{figure}
	\includegraphics[width=0.4\textwidth]{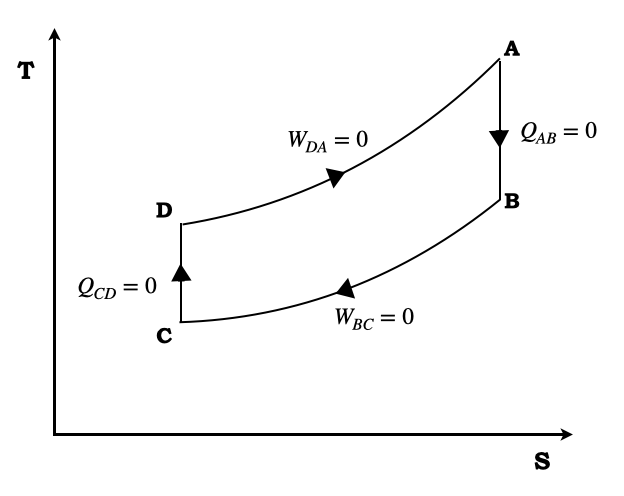}
	\caption{Schematic representation of the Otto cycle in the $T-S$ plane, where $T$ denotes the temperature and $S$ is the Von-Neumann entropy. The strokes AB and CD are isentropic strokes in which the energetic changes occur only in the form of work. On the other hand, the thermalization with the hot and cold baths occur in the strokes DA and BC, respectively, with no work performed during these strokes whatsoever. }\label{fig_otto}
\end{figure}

\subsubsection{The Otto Cycle}\label{subsubsec_otto}

For the Otto cycle {(see Fig.~\ref{fig_otto})}, we consider a quantum harmonic oscillator (QHO) as the working fluid with Hamiltonian,
\begin{equation}
H=\frac{\hat{p}^2}{2m}+\frac{1}{2}m\omega^2(t)\hat{x}^2=\omega(t)\left(a^\dagger a+1/2\right), 
\end{equation}
where $\hat{x}$ and $\hat{p}$ are the position and conjugate momentum operators, respectively. In addition, $a$ denotes the annihilation operator and $\omega$ is the natural frequency of the oscillator which can be controlled externally. The QHO can be coupled to a hot or a cold thermal bath, having temperatures $T_h$ and $T_c$ ($T_h>T_c$), respectively. Further, we assume that the initial frequency is $\omega(0)=\omega_A$, i.e., $H(0)=H_A=\omega_A \left(a^\dagger a+1/2\right)$, and the QHO is initialized in thermal equilibrium with bath $T_h$ so that, 
\begin{equation}\label{eq_otto_ini_den}
\rho_A = \frac{e^{-H_A/T_h}}{\mathcal{Z}(\omega_A,T_h)},
\end{equation}
where $\mathcal{Z}(\omega,T)={\rm Tr}(e^{-H_A/T})$. The energy expectation value is found to be,
\begin{equation}
\la E_A\ra=\Tr(\rho_AH_A)=\omega_A\left(\la n\ra+\frac{1}{2}\right)=\frac{\omega_A}{2}\coth \left(\frac{\omega_A}{2T_h}\right),
\end{equation}
where $\la n\ra$ is the mean occupation number.
The quantum Otto cycle is now constructed as follows:

\begin{enumerate}
	
\item \textit{Isentropic compression} -- The QHO is decoupled from the hot bath and its Hamiltonian is tuned from $H_A\to H_B$, where $H_B=\omega_B\left(a^\dagger a+1/2\right)$ and $\omega_B\leq\omega_A$. The unitary evolution in isolated conditions does not affect the Von-Neumann entropy of the QHO; hence it is an isentropic process. In the ideal cycle, the tuning occurs adiabatically, which ensures that the occupation probabilities of the instantaneous eigen-energy levels remain unchanged. The energy exchange, which importantly occurs only in the form of work, is  given by
\begin{equation}\label{eq_otto_W1}
W_{AB}=\frac{\omega_B-\omega_A}{2}\coth \left(\frac{\omega_A}{2T_h}\right).
\end{equation}

\item \textit{Cold isochore} -- In this stroke, the QHO is coupled to the cold bath and allowed to thermalize with its Hamiltonian held constant at $H=H_B$. By virtue of Eq.~\eqref{eq_qwork}, this ensures that no work is performed. The only energy exchange occurs in the form of heat transfer and is given by,
\begin{equation}
Q_c=\frac{\omega_B}{2}\left(\coth \left(\frac{\omega_B}{2T_c}\right)-\coth \left(\frac{\omega_A}{2T_h}\right)\right)
\end{equation}	

\item \textit{Isentropic expansion} -- As in the compression stroke, the QHO is decoupled from bath and the Hamiltonian is tuned as $H_B\to H_A$. Once again, in the ideal case, the populations remain invariant, and the work performed is found to be,

\begin{equation}\label{eq_otto_W2}
W_{BA}=\frac{\omega_A-\omega_B}{2}\coth \left(\frac{\omega_B}{2T_c}\right).
\end{equation}

\item \textit{Hot isochore} -- The cycle is completed by coupling the QHO to the hot bath so that it thermalises back to its initial state as given in Eq.~\eqref{eq_otto_ini_den}. The heat exchanged in the process is,
\begin{equation}
Q_h=\frac{\omega_A}{2}\left(\coth \left(\frac{\omega_A}{2T_h}\right)-\coth \left(\frac{\omega_B}{2T_c}\right)\right)
\end{equation}
\end{enumerate} 

We reemphasize that the work is performed only during the isentropes while heat is exchanged with the baths only during the isochores. As the QHO returns to its intial state after each cycle, the total energy is conserved, i.e. $Q_c+W_{AB}+Q_h+W_{BA}=0$. The efficiency can be calculated as,

\begin{equation}\label{eq_otto_eff}
\eta_o=-\frac{W_{AB}+W_{BA}}{Q_h}=\frac{Q_h+Q_c}{Q_h}=1-\frac{\omega_B}{\omega_A},
\end{equation} 
where we have used the convention that $W$ is positive if work is done on the system. When operating as a heat engine, the net work output is positive, i.e. $W_{AB}+W_{BA}\leq 0$. Using Eqs.~\eqref{eq_otto_W1} and~\eqref{eq_otto_W2}, this inequality is simplified to $\omega_B/\omega_A\geq T_c/T_h$. Hence the efficiency is limited by,

\begin{equation}
\eta_o\leq 1-\frac{T_c}{T_h}.
\end{equation}
Thus, we find that the efficiency of the quantum Otto engine is  bounded by the Carnot limit. As in the case of CTMs, one can check that at the Carnot point, $\omega_B/\omega_A=T_c/T_h$, the net work done as well as the heat exchanges vanish. For $\omega_B/\omega_A\leq T_c/T_h$, the machine operates as a quantum refrigerator, where the COP is also found to be upper-bounded by the COP of the Carnot refrigerator. 

\subsubsection{Efficiency at maximum power}
In practice, the power output of an ideal Otto engine, as in the case of Carnot engine, is zero. This is because of the fact that each of the adiabats as well as the  isochores ideally requires an infinite time to achieve perfect adiabatic evolution and thermalization, respectively. The efficiency at maximum power \cite{esposito_prl_130602, esposito_prl_150603, esposito_pre_031117, broeck_prl_190602, broeck_prl_210602,wang_pre_042119,erdman_prb_245432,erdman_njp_103049,dann_arxiv_02801} is therefore an important figure of merit to analyze the engine's performance. 

{The power of a heat engine can be maximized in two ways. In the first case, let us assume that each of the strokes are carried out over arbitrarily long but finite times so as to achieve near perfect adiabaticity and thermalization. Maximizing the power in this case thus amounts to maximizing the net work output. To evaluate the efficiency under such a restraint,} we expand the net work output in the high temperature limit as,
\begin{align}
W &= -\left(W_{AB} + W_{BA}\right)= \left(\omega_B-\omega_A\right)\left(\frac{T_c}{\omega_B}-\frac{T_h}{\omega_A}\right)\nn\\
&=\left(\frac{\omega_B}{\omega_A}-1\right)\left(\frac{T_c}{\left(\frac{\omega_B}{\omega_A}\right)}-T_h\right)
\end{align}
$W$ is maximum when the ratio $\omega_B/\omega_A$ satisfies $\omega_B/\omega_A=\sqrt{T_c}/\sqrt{T_h}$. The efficiency at maximum power $\bar{\eta}$ is thus,
\begin{equation}\label{eq_otto_max}
\bar{\eta}=1-\sqrt{\frac{T_c}{T_h}},
\end{equation}
which is surprisingly identical to the Curzon-Ahlborn (CA) efficiency \cite{curzon_ajp_22} of an ideal engine operating with finite power.

{However, a more meaningful way to achieve finite power {at maximum efficiency} is to devise ways that can lead to both perfect adiabaticity and thermalization within shorter times}. The time required in the former is in general comparatively much longer than the latter; a greater effort has thus been devoted to engineer methods to achieve adiabatic evolutions in finite time, which we shall be briefly discussing below. Nevertheless, we note that protocols to achieve fast thermalization have also been explored recently \cite{dann_prl_250402, dann_pra_052102, dann_njp_013055}.

\subsubsection{Quantum friction and shortcuts to adiabaticity}

The fact that the strokes associated with work extraction are required to be adiabatic is better explained in terms of quantum coherence. Diabatic excitations, associated with the build up of coherence in between the energy levels of the system, are generated when the time allocated to the isentropic processes is finite and $\left[H(t),H(t')\right]\neq 0$. The build up of coherence costs additional work, which effectively reduces the net useful work extracted from the heat reservoirs, thus undermining the efficiency of the engine. Coherence therefore plays the role of `quantum friction' \cite{plastina_prl_260601, francica_pre_042105} which hampers the engine's ability to extract useful work. 

A powerful technique which can mitigate work losses due to non-adiabatic driving is to utilize certain \textit{shortcuts to adiabaticity} \cite{chen_prl_063002, torrontegui_atmop_117, chen_pra_053403, masuda_prsa_1135, cui_jpca_2962, campo_sr_6208, dupays_prr_033178, funo_prl_150603}. One way in which this can be realized is by driving the working fluid along a certain path which ensures  that the final state reached at the end of the isentropic stroke does not have any coherence although they may exist at intermediate times and the populations of the energy eigen-states are thus the same at the beginning and at the end of the stroke. To illustrate, let us consider an isentropic stroke in which the frequency of the QHO is tuned from $\omega(0)=\omega_i$ to $\omega(\tau)=\omega_f$ in a finite duration of time $\tau$.  Now, consider the operator,
\begin{equation}
I(t) = \frac{1}{2}\left[\frac{m\omega_0^2}{b^2}\hat{x}^2+\frac{1}{m}\left(b\hat{p}-m\dot{b}\hat{x}\right)^2\right],
\end{equation}
where $b$ is a time-dependent parameter and $\omega_0=const>0$. This operator becomes an invariant of evolution if the parameter $b$ satisfies the Ermakov equation,
\begin{equation}\label{eq_ermakov}
\ddot{b} + \omega(t)^2b=\frac{\omega_0^2}{b^3}.
\end{equation}
Let us now impose the boundary conditions $b(0)=1$, $\dot{b}(0)=\ddot{b}(0)=0$ and $b(\tau)=\sqrt{\omega_0/\omega_f}$, $\dot{b}(\tau)=\ddot{b}(\tau)=0$. These choices of boundary conditions lead to $\omega_0=\omega_i$, $I(0)=H(0)$ and $I(\tau)=\omega_0H(\tau)/\omega_f$. The set of these six boundary conditions allows one to calculate a polynomial form of $b(t)$ from which the required time-dependent tuning of $\omega(t)$ can be determined. 

The above results imply that an initial set of eigen-states of the Hamiltonian $H(t)$ are also the eigen-states of the operator $I(t)$ at $t=0$. Let us consider a generic initial state $\ket{\psi(0)}=\sum c_n(0)\ket{\phi_n}$, where $\ket{\phi_n}$ are the eigen-states of both $I(0)$ and $H(0)$. If $\omega(t)$ is subsequently tuned following Eq.~\eqref{eq_ermakov}, $I(t)$ remains invariant and the state evolves as $\ket{\psi(t)}=\sum c_n(0)e^{i\theta_n(t)}\ket{\phi_n(t)}$, where $\ket{\phi_n(t)}$ are instantaneous eigen-states of $I(t)$ but not $H(t)$ ($0<t<\tau$) and $\theta(t)$ is some time-dependent phase. Finally, since  $\ket{\phi_n(\tau)}$ are also the eigen-states of $H(\tau)$, we conclude that the population of each of the eigen-states are restored at $t=\tau$. 

An alternative but similar approach to adiabatic shortcuts is provided by counter-diabatic (CD) protocols \cite{hartmann_arxiv_08689, abah_pre_032121, abah_prr_023120, abah_pre_022110, cakmak_pre_032108, funo_prb_035407}. These protocols involve adding additional interactions to the system Hamiltonian; the dynamics resulting from the inclusion of these additional CD interactions suppresses the diabatic excitations which would have been otherwise generated in the system due to fast driving of the bare Hamiltonian. A general form of the CD interactions is obtained as follows. Consider a system Hamiltonian of the form,
\begin{equation}
H_0(t) = \sum_n\varepsilon_n(t)\ket{n(t)}\bra{n(t)}.
\end{equation}
Under adiabatic driving, the system initialized in a given energy eigen-state at $t=0$ follows the same instantaneous state throughout the evolution. The time evolution operator is therefore required to be of the form,
\begin{equation}
U(t)=\sum_ne^{i\phi_n(t)}\ket{n(t)}\bra{n(0)}.
\end{equation}
where $\phi_n(t)$ is the phase acquired by the $n^{th}$ eigen-state during the unitary evolution and is given by,
\begin{equation}
\phi_n(t)=-\int_0^tdt'\left(\varepsilon_n(t')-i\braket{n(t')|\partial_{t'}n(t')}\right).
\end{equation}
We wish to find the Hamiltonian which mimics the above time evolution without any adiabatic approximations. This is easily done by substituting $U(t)$ in the (Schrodinger) equation, $H(t)=i\dot{U}(t)U^\dagger(t)$, which leads to,
\begin{equation}
H(t)=H_0(t)+H_{CD}(t)=\sum_n\varepsilon_n(t)\ket{n(t)}\bra{n(t)}+i\sum_n\Big(\ket{\partial_tn(t)}\bra{n(t)}-\braket{n(t)|\partial_tn(t)}\ket{n(t)}\bra{n(t)}\Big),
\end{equation}
where $H_{CD}$ encodes the CD interactions that are to be added to ensure an adiabatic evolution with reference to the bare Hamiltonian $H_0$. Of course, one must be careful to take into account the extra work done by the additional counter-diabatic terms while calculating the net work output and efficiency of the thermal machine. 

{Finally, it turns out that when coherence is generated in a quantum Otto engine during the isentropic strokes, an incomplete thermalization in between the two isentropic strokes can lead to a better engine performance.  An incomplete thermalization, typically realized by allowing the working fluid to interact with the hot bath for a short time during the isochoric strokes, fails to destroy the coherence generated in the energy basis during the preceeding isentropic stroke. This leads to an interference like effect between this residual coherence and the coherence generated in the subsequent isentropic stroke after the incomplete thermalization step. This has been explicitly demonstrated in Ref.~[\onlinecite{camati_pra_062103}], where the authors analyze a quantum Otto engine operating with a TLS as working medium, which is allowed to thermalize only partially during the hot isochoric stroke. The performance of this engine is compared to a second one which additionally incorporates a dephasing process after the incomplete thermalization step, so as to erase any residual coherence. Through numerical analysis, the authors demonstrate that a careful tuning of the finite times allocated to the isentropic and thermalization strokes can lead to a better power output and efficiency in the first engine when compared to the second one. }

\subsubsection{Non-thermal baths}\label{subsubsec_nonthermal}
It is important to realize that coherence is not always detrimental to the performance of QTMs. In fact, coherence plays the central role in boosting the work output of QTMs which utilize non-thermal baths as heat sources. In 2003, Scully \textit{et. al.}\cite{scully_sc_862} demonstrated that  it is possible to extract work effectively from a single `phaseonium' bath, which consist of three-level atoms that have small amount of coherence between almost- degenerate lower energy levels, . Dubbed as the photo-Carnot engine, the working fluid here is the radiation field generated by the atoms operating between a phaseonium \cite{scully_cambridge_book} bath with temperature $T_h$ and a bath $T_c$ with $T_h>T_c$. The working fluid relaxes to a thermal steady state with a temperature $T_{\phi}=T_h(1-\bar{n}\epsilon\cos\phi)$, where $T_h$ is the temperature of the  hot bath, $\bar{n}$ is the average photon number in the absence of coherence, $\epsilon$ is a measure of the magnitude of the coherence and $\phi$ is the associated phase. By appropriately tuning $\phi$, it is therefore possible for the working fluid to attain a higher temperature than the hot bath. This in turn allows work extraction even when $T_h=T_c$, thus effectively permitting work extraction from a single reservoir. Similar results were also reported in the context of quantum Otto engines, where the use of such `quantum coherent fuels' was shown to enhance performance \cite{hardal_sr_12953}.

A slightly different mechanism by which the use of non-thermal baths can boost engine performance \cite{abah_epl_20001}is when the working fluid itself is rendered non-thermal after interaction with the bath. This can be done, for example, by using  a squeezed thermal bath \cite{klaers_prx_031044, huang_pre_051105, rossnagel_prl_030602, manzano_pre_052120, niedenzu_njp_083012,oliveira_arxiv_02049} as the hot bath with squeezing parameter $r$. A QHO coupled to such a bath thermalizes to a squeezed thermal state with mean phonon number \cite{rossnagel_prl_030602}  $\la n\ra=\la n\ra_0+(2\la n\ra_0+1)\sinh^2(r)$. Returning to the analysis of the Otto cycle in Sec.~\ref{subsubsec_otto}, we note that the initial energy expectation $E_A$ is modified as, 
\begin{equation}
\la E_A\ra  = \frac{\omega_A}{2}\coth\left(\frac{\omega_A}{2T_h}\right)\Delta H_r,
\end{equation}
{where $\Delta H_r = 1+\left(2+1/\la n_0\ra\right)\sinh^2(r)$ with $\langle n_0\rangle=[\exp(\omega_A/T_h)-1]^{-1}$. Proceeding as in Sec.~\ref{subsubsec_otto}, the total work done is found to be,
\begin{subequations}
\begin{equation}
	W = \frac{\left(\omega_B-\omega_A\right)}{2}\left[\coth\left(\frac{\omega_B}{2T_c}\right)-\coth\left(\frac{\omega_A}{2T_h}\right)\Delta H_r\right],
\end{equation}
while the heat extracted is given by,
\begin{equation}
	Q_h = \frac{\omega_A}{2}\left[\coth\left(\frac{\omega_A}{2T_h}\right)\Delta H_r-\coth\left(\frac{\omega_B}{2T_c}\right)\right].
\end{equation}
\end{subequations}}

{The efficiency therefore turns out to be the same as the Otto engine efficiency $\eta_o=1-\omega_B/\omega_A$. However, in the high temperature limit, we have $\Delta H_r = 1+2\sinh^2(r)$ and the total work can be approximated as, 
\begin{equation}
	W = \left(\frac{\omega_B}{\omega_A}-1\right)\left(\frac{T_c}{\left(\frac{\omega_B}{\omega_A}\right)}-T_h\left(1+2\sinh^2(r)\right)\right)
\end{equation}
The efficiency at maximum power is then found by maximizing the work with respect to $\omega_B/\omega_A$ and is given by,
\begin{equation}
\bar{\eta} = 1-\sqrt{\frac{T_c}{T_h\left(1+2\sinh^2(r)\right)}}
\end{equation}}
For $r=0$, $\bar{\eta}$ reduces to the CA efficiency obtained in Eq.~\eqref{eq_otto_max} for the case of thermal reservoirs. On the other hand, for $r\to\infty$, we find $\bar{\eta}\to1$, which appears to surpass the Carnot limit. Nevertheless, this does not violate the second law as the non-thermal hot bath is found to have an effective higher temperature  $T_h^*>T_h$. On properly accounting for this effective temperature, the upper bound of the efficiency is found to be bounded by the generalized Carnot limit,
\begin{equation}
\eta_{gen}=1-\frac{T_c}{T_h\left(1+2\sinh^2(r)\right)}.
\end{equation}
Using squeezed thermal baths, it was also shown that work extraction is possible even from a single squeezed bath without violating the laws of thermodynamics \cite{klaers_prx_031044}. More recently, the efficiency bound on such quantum thermal machines has been quantitatively estimated using the notion of ergotropy, which quantifies the maximum amount of work that can be extracted from non-passive states through unitary protocols \cite{niedenzu_natcom_165}. 

\subsubsection{Four-stroke QTM based on many-body systems}
An increasing amount of research in recent times is focusing on deploying quantum many-body quantum systems as working fluid \cite{beau_ent_168, diaz_pra_032327, hartmann_arxiv_08689, campisi_natcom_11895, jaramillo_njp_075019, halpern_prb_024203, klimovsky_pra_022129, chen_npjqi_88, lekscha_pre_022142, ma_pre_022143, niedenzu_njp_113038,hartmann_arxiv_09327} in quantum stroke engines. This is because many-body systems, besides having the potential to naturally scale up the work output and efficiency per cycle of QTMs, are also capable of hosting certain novel phenomena which have no single-particle counterparts and  can serve as thermodynamic resources. As for example, finite size scaling theory predicts \cite{campisi_natcom_11895} that if the working fluid is operated closed to its critical point, the efficiency of the Otto engine can approach the Carnot limit at finite power. Similarly, using results known from energy-level statistics and localization properties  of many-body localized phases, it has been shown that a quantum Otto engine operated with the working fluid ramped between a localized and a thermal phase has significant advantages \cite{ halpern_prb_024203}. In particular, this engine exhibits lesser fluctuations in work output and can be easily scaled up in size as the localization ensures that different `sub-engines' work independently of each other. At the same time, it is to be noted that ensuring adiabatic driving protocols is a more challenging task in many-body systems as compared to single-particle systems. Recent works \cite{beau_ent_168, hartmann_arxiv_08689} have therefore focused on exploring viable shortcuts to adiabaticity protocols for many body quantum heat engines.

A remarkable feature which emerges in many-body quantum engines is that non-adiabatic affects in some cases may even lead to enhancement of the engine's performance. Such an enhancement has been demonstrated in the case of an Otto engine where the working fluid is an interacting Bose gas confined in a time-dependent harmonic trap \cite{jaramillo_njp_075019}. The efficiency achieved using the many-particle system is greater than the efficiency of an ensemble of single particle heat engines that have the same amount of thermodynamic resources at their disposal. Another mechanism \cite{klimovsky_pra_022129} though which non-adiabatic affects can be exploited to tap into the cooperative resources of many-body systems is tied to the notion of passive states \cite{pusz_cmp_273, lenard_jsp_575, allahverdyan_epl_565, campaioli_springer_book}. Technically, passive states are characterized by density matrices which are diagonal in energy basis and the population decrease with increase in energy. If a system is initially prepared in a passive state, then no work can be extracted out of it through cyclic unitary protocols. In a single-particle Otto engine, the final states after the completion of the isentropic strokes are passive states and the efficiency, as we have seen, is maximized when the strokes are adiabatic.  However, it can be shown that the direct product state of multiple identical copies of a passive state, which is not additionally thermal, need not be a passive \cite{alicki_pre_042123}. This opens up the possibility of extracting extra amount of work in many-body systems. For maximizing efficiency, the direct product state at the end of the isentropic strokes is required to be passive, which in turn necessitates non-adiabatic excitations so that the populations among different copies can be interchanged. We will return to passive states in more detail when we address quantum batteries in Sec.~\ref{sec_battery}.

\subsubsection{Non-Markovian QTMs}

In all the examples of four-stroke thermal machines considered thus far, the underlying assumption is that the dynamics of the QTM is strictly Markovian in nature. However, significant new results have also been reported recently for QTMs operating in the non-Markovian regime \cite{pezzutto_qst_025002, thomas_pre_062108, klimovsky_pra_022112, abiuso_pra_052106, erez_nat_724,kerstjens_njp_043034,  das_prr_033083}. In general, non-Markovian dynamics can result in a number of scenarios; such as in the limit of strong system-bath couplings and long decay times of bath correlation functions which implies a bath with memory. 

In Ref [\onlinecite{klimovsky_pra_022112}], it was shown that frequent quantum nondemolition measurements can lead to extraction of useful work from the system-bath correlation energy if the cycle is operated within the bath memory time. Similarly, it was pointed out in Ref [\onlinecite{thomas_pre_062108}] that in an Otto cycle with a TLS as the working fluid, thermalizing with a non-Markovian bath is not necessarily accompanied by a monotonous increase or decrease in the effective temperature of the TLS. This in turn allows the Otto engine to attain an efficiency {which can apparently} exceed the Carnot limit when operated in finite time. {However, this apparent violation of the second law is resolved when the heat exchanged with the non-Markovian reservoirs is appropriately redefined by incorporating the minimum costs associated with  the non-Markovianity of the dynamics.} 

{Another key result demonstrated in Ref.~[\onlinecite{abiuso_pra_052106}] was that even if the working fluid shares correlations with only some degrees of freedom with the baths with the overall evolution remaining Markovian, the power output can still be boosted. In this regard, it is interesting to note a different model of a quantum Otto engine explored in Ref.~[\onlinecite{newman_pre_032139}]. This engine consists of a TLS strongly coupled with multimode harmonic oscillator reservoirs and can be mapped to a similar setup as studied in Ref.~[\onlinecite{abiuso_pra_052106}]. Through the so-called reaction-coordinate mapping \cite{smith_pra_032114, smith_jcp_044110}, the total system of the TLS and the harmonic oscillator environment can be mapped to an enlarged system in which the TLS couples strongly with a single collective mode (called RC) of the environment. Further, the collective mode in turn couples weakly to the residual environment. This allows the strongly coupled TLS-RC system to be treated exactly by tracing out the weakly coupled residual environment using the Born-Markov approximations. Using this formalism,  Ref.~[\onlinecite{newman_pre_032139}] investigated the consequences of strong-coupling between the TLS and the reservoirs, incorporating the additional costs associated with sudden coupling/decoupling between the TLS and the reservoirs which becomes prominent in a strongl coupled setup. Indeed, the authors find that the performance of the engine is lowered due to these additional costs.} 

In conclusion, it is fair to say that the impact of non-Markovian dynamics on the operation of QTMs is far from being fully explored and more research in this direction is expected in near future.

\subsection{Two stroke devices}\label{subsec_2stroke}
Apart from four-stroke thermal machines, it is also possible to construct reciprocating thermal machines based on \textit{two-stroke} cycles \ct{allahverdyan_pre_041118, allahverdyan_pre_051129,campisi_njp_035012, uzdin_prx_031044, campisi_jpa_345002, buffoni_prl_070603,piccione_pra_032211}. The reduction in number of strokes per cycle is compensated by increasing the number of systems or working fluids to two. As for example, consider two qubits $S_1$ and $S_2$ with energy gaps $\omega_1$ and $\omega_2$, respectively.  These qubits can be individually coupled to two thermal baths with temperatures $T_1$ and $T_2$, such that $T_1>T_2$. The two-stroke cycle consists of the following sequential strokes -- (1) a thermalization stroke in which the qubits $S_1$ and $S_2$ are coupled with the baths having temperatures $T_1$ and $T_2$, respectively, and allowed to thermalize; (2) a unitary stroke in which the two qubits interact with each other, with no contact whatsoever with the baths, and work is performed on the composite system. The thermal machine based on this two-stroke cycle can work in three different modes depending on the ratio of the energy gaps $\omega_1$ and $\omega_2$. The different modes are characterized by the relative sign of the heat gained from the hot (cold) bath $Q_h$ ($Q_c$) and the work performed $W$. These modes and their regime of operation are listed below:
\begin{itemize}
	\item Refrigerator ($Q_h<0$, $Q_c>0$, $W>0$), when $\omega_1/\omega_2>T_1/T_2$.
	\item Engine ($Q_h>0$, $Q_c<0$, $W<0$), when $1<\omega_1/\omega_2<T_1/T_2$.
	  
	\item Accelerator ($Q_h>0$, $Q_c<0$, $W>0$), when $\omega_1/\omega_2<1$.
\end{itemize}   

In the engine mode of operation, the efficiency is identical to that of the Otto  cycle and is upper bounded by the Carnot efficiency, $\eta_{two-stroke}=1-\omega_2/\omega_1< 1-T_2/T_1$. In Sec.~\ref{subsec_magneto}, we will discuss the working of the two-stroke cycle in more detail with an application of the same to quantum magnetometry \cite{bhattacharjee_njp_013024}.

{\subsection{Quantum enhancement in performance of outcoupled QTMs}
The purpose of a heat engine is to supply power to an external system. It is therefore plausible that any enhancement of quantum origin in the performance of quantum heat engines may manifest itself in the dynamics of the external system on which the engine performs work. This motivated the investigation of `outcoupled' quantum heat engines \cite{watanabe_prl_050601, watanabe_prl_210603, holmes_prl_210601} in which the engine acts as a work resource for an external system and the performance of the engine is analyzed through energy measurements on the external system.} 

\subsubsection{Enhancement in performance from inter-cycle coherence}\label{subsubsec_cycles}
{Using such an outcoupled quantum heat engine, it was shown in Ref.~[\onlinecite{watanabe_prl_050601}] that the average work performed by the engine over $n$ cycles of the engine operation can be greater than $n$ times the work performed over a single cycle. To this end, the authors considered a total Hamiltonian of the form,
	\begin{equation}\label{eq_ham_outcouple}
		H(t) = H_E(t) + H_B + H_{EB}(t) + H_S + H_{SE}(t),
	\end{equation}
	where $H_E(t)$, $H_B$ and $H_{EB}(t)$ represent the Hamiltonians of the engine (working fluid), baths and the engine-baths interaction respectively. The external system on which the engine perform work is represented by the time-independent Hamiltonian $H_S$ and it is coupled with the engine through the interaction $H_{SE}(t)$. We would like to point out that, following the convention used in Ref.~[\onlinecite{watanabe_prl_050601}], the term `engine' here corresponds to only the working fluid and the baths are treated separately from the engine. All the time-dependent terms in the above Hamiltonian are periodic over a time $T$ such that the total Hamiltonian also satisfies $H(t+T)=H(t)$. The unitary time evolution operator over $n$ cycles is thus given by $U_{nT}=U_T^n$ where $U_T=\mathcal{T}\int_0^T\exp{[-iH(t)t]}dt$.}

{In the above setup, the average work done by the engine is measured by projective energy measurements on the external system and calculating their differences. This can be done in two ways over $n$ cycles. In the first case, the projective measurements are made at the start ($t=0$) and at the end of $n$ cycles ($t=nT$). In the other case,  $n+1$ projective measurements are made, the first one at $t=0$ followed by subsequent measurements at the end of each cycle. Note that in the second case, the per-cycle measurements  destroy any coherence built up in the energy basis after each cycle.} 
	
{To compare the two cases discussed above, consider the Hamiltonian of the external system to be of the form $H_S=\sum_{i}E_i\ket{i}$ and assume that the external system is initially prepared in the ground state $\ket{i=0}$. Thus, starting from a total initial state of the form $\rho_0=\rho_0^{EB}\otimes\ket{0}\bra{0}$, where  $\rho_0^{EB}$ is the initial state of the bath-engine composite, it follows that the average work done in the first case is given by,
	\begin{equation}\label{eq_work_many}
		\la w\ra_n = \sum_i\left(E_i-E_0\right)\sum_{\bf {k}, \bf{k'}}\mathcal{U}_{i,0}^{\bf{k};\bf{k'}},
	\end{equation}
	where $\mathcal{U}_{i,0}^{\bf{k};\bf{k'}}$ is of the form,
	\begin{equation}
		\mathcal{U}_{i,0}^{\bf{k};\bf{k'}}=\Tr_{EB}\left[\bra{i}U_T\ket{k_{n-1}}\dots\bra{k_2}U_T\ket{k_1}\bra{k_1}U_T\ket{0}\rho_0^{EB}\bra{0}U_T^\dagger\ket{k'_1}\bra{k'_1}U_T^\dagger\ket{k'_2}\dots\bra{k'_{n-1}}U_T^\dagger\ket{i}\right].	
	\end{equation}
	In the above equations, $\ket{k_j}, \ket{k'_j}$ are the eigenstates of $H_S$ with  ${\bf k}=\{k_1,k_2,k_3,\dots,k_{n-1}\}$ and ${\bf k'}=\{k'_1,k'_2,k'_3,\dots,k'_{n-1}\}$. On the contrary, the average work done in the second case when projective measurements are performed at the end of each cycle is given by,
	\begin{equation}\label{eq_work_proj}
		\la \tilde{w}\ra_n = \sum_i\left(E_i-E_0\right)\sum_{\bf {k}. \bf{k}}\mathcal{U}_{i,0}^{\bf{k};\bf{k}},
	\end{equation}
	Note that  the sum over the intermediate states in the above equation runs only over $\bf k$, unlike that in Eq.~\eqref{eq_work_many} where the sum runs over $\bf k$ and $\bf k'$. This is an artefact of the fact that the `inter-cycle' coherence is suppressed in the second case.
}  

\begin{figure}
	\includegraphics[width=0.6\textwidth]{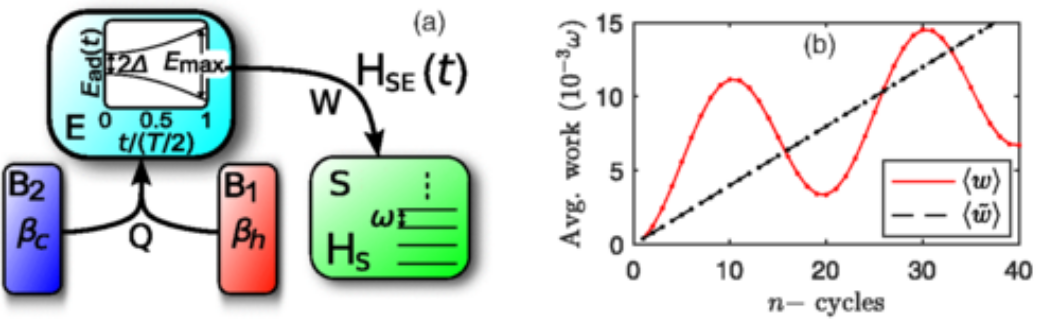}
	\caption{{(a) Schematic representation \cite{watanabe_prl_050601} of the outcoupled quantum Otto engine  discussed in Sec.~\ref{subsubsec_cycles}. The inset represents the adiabatic increase in the energy gap of the TLS engine during the isentropic compression stroke. (b) Numerical comparison (dots) of the average work performed by an Otto engine using Eqs.~\eqref{eq_work_many} and~\eqref{eq_work_proj}. The solid lines are analytical values calculated from a perturbative treatment with $g\ll 1$ (see Ref.~\onlinecite{watanabe_prl_050601} for details). The other relevant parameters chosen for  the numerics are $g=0.02$, $b=0.1/\Delta$, $\omega T=2\pi\times0.05$, $v=0.5\Delta^2$, $T=20/\Delta$, $\beta_c=1/\Delta$ and $\beta_h=1/4 E_{max}$ where $E_{max}=2\sqrt{\Delta^2+(vT)^2/4}$.}}\label{fig_campo_cycles}
\end{figure}

{An Otto engine is now constructed choosing a TLS as the working fluid and a harmonic oscillator as the external system as follows (see Fig.~\ref{fig_campo_cycles} (a) for a schematic representation). In the thermalization steps, the TLS with Hamiltonian $H_E = \Delta\sigma_x$ is decoupled from the external system and is allowed to thermalize with one of the (thermal) baths. In the isentropic steps, the engine is decoupled from the baths and interacts with the external system, with Hamiltonian $H_S=\omega a^\dagger a$, via an impulse type interaction of the form $H_{SE}(t)=g_{SE}(t)\sigma_x(a+a^\dagger)$ where,
	\begin{equation}
		g_{SE}(t)=g\sum_{m=0}^\infty\delta\left[t-(m+b)T\right],
	\end{equation}
	with $g\ll 1$ and $0<b<1$. The unitary evolution in the isentropic steps is driven by the Hamiltonian $H_1(t) = H_E(t)+H_S+H_{SE}(t)$ where,
	\begin{equation}\label{eq_int_cycles}
		H_E(t)=\Delta\sigma_x+\Omega(t)\sigma_z,
	\end{equation}
	where $\Omega(t)=-vt$ during the isentropic compression stroke ($0<t<T/2$) and $\Omega(t)=-v(T-t)$ in the isentropic expansion stroke ($T/2<t<T$), with $v$ being a constant. Each of the isentropic strokes takes a duration of time $T/2$ while the thermalization strokes are assumed to be completed in negligible times.}

{A numerical comparison of the average work performed over $n$ cycles using Eqs.~\eqref{eq_work_many} and~\eqref{eq_work_proj} is given in Fig.~\ref{fig_campo_cycles} (b). It can be clearly seen that $\la w\ra_n$ can be greater than $\la \tilde{w}\ra _n$ for appropriate choices of $n$, thus highlighting the possible enhancement in performance when inter-cycle coherence is preserved. Similar enhancements are also found to occur if one considers more generic interactions $g_{SE}(t)$ instead of the impulse type coupling considered in Eq.~\eqref{eq_int_cycles} (see Ref.~\onlinecite{watanabe_prl_050601} for details)}. 

{\subsubsection{Enhancement in performance from quantum statistical effects}\label{subsubsec_multiple}
In a classical setting, it is reasonable to assume that the work performed on an external system by multiple heat engines working in parallel between a set of heat baths is proportional to the number of heat engines (Once again, `engine' here refers to the working fluid and the baths are treated separately). However, it was shown in  Ref.~[\onlinecite{watanabe_prl_210603}] that this need not be true in a quantum setting, particularly when the engines are composed of indistinguishable systems. In fact, the bosonic statistics arising in such a scenario can boost the performance as we discuss below. }

\begin{figure}
	\includegraphics[width=0.8\textwidth]{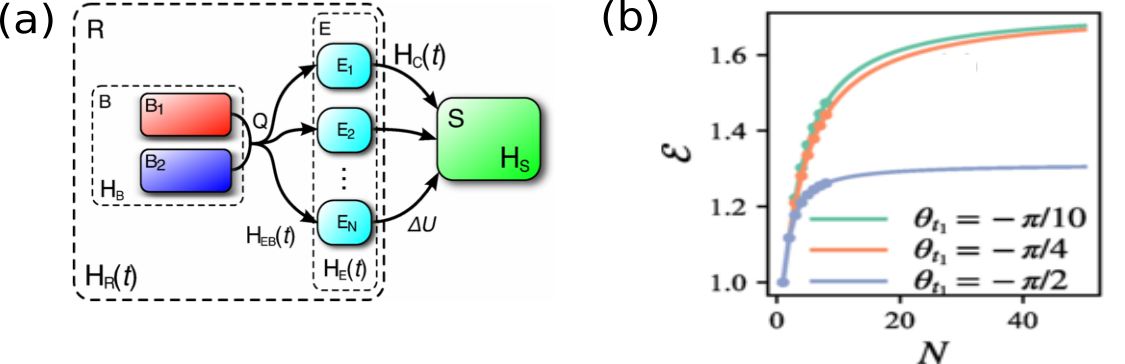}
	\caption{{(a) Schematic representation \cite{watanabe_prl_210603} of the mutliple atoms outcoupled heat engine discussed in \ref{subsubsec_multiple}. (b) Ratio of the work done $\mathcal{E}=\la w\ra_{indist}/\la w\ra_{dist}$ as a function of number of atoms $N$. The solid lines correspond to the analytical values obtained using Eq.~\eqref{eq_multiple_ana} and the dots correspond to numerically calculated values.  Here $\theta_t=\tan^{-1}\left(\Omega(t)/\Delta\right)$ and the parameters chosen for numerics are $g=0.01$, $v=0.1\Omega(0)^2$, $T=20/\Omega(0)$, $t_1=0.35T/2$, $\omega T=2\pi\times0.05$,$\beta_c=2/\epsilon_0$ and $\beta_h=1/4\epsilon_{T/2}$ where $\epsilon_t=\sqrt{\Omega(t)^2+\Delta^2}$ (see Ref.~[\onlinecite{watanabe_prl_210603}] for details).} }\label{fig_campo_multiple}
\end{figure}

{In Ref.~[\onlinecite{watanabe_prl_210603}], the authors considered an outcoupled quantum heat engine constructed from $N$ two-level atoms  as the working fluid (see Fig.~\ref{fig_campo_multiple}~(a) for a schematic representation). The external system is once again chosen to be a harmonic oscillator with $H_S=\omega a^\dagger a$. As in Sec.~\ref{subsubsec_cycles}, the total Hamiltonian  is given by Eq.~\eqref{eq_ham_outcouple}, where $H_E(t)$ is the total Hamiltonian of all the atoms. Further, the interaction between the engines and the external system is of the form, 
\begin{equation}	
	H_{SE}(t)=g_{SE}(t)V_{E}\otimes (a+a^\dagger)
\end{equation}
where $V_{E}$ is a local operator defined on the Hilbert space of the engines. The coupling $g_{SE}(t)=g\delta(t-t_1)$ with $0<t_1<T/2$ is chosen to be an impulse type coupling for simplicity in calculations. When the working fluid is composed of $N$ distinguishable atoms, we have $V_{E}=\sum_{j=1}^N\sigma_x^j$. Similarly, the Hamiltonian of the engines during the isentropic strokes are given by,
\begin{equation}
	H_E(t) = \Delta\sum_{j=1}^N\sigma_x^j + \Omega(t)\sum_{j=1}^N\sigma_z^j.
\end{equation}
where $\Omega(t)=\Omega(0)+vt$ for the isentropic compression stroke and $\Omega(t)=\Omega(0)+v(T-t)$ for the isentropic expansion stroke.} 

{On the other hand, when the atoms are considered to be bosonic and indistinguishable, we have $V_{E}=S_x$ and $H_E(t)$ during the isentropic strokes assume the form,
\begin{equation}
	H_E(t)=\Delta S_x+\Omega(t)S_z,
\end{equation}
where $S_x=b_0^\dagger b_1+b_1^\dagger b_0$ and $S_z=b_0^\dagger b_0-b_1^\dagger b_1$, where $b_0(b_0^\dagger)$ and $b_1(b_1^\dagger)$ are the bosonic annhilation (creation) operators of the ground-state and excited-state atoms, respectively.}

{The Otto engine is constructed in the same way as the mutli-cycle Otto engine discussed in \ref{subsubsec_cycles}, except that the measurements are made only over a single cycle. Therefore, the work performed can be evaluated by calculating the difference between the average energy of the external system at the start and the end of the cycle. Setting $E_0=0$, we have,
\begin{equation}
	\la w\ra = \sum_iE_i\Tr_{EB}\left[\bra{i}U_T^I\rho_0U_T^{I\dagger}\ket{i}\right],
\end{equation}
where $\rho_0$ is the initial state of the total system and $U_T^I=\mathcal{T}\exp{\left(-i\int_0^TH_{SE}(t)dt\right)}$ is the unitary evolution operator in the rotating frame with respect to $H_0(t)=H_E(t)+H_B+H_{EB}(t)+H_S$. After extensive calculations (see Ref.~[\onlinecite{watanabe_prl_210603}] for details), an approximate analytical expression for the work done is found to be,
\begin{equation}\label{eq_multiple_ana}
\la w\ra\simeq g^2\la [V_{E}^I(t_1)]^2\ra_{\rho_0^{EB}}\sum_{i\neq 0}E_i\left|\bra{i}V_S^I(t_1)\ket{0}\right|^2,
\end{equation} 
where $V_S^I(t)=e^{i\omega a^\dagger at}(a+a^\dagger)e^{-i\omega a^\dagger at}$, $V_{E}^I(t)=U_{EB}^{\dagger}(t)V_{E}U_{EB}^{\dagger}(t)$ with $U_{EB}(t)=\mathcal{T}\exp{[-i\int_0^t(H_E(t')+H_B+\\H_{EB}(t'))dt']}$ and $\la\dots\ra_{\rho_0^{EB}}=\Tr_{EB}\left[\dots\rho_0^{EB}\right]$. It is important to note that in the distinguishable case, all
the possible $2^N$ configurations of the atomic pseudospins
are taken into account while calculating the trace over the eigenstates of the engine. On the contrary, the trace is taken only over the $N+1$ symmetrized eigenstates of $S_z$ in the case of indistinguishable atoms.}

{Defining $\mathcal{E}=\la w\ra_{indist}/\la w\ra_{dist}$ where $\la w\ra_{indist}$ and $\la w\ra_{indist}$ are the work performed with indistinguishable and distinguishable atoms, respectively, the authors of Ref.~[\onlinecite{watanabe_prl_210603}] compared the work performed in the two cases. The results are illustrated in Fig.~\ref{fig_campo_multiple}~(b) which clearly shows that the Bosonic statistics of the indistinguisable particles result in greater work output in a cycle as compared to the work output with distinguishable particles.}

\section{Equivalence of thermal machines}\label{sec_equiv}
Naively, the two broad class of quantum thermal machines we have discussed, namely the continuous and reciprocating thermal machines, may appear to be vastly different in terms of their construction and operation. However, it was pointed out that in the limit of small bath action, they are indeed thermodynamically equivalent \cite{uzdin_prx_031044}. The equivalence is valid within the Markovian and rotating wave approximations. To elaborate, consider the mapping $\rho_{N\times N}\to\ket{\rho}_{1\times N^2}$ of the density matrix of a $N$-level system. The GKLS master equation, governing the evolution of the density matrix can therefore be represented as,
\begin{equation}
i\hbar\frac{d\ket{\rho}}{dt}=\mathcal{H}(t)\ket{\rho},
\end{equation}
where $\mathcal{H}_{N^2\times N^2}$ is the super-operator which consists of terms arising from the system Hamiltonian as well as Lindblad operators. The bath action is then defined as \cite{uzdin_prx_031044},
\begin{equation}
s=\int_0^{\tau_{cyc}}\|\mathcal{\tilde{H}}(t)\|dt,
\end{equation}
where $\tau_{cyc}$ is the duration of one full cycle of operation, $\tilde{\mathcal{H}}$ denotes the super-operator $\mathcal{H}$ in the interaction picture and $\|\cdot\|$ is the operator norm defined as $\|\cdot\|=\mathrm{max}\sqrt{eig(\cdot^\dagger\cdot)}$. In the regime of small bath action with respect to the Planck's constant, $s\ll\hbar$, it can be shown that the state of the system in the continuous and reciprocating thermal machines differs by order $\mathcal{O}(s/\hbar)$ before completion of the cycle and by $\mathcal{O}((s/\hbar)^3)$ at the end of the cycle. In fact, the  work and heat transferred also differ by the same order of magnitude. Physically, the emergence of the equivalence is explained as follows. In general, work can be extracted through a coherent mechanism which involves alteration of the off-diagonal terms of the density matrix in energy eigen basis, as well as a  stochastic mechanism which involves alteration of populations. In continuous machines, only the coherent mechanism is present while in the reciprocating machines, the stochastic mechanism is also present. In the limit of small bath action, the coherent mechanism strongly dominates and hence the thermal machines types become thermodynamically equivalent. Later, this equivalence was also extended to non-Markovian systems \cite{uzdin_ent_124}.

\begin{figure}
	\includegraphics[width=0.5\textwidth]{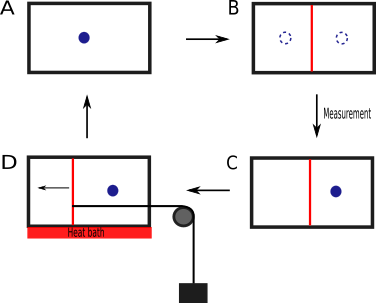}
	\caption{{Schematic representation of the Szilard engine. In the first step (step A), a single molecule gas is assumed to be in thermal equilibrium within a cylinder. A piston is then inserted (step B) in the middle of the cylinder and the molecule can be present in either part of the cylinder. An intelligent demon then measures the position of the gas molecule, depending on whose outcome, it attaches a pulley on the appropriate side of the piston. Finally, in step D, the gas expands by absorbing heat from a heat bath which pulls the load, thus performing useful work.}}\label{fig_szilard}
\end{figure}

\section{Quantum Szilard Engine}\label{sec_szilard}
Before concluding our discussion on QTMs, let us briefly mention the Szilard engine which, though not as technologically relevant, is tremendously important to gain a better understanding of the rapport between information and thermodynamics \cite{goold_jpamt_143001}. When Maxwell proposed using `information' as a resource to conceive what came to be famously known as the Maxwell's demon \cite{maxwell_demon,maruyama_rmp_1,quan_prl_180402}, the second law appeared to be under some serious challenge. Building on his work, Szilard \cite{szilard} proposed an engine in which a feedback assisted cyclic process appeared to allow conversion of all the heat extracted from a single reservoir into work. This conceptual and ideal engine, operating with a single molecule gas as the working fluid, works as follows {(see Fig.~\ref{fig_szilard} for a schematic representation)}. Consider a single molecule gas confined in a cylinder which is in contact with a heat reservoir. A piston, having a wide opening in its centre that can be closed with a friction less shutter, is placed {in the middle of} the cylinder. When the shutter is closed, the piston divides the cylinder volume into two equal parts {and the molecule can be present on either side of the piston with equal probability}. An intelligent demon then performs a \textit{measurement} to determine on which side of the piston the molecule is located. Depending on the measurement outcome, the demon attaches a string to the right or left side of the piston such that the isothermal expansion of the single molecule gas pulls the string, thus lifting any weight attached to the other end of the string. Once the required work is completed, the shutter is opened so that the single molecule gas can once again occupy the whole volume of the cylinder. Therefore, the expansion of the gas driven by the heat extracted from the reservoir is completely converted into work.

{
	Classically, the widely accepted solution to this paradox is given by the Landauer's erasure principle \cite{landauer_ibm_183, landauer_pla_188}, which associates an energetic cost with any \textit{logically irreversible manipulation of information}. As for example, resetting the information stored in a memory bit amounts to an increase of entropy $\sim k_B\log 2$ where $k_B$ is the Boltzmann constant and hence, a minimum amount of heat $\sim k_BT\log 2$ is dissipated to the environment at temperature $T$. A truly cyclic process demands that the memory of the demon is also restored at the end of the complete cycle. This requires an erasure of the information acquired by the demon during the measurement which results in heat dissipation, thus accounting for a second heat reservoir or sink. Thus it became apparent that \textit{information} should be visualized as a physical entity that has a direct bearing on thermodynamic processes \cite{penrose_book_oxford, bennett_ijtp_905,bennett_shpmp_501}.
}

{
	Note that the Szilard engine, by construction, is a microscopic engine working with a single molecule and hence, a rigorous analysis must also incorporate the quantum effects in play. With this motivation, numerous quantum models of the Szilard engine has been investigated over the years \cite{zurek_book_springer, biedenharn_fp_1221, lloyd_pra_3374, parrondo_chaos_725, kim_prl_070401, li_ap_2955, cai_pre_031114, park_prl_230402, plesch_sr_6995,aydin_ent_294,zurek_pr_21,bengtsson_prl_100601,elouard_npjq_9, elouard_prl_260603,mohammady_njp_113026}. While early works had explored the consequences of endowing a quantum nature to the measurement and information erasure processes \cite{zurek_book_springer, biedenharn_fp_1221,lloyd_pra_3374}, many recent works have also investigated exotic variations of the Szilard engine. Similarly, it was pointed out in Ref [\onlinecite{kim_prl_070401}], that an energetic cost should be associated with the insertion or removal of piston in the quantum case, resulting from changes of boundary conditions. Recently, it has also been shown that it is possible to operate a Szilard engine working without any thermal source by drawing energy from projective measurements \cite{elouard_npjq_9, elouard_prl_260603,mohammady_njp_113026}, although it may lose some of its characteristics in the process \cite{mohammady_njp_113026}. Nevertheless, there is no universal consensus regarding the significance and implications of a fully quantized Szilard engine and much remains to be understood. 
}

\section{Applications in quantum metrology}\label{sec_appl}

In this section, we review a couple of recently proposed applications of quantum thermal machines in the field of quantum metrology \cite{paris_ijqi_125, correa_prl_220405, zhou_natcom_78, giovannetti_np_222, degen_rmp_035002, kurizki_tech_1, pezze_rmp_035005}. The basic idea behind the protocols is to use quantum thermal machines as sensitive probes, which can be coupled to a given system and some parameter of the system is then estimated through an indirect measurement on the probe. The precision attained by using a particular protocol is assessed through a comparison with the minimum bound on the relative error dictated by the quantum Cramer-Rao bound \cite{paris_ijqi_125, braunstein_prl_3439, ma_pr_89, zhang_pra_043832, hovhannisyan_prb_045101,  pasquale_natcom_12782, gefen_pra_032310, giovannetti_np_222, giovannetti_prl_010401}. This bound is quantified by the so called quantum Fisher information, which unlike the classical Fisher information, has a geometrical origin and is therefore uniquely determined by the state of the system on which the measurement is to be performed. It is therefore imperative that we present a brief summary of the QFI before moving on to discuss the protocols in detail.
 
\subsection{Quantum Fisher Information}\label{subsec_qfi}
Let us consider that we wish to estimate a parameter $\theta$ through an indirect measurement on a random variable $X$. The probability that the outcome of a measurement on $X$ is $x_i\in X$, is determined by the conditional probability $(x|\theta)$. The parameter $\theta$ is read off from the measurement on $X$ through the estimator $\hat{\theta}(X)$. Let us assume that the true value of the parameter is $\theta=\theta_0$. Further, we assume an \textit{unbiased} estimator, i.e.
\begin{equation}
 \la\hat{\theta}(x)-\theta_0\ra=0. 
\end{equation}
Taking a partial derivative w.r.t. $\theta_0$, we get,
\begin{equation}
\frac{\partial}{\partial\theta_0}\la\hat{\theta}(x)-\theta_0\ra=\int\left(\hat{\theta}(x)-\theta_0\right)\frac{\partial p(x|\theta_0)}{\partial\theta_0}dx-\int p(x|\theta_0)dx=0.
\end{equation}
or,
\begin{equation}\label{eq_cfi_deriv}
\int\left(\hat{\theta}(x)-\theta_0\right)p(x|\theta_0)\frac{\partial}{\partial\theta_0} \log p(x|\theta_0)dx=1,
\end{equation}
where we have used the equalities $\partial_{\theta_0} p(x|\theta_0)= p(x|\theta_0)\partial_{\theta_0} \log p(x|\theta_0)$ and $\int p(x|\theta_0)dx=1$. Next, consider the following relation by virtue of the  Cauchy-Schwarz inequality,
\begin{widetext}
\begin{multline}\label{ineq_cauchy}
\int\Big[\left(\hat{\theta}(x)-\theta_0\right)\sqrt{p(x|\theta_0)}\Big]\Big[\sqrt{p(x|\theta_0)}\frac{\partial}{\partial\theta_0} \log p(x|\theta_0)\Big]dx\\
\leq\sqrt{\Big[\int\left(\hat{\theta}(x)-\theta_0\right)^2p(x|\theta_0)dx\Big]\Big[\int p(x|\theta_0)\left(\frac{\partial}{\partial\theta_0} \log p(x|\theta_0)\right)^2dx\Big]}.
\end{multline}
\end{widetext}
Recognizing,
\begin{equation}
\int\left(\hat{\theta}(x)-\theta_0\right)^2p(x|\theta_0)dx=\mathrm{Var}(\theta),
\end{equation} 
and rearranging Eq.~\eqref{eq_cfi_deriv} appropriately, we arrive at the (classical) Cramer-rao bound,
\begin{equation}
\mathrm{Var}(\theta)\geq\frac{1}{\mathcal{I}(\theta_0)},
\end{equation}
where $\mathcal{I}(\theta_0)$ is the classical Fisher information (CFI) given by,
\begin{widetext}
\begin{equation}\label{eq_fi}
\mathcal{I}(\theta_0)=\int p(x|\theta_0)\left(\frac{\partial}{\partial\theta_0} \log p(x|\theta_0)\right)^2dx=\int\frac{1}{p(x|\theta_0)}\left(\frac{\partial p(x|\theta_0)}{\partial\theta_0}\right)^2dx.
\end{equation}
\end{widetext}
The CFI, as defined above, depends on the probability distribution $p(x|\theta)$. In other words, the CFI is influenced by the choice of the random variable $X$ used for estimating the parameter $\theta$.

From a quantum mechanical viewpoint, the probability distribution $p(x|\theta)$ is obtained as $p(x|\theta)=\Tr\left[\rho_\theta\Pi_x\right]$, where $\{\Pi_x\}$ refers to the set of elements of a positive-operator-value-measure (POVM) and $\rho_{\theta}$ is the state of the quantum system which naturally depends on the parameter $\theta$. The CFI can therefore be written as,
\begin{equation}\label{eq_cfi_deriv2}
\mathcal{I}(\theta_0)=\int\frac{\Tr\Big[\partial_{\theta_0}\rho_{\theta_0}\Pi_x\Big]^2}{\Tr\Big[\rho_{\theta_0}\Pi_x\Big]}dx=\int\frac{\mathrm{Re}\left(\Tr\Big[\rho_{\theta_0}\Pi_xL_{\theta_0}\Big]\right)^2}{\Tr\Big[\rho_{\theta_0}\Pi_x\Big]}dx,
\end{equation} 
where we have introduced the symmetric logarithmic derivative (SLD) $L_{\theta_0}$ which is defined as,
\begin{equation}\label{eq_sld}
\frac{\partial\rho_{\theta_0}}{\partial\theta_0}=\frac{1}{2}\left(L_{\theta_0}\rho_{\theta_0}+\rho_{\theta_0}L_{\theta_0}\right).
\end{equation}
Using the inequality $\mathrm{Re}(z)\leq|z|$, we therefore obtain,
\begin{widetext}
\begin{equation}
\mathcal{I}_{\theta_0}\leq \int\left|\frac{\Tr\Big[\rho_{\theta_0}\Pi_xL_{\theta_0}\Big]}{\sqrt{\Tr\Big[\rho_{\theta_0}\Pi_x\Big]}}\right|^2dx=\int\left|\Tr\left[\frac{\sqrt{\rho_{\theta_0}}\sqrt{\Pi_x}}{\sqrt{\Tr\Big[\rho_{\theta_0}\Pi_x\Big]}}\sqrt{\Pi_x}L_{\theta_0}\sqrt{\rho_{\theta_0}}\right]\right|^2dx,
\end{equation}
\end{widetext}
Using the Cauchy-Scwarz inequality again, we finally arrive at,
\begin{equation}\label{eq_qfi_pre}
\mathcal{I}_{\theta_0}\leq\int\Tr\left[\Pi_xL_{\theta_0}\rho_{\theta_0}L_{\theta_0}\right]dx=\Tr\left[\rho_{\theta_0}L^2_{\theta_0}\right]=\mathcal{H}_{\theta_0},
\end{equation}
where we have used the completeness of the POVM $\int\Pi_xdx=\mathbb{I}$ in obtaining the second equality. The quantity $\mathcal{\theta_0}$ is the quantum Fisher information, which is purely a geometrical quantity and does not depend upon the choice of measurement parameter $X$. The QFI therefore sets a stricter bound on the relative error bound,
\begin{equation}\label{eq_cramer}
\mathrm{Var}(\theta)\geq\frac{1}{\mathcal{H}_{\theta_0}}.
\end{equation}

To derive a more explicit form of the QFI, we write the solution pf the SLD obtained from Eq.~\eqref{eq_sld} as,
\begin{equation}
L_{\theta_0}=2\int_0^\infty\left(e^{-\rho_{\theta_0}t}\frac{\partial\rho_{\theta_0}}{\partial\theta_0}e^{-\rho_{\theta_0}t}\right)dt.
\end{equation} 
Substituting $\rho_{\theta_0}=\sum_ip_i\ket{\phi_i}\bra{\phi_i}$, the above equation simplifies to,
\begin{equation}
L_{\theta_0}=\sum_i\frac{1}{p_i}\frac{\partial p_i}{\partial\theta_0}\ket{\phi_i}\bra{\phi_i}+\sum_{i\neq j}\frac{p_i-p_j}{p_i+p_j}\bra{\phi_j}\frac{\partial}{\partial\theta_0}\ket{\phi_i}\ket{\phi_j}\bra{\phi_i}.
\end{equation}
The QFI obtained in Eq.~\eqref{eq_qfi_pre} hence assumes the form,
\begin{equation}\label{eq_qfi}
\mathcal{H}_{\theta_0}=\sum_i\frac{1}{p_i}\left(\frac{\partial p_i}{\partial\theta_0}\right)^2+2\sum_{i\neq j}\frac{\left(p_i-p_j\right)^2}{p_i+p_j}\left|\bra{\phi_j}\frac{\partial}{\partial\theta_0}\ket{\phi_i}\right|^2.
\end{equation} 
The geometrical origin of the QFI is now explicitly seen as the above expression derived for the QFI is identical to that of fidelity susceptibility $\mathcal{F}_I(\theta_0)=-2\lim_{\epsilon\to 0}\partial^2\mathcal{F}\left(\rho(\theta_0+\epsilon),\rho(\theta_0)\right)/\partial\epsilon^2$, where $\mathcal{F}(\rho_1,\rho_2)=\Tr\left[\sqrt{\sqrt{\rho_1}\rho_2\sqrt{\rho_1}}\right]$ is the fidelity between the states $\rho_1$ and $\rho_2$. {We note in passing that for a pair of pure states, $\mathcal{F}(\rho_1,\rho_2)$ reduces to the pure state fidelity $\mathcal{F}(\ket{\psi_1},\ket{\psi_2})=\left|\braket{\psi_1|\psi_2}\right|$. Further, the (ground-state) fidelity in itself exhibits many interesting behavior, particularly those associated with quantum critical phenomena \cite{dutta_cambridge_book}.}

\begin{figure}
\subfigure[]{
\includegraphics[width=0.36\textwidth]{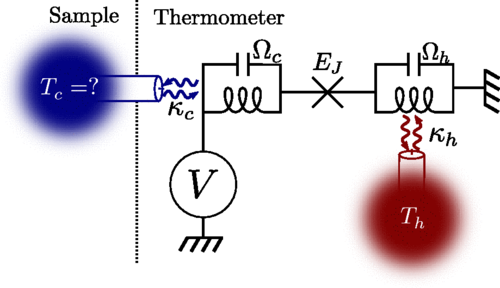}
\label{fig_therm_schem}}\hspace{2cm}
\subfigure[]{
	\includegraphics[width=0.45\textwidth, height=3.5cm]{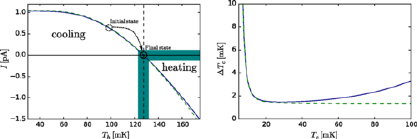}
\label{fig_therm_res}}
\caption{{(a) Schematics of the circuit QED implementation of the thermometer introduced in Ref.~[\onlinecite{hofer_prl_090603}] and discussed in Sec.~\ref{subsec_thermo}. Two harmonic oscillators representing resonant cavities are coupled through a Josephson junction and are also coupled individually to separate heat baths. An external bias voltage provides the power source. (b) \textbf{Left panel} Charge current $I$ as a function of $T_h$ for a fixed $T_c=15$ mK. The current vanishes at the Carnot point (solid circle). This point can only be determined up to a certain error $\Delta I$ and $\Delta T_h$ (shaded area). \textbf{Right panel} A demonstration \cite{hofer_prl_090603} of the error in the temperature estimation. Assuming  $\Delta I=0.3$ pA and $\Delta T_h=10$ mK lead to $\Delta T_c<2$ mK down to temperatures of $15$ mK. The blue (solid) lines represent numeric solutions while the green (dashed) lines are  analytic solutions obtained using a simplified model (see Eq.~\eqref{eq_err_therm} and the following discussion). The other relevant parameter values chosen for the calculations are: $\Omega_h=2\pi\times 8.5$ GHz, $\Omega_c=2\pi\times1$ GHz, $\kappa_h=\kappa_c=2\pi\times0.06$ GHz, $E_J=2\pi\times0.2$ GHz, $\lambda_h=\lambda_c=0.3$. } }
\end{figure}

\subsection{Application of QTMs in Quantum thermometry}\label{subsec_thermo}
A QTM operating at Otto efficiency was shown to be capable of measuring very low temperatures with high precision in Ref.~[\onlinecite{hofer_prl_090603}], mimicking the role of a nano-scale thermometer. The thermal machine is constructed using a circuit-QED setup as follows {(see Fig.~\ref{fig_therm_schem}). The working fluid of the QTM consists of two resonant cavities with characteristic frequencies $\Omega_h$ and $\Omega_c$, coupled to each other through a voltage-biased Josephson junction. The voltage-bias acts as the external power source for the QTM. In addition, each of the cavity is coupled through transmission lines to distinct thermal baths having temperatures $T_h$ and $T_c$, respectively, where the temperature $T_c$ is unknown and is to be measured. Since Cooper pairs flowing across the junction carries no entropy, the charge current is completely separated from the heat currents.} Under the resonance condition $2eV=\Omega_h-\Omega_c$ the Hamiltonian reads \cite{hofer_prb_041418},
\begin{equation}
H=\frac{E_J}{2}\left(a_h^\dagger A_hA_ca_c+h.c.\right),
\end{equation} 
where $E_J$ is the Josephson energy and $a_{h(c)}$ is the annihilation operator acting on the Fock space of the oscillator with frequency $\Omega_{h(c)}$ and the operators. The operators $A_{\alpha}$ are of the form \cite{hofer_prb_041418},
\begin{equation}
A_{\alpha}=2\lambda_{\alpha}e^{-2\lambda_{\alpha}^2}\sum_{n_{\alpha}}\frac{L_{n_\alpha}^{(1)}(4\lambda_{\alpha}^2)}{n_{\alpha}+1}\ket{n_\alpha}\bra{n_{\alpha}},
\end{equation}
where $\lambda_{\alpha}$ are the amplitude of oscillator zero-point phase fluctuations, $L_n^k(x)$ are the generalized Laguerre polynomials and $\ket{n_{\alpha}}$ are Fock-states associated with the oscillator with frequency $\Omega_{\alpha}$. Similarly, the current operator assumes the form,
\begin{equation}\label{eq_thermo_current}
I=-\frac{eE_J}{i}\left(a_h^\dagger A_hA_ca_c-h.c.\right).
\end{equation}
 
The dynamical evolution of the system is assumed to be governed by the Lindblad master equation. In the steady state operation, the heat currents are given by,
\begin{subequations}
\begin{equation}
J_{h(c)}=\Omega_{h(c)}\kappa_{h(c)}\left(\la n_{h(c)}\ra-n_B^{h(c)}\right)
\end{equation}	
and the power is calculated as,
\begin{equation}
P=V\la I\ra,
\end{equation}
\end{subequations}
where $\kappa_{h(c)}$ is the decay rate associated with the bath with temperature $T_{h(c)}$, $n_B^{h(c)}$ is the mean occupation number of the corresponding baths and $\la\cdot\ra$ denotes expectation values in the steady state.

Using the above equations, one can show that the steady state operation resembles that of a quantum heat engine or refrigerator. Further, the engine like operation is characterized by an efficiency equal to the Otto efficiency $\eta=1-\Omega_c/\Omega_h$, which approaches the Carnot efficiency {when the following relation is satisfied,
\begin{equation}\label{eq_null_thermo}
	\frac{\Omega_c}{\Omega_h}=\frac{T_c}{T_h}.
\end{equation}
Both the heat currents and the power vanish when the above equality is reached. The crucial point to note here is that the unknown temperature $T_c$ depends only on $T_h$, $\Omega_c$ and $\Omega_t$ when Eq.~\eqref{eq_null_thermo} is satisfied and becomes independent of any other system parameters. This allows $T_c$ to be estimated as follows.} For a fixed set of $\Omega_c$ and $\Omega_h$, whose values are assumed to be known with high precision, one can vary $T_h$ until the steady state power vanishes at a certain $T_h^*$, {as demonstrated in the left panel of Fig.~\ref{fig_therm_res}}. It then follows from Eq.~\eqref{eq_null_thermo} that the unknown temperature satisfies $T_c=T_h^*\Omega_c/\Omega_h$.

{To estimate the error in determining $T_c$, note that the protocol requires only two measurements-- one of the power output, (i.e. the current $\la I\ra$ as $V$ is already known) and the second of the temperature $T_h$ at which the power vanishes. Both of these measurements are error-prone and hence the net error in determining $T_c$ is estimated using the error propagation formula as,}
\begin{equation}\label{eq_err_therm}
\Delta T_c=\sqrt{\left(\frac{\partial T_c}{\partial \la I\ra}\right)^{2}\left(\Delta I\right)^2+\left(\frac{\partial T_c}{\partial T_h}\right)^{2}\left(\Delta T_h\right)^2}
=\sqrt{\left(\frac{\partial \la I\ra}{\partial T_c}\right)^{-2}\left(\Delta I\right)^2+\left(\frac{\Omega_c}{\Omega_h}\right)^2\left(\Delta T_h\right)^2},
\end{equation}
where $\Delta X$ denotes the root mean square error in the measurement of parameter $X$. {The second equality in the above equation follows from the equality $T_c=T_h\Omega_c/\Omega_h$ which is satisfied when $T_h=T_h^*$. By choosing $\Omega_h\gg\Omega_c$, one can therefore neglect the error contribution from $\Delta T_h$.} To evaluate the remaining term in the above equation, an approximate model is considered by substituting $E_JA_hA_c/2=g$. In this case, the error evaluates to $\Delta T_c=\alpha T_c^2\sinh\left(\Omega_c/2T_c\right)/\Omega_c$. At the same time, the minimum possible error in measuring the temperature as determined from the Cramer-Rao bound (see Eq.~\eqref{eq_cramer}) is found to be $\Delta T_c^{cr}=1/\mathcal{H}_{T_c}= C_1\Delta T_c/C_2$, where $C_1$ and $C_2$ depend on the coupling constants as follows:
{\begin{subequations}
\begin{equation}
	C_1 = \frac{2\left(\kappa_h+\kappa_c\right)\left(\kappa_h\kappa_c+4g^2\right)}{\kappa_c\sqrt{8g^2\kappa_c\kappa_h+\kappa_h^2(\kappa_c^2+16g^2)+2\kappa_c\kappa_h^3+32g^4+\kappa_h^4}},
\end{equation}
\begin{equation}
C_2	= \frac{\left(\kappa_h+\kappa_c\right)\left(\kappa_h\kappa_c+4g^2\right)}{\sqrt{2}\kappa_h\kappa_cg}.	
\end{equation}
\end{subequations}
}

For any choice of the coupling constants, one always finds $C_2\geq C_1$ which is expected as the minimum error can not be lower than that of the Cramer-Rao bound. The optimal value is found to be $C_2/C_1\approx2.55$ which shows that the proposed thermometer is capable of measuring temperatures with a precision close to the  maximum theoretically possible precision. {A comparison of the error arising from numerical result and that obtained analytically using the approximate model is shown for a certain set of parameters in the right panel of Fig.~\ref{fig_therm_res}.}
\begin{figure*}
	\centering
	\begin{center}
		\subfigure[]{
			\includegraphics[width=0.4\textwidth]{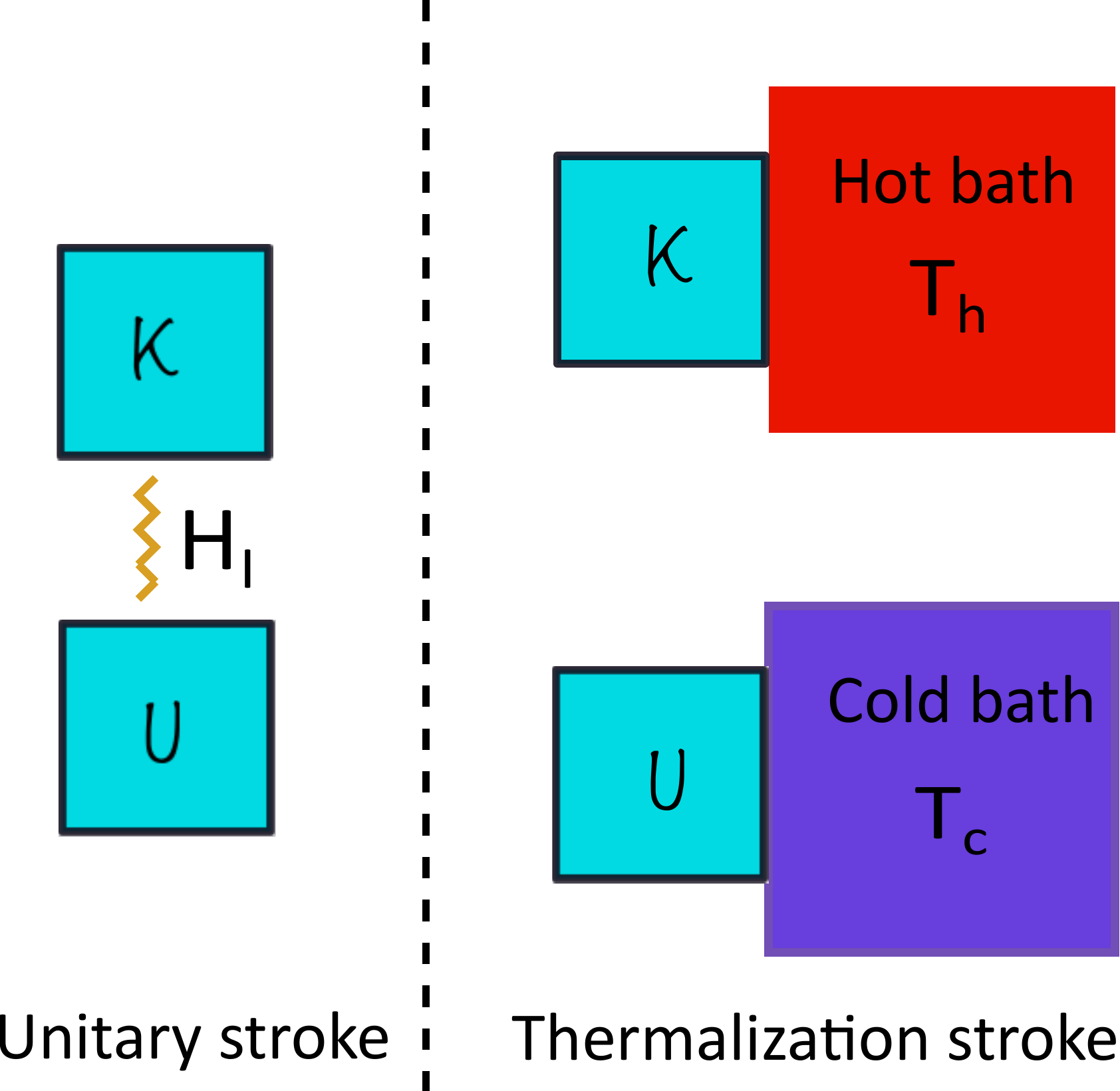}	
			\label{fig1}
		}\hspace{2cm}
		\subfigure[]{
			\includegraphics[width=0.4\textwidth]{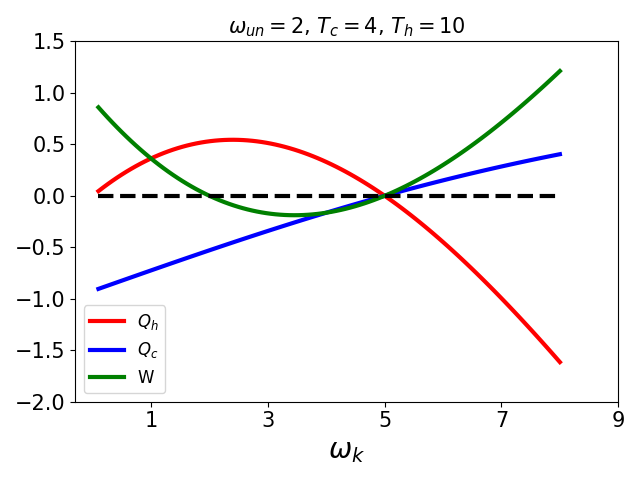}
			\label{fig2}	
		}
	\end{center}
	\caption{ (a) Schematic representation \cite{bhattacharjee_njp_013024} of the two stroke quantum thermal machine. (b) The heat exchanged with the baths and work done as a function of $\omega_k$. At the Carnot point $\omega_k=5$, all of the quantities reverse their sign signaling the transition from engine-like to refrigerator-like operation or vice-versa. For these particular numerical results, the interaction Hamiltonian is assumed to induce a swap of populations between the two TLSs at the end of the unitary stroke which corresponds to choosing  $\theta=\pi/2$ in Eq.~\eqref{eq_two_stroke}.  }
\end{figure*}
\subsection{Application of QTMs in Quantum Magnetometry}\label{subsec_magneto}
As in thermometry, it has also been proposed that it is possible to boost precision of weak field magnetometry measurements by utilizing a two-stroke QTM acting as a quantum probe \cite{bhattacharjee_njp_013024}. To elaborate, let us consider two qubits labeled as $\mathcal{K}$ and $\mathcal{U}$, which are coupled to two separate magnetic fields of known and unknown intensities, respectively. Further, we assume that the known magnetic field is relatively stronger than the unknown field. This setting therefore gives rise to two TLSs, whose energy levels are denoted as  $\pm\omega_k$ and $\pm\omega_{un}$, with $\omega_k>\omega_{un}$. In addition, we also have two thermal baths with temperatures $T_h$ and $T_c$, such that $T_h>T_c$. Initially, $\mathcal{K}$ and $\mathcal{U}$ are assumed to be in thermal equilibrium with the hot bath $T_h$ and cold bath $T_c$, respectively, with no interactions existing between the two. The density matrix of the total system (excluding the baths) is therefore given by,
\begin{equation}\label{eq_rho_tot}
\rho_\mathrm{tot}(0)=\rho_\mathrm{k}(0)\otimes\rho_\mathrm{un}(0),
\end{equation}
where
\begin{subequations}\label{eq_denTLS}
	\begin{equation}
	\rho_\mathrm{un}(0)=
	\begin{pmatrix}   
	n_\mathrm{un} & 0\\
	0 & 1-n_\mathrm{un}
	\end{pmatrix}
	=\frac{1}{\mathcal{Z}_\mathrm{un}}
	\begin{pmatrix}   
	e^{-{\frac{\omega_{un}}{T_c}}} & 0\\
	0 & e^{{\frac{\omega_{un}}{T_c}}}
	\end{pmatrix},
	\end{equation}
	\begin{equation}
	\rho_\mathrm{k}(0)=\begin{pmatrix}   
	n_\mathrm{k} & 0\\
	0 & 1-n_\mathrm{k}
	\end{pmatrix}=\frac{1}{\mathcal{Z}_\mathrm{k}}\begin{pmatrix}   
	e^{-{\frac{\omega_k}{T_h}}} & 0\\
	0 & e^{{\frac{\omega_k}{T_h}}}
	\end{pmatrix}.
	\end{equation} 
\end{subequations}
Here, $n_\mathrm{un(k)}$ denotes the excited state population of $\mathcal{U}$~($\mathcal{K}$) and $\mathcal{Z}_\mathrm{un(k)}$ are the respective partition functions. The device is then operated in a two-step cycle as per the following strokes (see Fig.~\ref{fig1}):

\begin{enumerate}
	
\item \textit{Unitary Stroke}: The TLSs are decoupled from their respective baths and allowed to interact with each other; this interaction lasts for a duration of time $\tau_U$. We specifically consider an interaction of the form,
\begin{equation}
  H_\mathrm{I}(t) = 2\omega_\mathrm{I}(t) \left(\ket{\uparrow_\mathrm{k}}\ket{\downarrow_\mathrm{un}}\bra{\downarrow_\mathrm{k}}\bra{\uparrow_\mathrm{un}} + \mathrm{h.c.} \right)
\end{equation}
where \{$\ket{\uparrow_\mathrm{k(un)}},\ket{\downarrow_\mathrm{k(un)}}$\} is the energy eigenbasis of the TLS $\mathcal{K}(\mathcal{U})$ and $\omega_\mathrm{I}(t)$ is a time-dependent modulation. This choice of $H_\mathrm{I}(t)$ results in a unitary evolution in which only the projection of $\rho_{tot}(0)$ on the subspace spanned by the states $\ket{\uparrow_\mathrm{k}}\ket{\downarrow_\mathrm{un}}$ and $\ket{\downarrow_\mathrm{k}}\ket{\uparrow_\mathrm{un}}$ is rotated. We denote this rotation by an angle $\theta$. 
	
\item \textit{Thermalization Stroke}: At the end of the unitary stroke,  the interaction $H_\mathrm{I}(t)$ is switched off and each TLS is again coupled to its respective thermal bath, i.e., the bath with which it was initially in equilibrium before the unitary stroke. This thermalization stroke last for a duration of time $\tau_\mathrm{T}$ with the assumption that $\tau_\mathrm{T}$ is sufficiently long so that each TLS returns to its initial configuration given by Eq.~\eqref{eq_denTLS}. Note that heat exchanges between the TLSs and the baths only occur during this second stroke.
\end{enumerate}

The work done and the heat exchanged during the cycle are calculated as follows. Since all heat exchanges occur during the thermalization stroke, one can calculate the same using \eqref{eq_qheat} as, 	
\begin{equation}\label{eq_mix_heat}
Q_\mathrm{h(c)}={\Tr}[\rho_\mathrm{k(un)}(\tau_\mathrm{U}+\tau_\mathrm{T})H_\mathrm{k(un)}] - {\Tr}[\rho_\mathrm{k(un)}(\tau_\mathrm{U})H_\mathrm{k(un)}]
\end{equation}
where, $\rho_\mathrm{k(un)}(t)={\Tr}_{\mathrm{un}(\mathrm{k})}[\rho_{tot}(t)]$. On simplification, this evaluates to,
\begin{subequations}\label{eq_two_stroke}
\begin{align}
Q_c=2\omega_{un}(n_\mathrm{un}-n_\mathrm{k})\sin^2{\theta},
\end{align}
\begin{align}
Q_h=2\omega_k(n_\mathrm{k}-n_\mathrm{un})\sin^2{\theta}\label{eq_mix_heat2}.
\end{align}
Using the first law, the work done evaluates to
\begin{equation}
W=-2(\omega_k-\omega_{un})(n_\mathrm{k}-n_\mathrm{un})\sin^2{\theta}.
\end{equation}
\end{subequations}
It is easy to see from the above expressions that if $n_k>n_{un}$, the machine acts as an engine while it acts as a refrigerator if $n_k<n_{un}$ (see Fig.~\ref{fig2}). In the regime of engine-like operation, the efficiency is found to be equal to the Otto efficiency,
\begin{equation}
\eta=\frac{-W}{Q_h}=1-\frac{\omega_{un}}{\omega_k},
\end{equation}
Importantly, the transition from engine-like to refrigerator-like operation occurs at $n_k=n_{un}$. It follows from Eq.~\eqref{eq_denTLS} that this is possible when,
\begin{equation}\label{eq_carnot_point}
\frac{\omega_{un}}{\omega_k}=\frac{T_c}{T_h}.
\end{equation}
At this transition point, the efficiency equals the Carnot efficiency accompanied by the vanishing of heat exchanged and work performed during the cycle. 

\begin{figure}
	\includegraphics[width=0.5\textwidth,height=0.3\textheight]{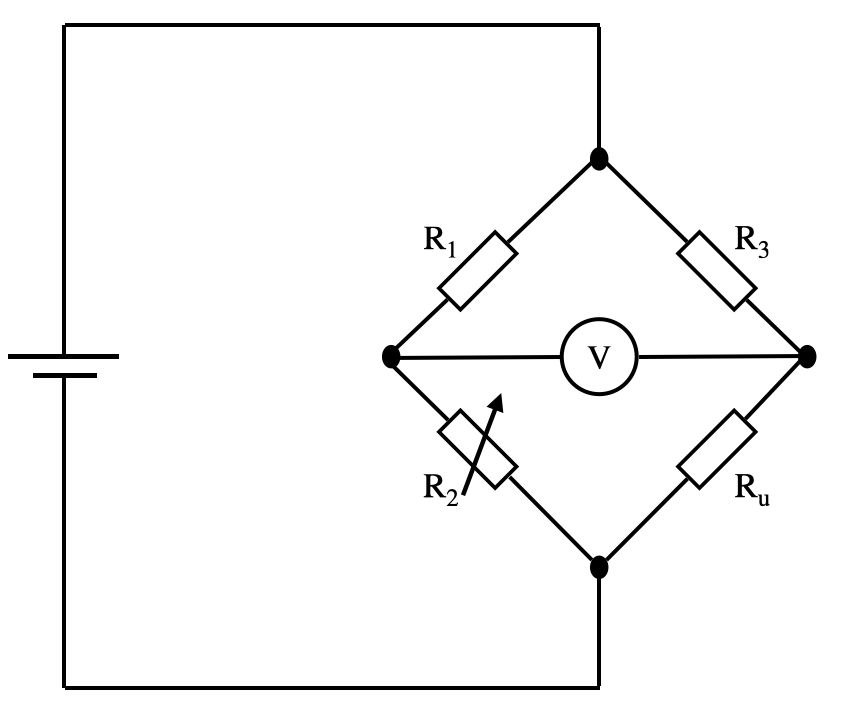}
	\caption{The Wheatstone bridge setup is used to measure electrical resistance with high precision. The variable resistance $R_2$ with known values is tuned until the voltage (measured by a voltmeter $V$) across the two arms of the bridge are balanced. At this balanced position, the unknown resistance is calculated using the formula $R_u=R_2R_3/R_1$. The thermometry (Sec.~\ref{subsec_thermo}) and the magnetometry (Sec.~\ref{subsec_magneto}) protocol discussed in the text are also based on the Wheatstone bridge principle as can be seen by comparing the above equation with Eqs.~\eqref{eq_null_thermo} and~\eqref{eq_carnot_point}. }\label{fig_wheatstone}
\end{figure}

The intensity of the unknown magnetic field to which the qubit $\mathcal{U}$ was coupled can now be determined using Eq.~\eqref{eq_carnot_point}. Experimentally, the Carnot point can be identified by carefully tuning $\omega_k$ (the known magnetic field) and observing when the heat exchanged and the work done vanishes. Indeed, one only needs to look for a sign reversal in the magnitude of these quantities, which in general, can be done much more accurately than determining exact values of the same. Provided that the temperatures of the baths are accurately known, the error in determining $\omega_{un}$ is therefore $\Delta \omega_{un}=\Delta\omega_k T_c/T_h$. Further, the error $\Delta \omega_k$ has contributions arising from two sources -- (i) from measurement errors in determining the null of the heat exchanges or the work and (ii) the error in directly measuring $\omega_k$ through standard methods. The former can be shown to be negligibly small under certain feasible conditions \ct{bhattacharjee_njp_013024}. Consequently, the error $\Delta \omega_k$ arises solely from its direct measurement and this is equal to the error $(\Delta \omega_{un})^{dir}$ that would have been present if $\omega_{un}$ was measured directly. Hence, we can write,
\begin{equation}
\Delta\omega_{un}=(\Delta\omega_{un})^{dir}\frac{T_c}{T_h}.
\end{equation} 
Therefore, the indirect measurement by using the two-stroke thermal machine helps in reducing the error in measuring the weaker magnetic field by a magnitude of $T_c/T_h$ as compared to a direct measurement. 

It is important to note that the protocols discussed above are based on the Wheatstone bridge {(see Fig.~\ref{fig_wheatstone})} principle, where an unknown parameter is measured by tuning another auxiliary parameter and searching for the zero-point where some observable such as the currents vanish. At this point, the values of the known and the unknown parameters must satisfy a certain preset ratio which therefore enables one to estimate the unknown parameter. The practical advantage of these protocols resides in the fact that such zero-point measurements in general can be carried out with very high precision as compared to absolute value measurements. Further, knowledge of other microscopic parameters in the set up is not required which eliminates possible sources of error arising from the uncertainty in the known value of such microscopic parameters.

\section{Quantum batteries}\label{sec_battery}
As much as it is important to engineer devices that extract energy or useful work from heat sources, it is also equally important to realize ways of efficiently storing this energy. As with thermal machines, it is natural to ponder whether quantum effects can be utilized to our advantage so as to facilitate a better storage of useful energy, or faster charging as well as discharging of batteries operating in the quantum regime. The study of so called \textit{quantum batteries} \cite{alicki_pre_042123, hovhannisyan_prl_240401, campaioli_prl_150601, farre_prr_023113, binder_njp_075015, le_pra_022106, friis_q_61, andolina_prl_047702, rossini_prb_115142, ferraro_prl_117702, zhang_pre_052106, andolina_prl_047702, andolina_prb_205423, farina_prb_035421, andolina_prb_205437,crescente_arxiv_09791, ghosh_pra_032115, rossini_arxiv_07234, rosa_arxiv_07247, carrega_arxiv_03551, caravelli_prr_023095, barra_prl_210601,hovhannisyan_njp_052001, liu_jpc_18303, crescente_njp_063057, quach_arxiv_10044, carrega_arxiv_14034, zakavati_arxiv_09814, ghosh_arxiv_12859, hovhannisyan_prr_033413,bai_arxiv_06982,purves_arxiv_09065,mitchison_quantum_500} deal precisely with these questions and is currently one of the most active fields of research (see Ref.~[\onlinecite{quach_arxiv_06026}] for a recent experimental realization).  In this section, we introduce the fundamentals of quantum batteries and outline some of the recent advances in this rapidly developing field. We shall first review the concept of passive states, which are the target states of any unitary protocol that aims to facilitate maximal energy extraction from a quantum system. Subsequently, we shall also see how the presence of entanglement may speed up the process of energy deposition or energy extraction from the battery,  commonly referred to as \textit{charging} or \textit{discharging} the quantum battery, respectively. Before proceeding with the detail, we would like to remark here that majority of the work in this area so far has focused on unitary charging and discharging protocols. These include case studies  where the `charging' corresponds to  a unitary evolution of the battery under a time-dependent driving Hamiltonian as well as other cases in which energy is transferred from an external source (charger) to the battery under a global unitary evolution. We shall therefore limit the discussions mostly to unitary protocols and only highlight a couple of works dealing with dissipative protocols at the end of this section.

\subsection{Passive states and maximal work extraction}\label{subsec_maxwork}
In a unitary evolution of a quantum system, the von-Neumann entropy remains invariant, which follows directly from the fact that the eigen-values of the density matrix is not altered through a unitary transformation. However, the converse is not true -- two states having the same von-Neumann entropy are not necessarily connected by a unitary transformation, except in two-dimensional Hilbert spaces. This simple observation suggested that the maximum work (per unit cell) that can be extracted  from a  `battery' of cells can be higher than that possible from a single cell \cite{alicki_pre_042123}. To elaborate further, we first recall the notion of passive states \cite{pusz_cmp_273, lenard_jsp_575, allahverdyan_epl_565, campaioli_springer_book} which we had briefly touched upon previously in Sec.~(\ref{subsec_4stroke}).  

Let us consider a system with Hilbert space dimension $d$ and Hamiltonian $H_0$ such that $h_0=\sum_{j=1}^d\varepsilon_j\ket{j}\bra{j}$ with $\varepsilon_j\leq\varepsilon_{j+1}$. We are interested in the unitary evolution of the system when an arbitrary local interaction is switched on, so that the total Hamiltonian reads,
\begin{equation} 
h(t)=h_0+h_1(t).
\end{equation}
Note that this can also be interpreted as \textit{quenching} the system. The Hamiltonian $h_1(t)$ remains finite for $0\leq t\leq\tau$ and vanishes otherwise; the operation is thus cyclic in nature. Denoting the initial state  of the system as $\rho$, the maximum amount of work which can be extracted from it during this cyclic process is known as \textit{ergotropy} \cite{allahverdyan_pa_542, allahverdyan_epl_565, campaioli_springer_book, francica_npj_12, monsel_prl_130601,francia_prl_180603}, defined as,
{\begin{align}\label{eq_ergo}
\mathcal{E}^1=\overline{W}^1_{U,max}&=\Tr\left[\rho h_0\right] -\min_{U(\tau)\in SU(d)}\{\Tr\left[U(\tau)\rho U(\tau)^\dagger h_0\right]\}\nn\\
		&=\Tr\left[\rho h_0\right] -\Tr\left[\sigma_{\rho} h_0\right],
\end{align}}
where the superscript `$1$' denotes that we are working with a single \textit{cell} or single copy of the system, {$\overline{W}_{U,max}^1$} denotes the maximum work that can be extracted {(negative of the work done, $\overline{W}=-W$)} using unitary operations and $U(\tau)$ are unitary time-evolution operators acting for the duration of time $\tau$. Note that the minimization in the first equality above is over all $d$-dimensional unitary matrices (belonging to the $SU(d)$ group) and hence an implicit minimization over $\tau$ is also implied. In the second equality, $\sigma_{\rho}$ is the passive state corresponding to $\rho$, defined as the state having zero ergotropy. Hence, no energy can be extracted from $\sigma_\rho$ through cyclic unitary processes, i.e.
\begin{equation}
\Delta E=-\overline{W}^1_U=\Tr\left[U\sigma_\rho U^\dagger h_0\right]-\Tr\left[\sigma_\rho h_0\right]\geq 0, \quad \forall~U\in SU(d).
\end{equation}
where a positive value of $\Delta E$ corresponds to a negative work extraction, or equivalently, a work deposition on the system. In general, it can be shown that a state is passive if it is diagonal in the energy eigen-basis of the system with non-decreasing diagonal elements (populations), when arranged in the order of non-increasing energies, i.e. $\sigma=\sum_{j=1}^ds_j\ket{j}\bra{j}$ with $s_{j}\geq s_{j+1}$ for  $\varepsilon_{j}\leq\varepsilon_{j+1}$ \cite{pusz_cmp_273, lenard_jsp_575}. Consequently, the passive state $\sigma_\rho$ which can be attained by means of local unitary operations on $\rho$ is unique in nature. {Unless otherwise mentioned, we shall use the notation $\sigma_\rho$ to denote the unique passive state corresponding to $\rho$ in the rest of the article.}

\begin{figure}
	\includegraphics[width=0.6\textwidth]{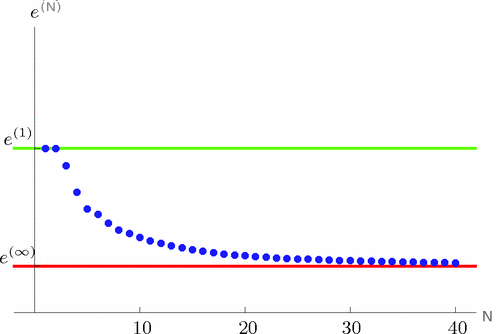}
	\caption{{Energy per copy $e^{(n)}=\Tr[\sigma_{\otimes^n\rho}H_0]/n$, of the passive state $\sigma_{\otimes^n\rho}$ obtained from a product state $\otimes^n \rho$ as a function of $n$. The Hamiltonian of the individual cells $h_{0,i}$ are chosen to be identical, each having eigen energies $\{0,0.579,1\}$ while the passive state $\sigma$ of a single cell is set to be $\{0.538,0.237,0.224\}$. (see Ref.~\onlinecite{alicki_pre_042123})}}\label{fig_batt_alicki}
\end{figure}

Since $\rho$ and $\sigma_{\rho}$ are connected by a unitary  transformation, they have the same Von-Neumann entropy, $S(\rho)=S(\sigma_\rho)$. However, there may exist other states that have the same entropy as $\rho$ but with  energies even lower than $\sigma_\rho$, although not accessible by unitary transformations alone. Using the fact that  the thermal state $\zeta_{\rho}=e^{-\beta h_0}/\Tr\left[e^{-\beta h_0}\right]$  minimizes the free energy,
\begin{equation}
\Tr(\rho h_0)-\frac{1}{\beta}S(\rho)\geq \Tr(\zeta_\rho h_0)-\frac{1}{\beta}S(\zeta_\rho),
\end{equation}
where the inverse temperature $\beta$ is determined by the entropy constraint $S(\zeta_\rho)=S(\rho)$, we have $\Tr(\rho h_0)\geq\Tr(\zeta_\rho h_0)$. Hence, the maximum work that can be extracted from $\rho$ is \cite{alicki_pre_042123},
\begin{equation}
\overline{W}_{max}^1=\Tr\left[\rho h_0\right]-\Tr[\zeta_\rho h_0]
\end{equation}
Hence, there remains an unattainable work of magnitude $\overline{W}_{max}^1-\overline{W}^1_{U,max}$ within a constraint of constant entropy, which can not be extracted by local unitary protocols acting on the system. 
 
However, the above scenario completely changes when one considers a \textit{battery} of cells where {each cell corresponds to an identical copy of $\rho$}. In this case, it becomes possible to attain a higher work extraction capacity per system than that possible from a single copy of the system. {To elaborate, let us denote the battery Hamiltonian as $H_0=\sum_i^N h_{0,i}$ where  $h_{0,i}$ is the Hamiltonian of the $i^{th}$ cell. Fig.~\ref{fig_batt_alicki} shows the energies  per copy, $e^{(N)}=\Tr[\sigma_{\otimes^N\rho}H_0]/N$, of the passive state $\sigma_{\otimes^N\rho}$ obtained from a product state $\otimes^N \rho$ as a function of $N$. As can be clearly seen, the energy per unit cell diminishes and saturates asymptotically to a value which is lower than the energy of a single cell}. 

The possibility arises because in general, for $N$ copies of the system, $\otimes^N \sigma_{\rho}$ may not necessarily be the same as $\sigma_{\otimes^N\rho}$.{  In fact, the equality holds true only for thermal states. In other words, if the passive state corresponding to $\rho$ is a thermal state, $\sigma_{\rho}=\zeta_{\rho}$, then it can be easily shown that $\otimes^N \zeta_{\rho}=\sigma_{\otimes^N\rho}$. Hence, thermal states are also known as completely passive states.} Let us now define the maximum extractable work per unit cell,
\begin{equation}
\overline{w}_{max}=\frac{\overline{W}^N_{max}}{N}=\frac{1}{N}\Big(\Tr\left[\otimes^N\rho H_{0}\right] -\Tr\left[\sigma_{\otimes^N\rho} H_{0}\right]\Big),
\end{equation}
In the limit $N\to\infty$, it can be shown that \cite{alicki_pre_042123},
\begin{equation}
\lim_{N\to\infty}\overline{w}_{max}=\overline{W}_{max}^1,
\end{equation}
In other words, the maximum work extracted per system or `cell' in a quantum battery can saturate the maximum available thermodynamic energy within the constraint of a constant entropy. {Another way to look at the higher energy extraction per cell is that, although the battery as a whole is driven unitarily, the dynamics of the individual cells are no longer necessarily unitary. This allows them to attain final states which have lesser energy than their corresponding passive states.}

\subsection{Entangling vs non-entangling protocols}

Let us suppose that we wish to extract the maximum work available from a battery through a unitary evolution $U$ such that $U(\otimes^N\rho) U^\dagger=\sigma_{\otimes^N\rho}$. The initial state of the battery $\otimes^N\rho$, by construction, is separable as it exists as a direct product of the states of the individual cells. The final state $\sigma_{\otimes^N\rho}$ is diagonal in the eigen-basis of the battery Hamiltonian,  $H_0=\sum_{i=1}^Nh_{0,i}$ and hence is also separable. We can now categorize the unitary operators $U$ facilitating the required work extraction into two groups -- one in which the battery remains separable at all intermediate times and another in which the individual cells are allowed to get entangled at intermediate times. 

As an illustration \cite{hovhannisyan_prl_240401}, consider the simple case in which the initial state $\otimes^N\rho$ is also diagonal in the energy basis but not passive, $\otimes^N\rho = \mathrm{diag}\left(p_1,~p_2,\dots,~p_{d^N}\right)$. The required unitary protocol for maximum work extraction therefore corresponds to a set of permutation operations which rearrange the populations of the density matrix. Note that permutation operations are non-local in nature and is capable of generating entanglement. Let us consider one such operation $\bra{\alpha}\otimes^N\rho\ket{\alpha}\equiv p_\alpha\leftrightarrows p_\beta\equiv\bra{\beta}\otimes^N\rho\ket{\beta}$, where $\ket{\alpha(\beta)}=\ket{i_1^{\alpha(\beta)},i_2^{\alpha(\beta)},\dots,i_N^{\alpha(\beta)}}$ with $\ket{i_j^{\alpha(\beta)}}$ denoting the state of the $j^{th}$ cell. One possible way to carry out this transposition while preserving the separability of the state at all times is to carry out a sequence of $2N-1$ operations as follows,
\begin{align}
	\ket{i_1^\alpha,i_2^\alpha,\dots,i_N^\alpha}\leftrightarrows\ket{i_1^\beta, i_2^\alpha,\dots,i_N^\alpha}&\leftrightarrows\ket{i_1^\beta, i_2^\beta,\dots,i_N^\alpha}\nn\\
	&\leftrightarrows\ket{i_1^\beta, i_2^\beta,\dots,i_N^\beta}.
\end{align}
On the other hand, if entanglement generation is permitted, than the above operation can be completed in a single step with a unitary operation of the form $U=\sum_{\mu\neq\alpha,\beta}\ket{\mu}\bra{\mu}+\ket{\alpha}\bra{\beta}+\ket{\beta}\bra{\alpha}$ \cite{hovhannisyan_prl_240401}. The second approach can be considered equivalent to taking a shortcut through the subspace of entangled states in the Hilbert space, as opposed to restricting to the subspace of separable states.

The above illustration provides two important results \cite{hovhannisyan_prl_240401}. Firstly, maximal work can be extracted from a battery without requiring any generation of entanglement between the individual cells. Secondly, a large number of operations are required to extract the work if the battery is to remain in a separable state at all instants of time, thus requiring a long time in the process. These observations naturally give rise to the question -- can quantum entanglement  provide an advantage in terms of the speed of the charging  or discharging of quantum batteries.

\subsection{Quantum speed limits}\label{subsec_qsl}
It is important to realize that for a set of fixed initial and final states (and hence a fixed energy difference), the speed of charging is intricately tied to the concept of quantum speed limits (QSL) \cite{deffner_jpa_335302,deffner_jpa_453001}. QSLs put a fundamental lower bound on the minimum time of evolution between two given states; these bounds are consequences of the energy time uncertainty relation $\Delta E\Delta t\geq1$. Various such bounds \cite{deffner_jpa_453001} have been proposed in literature depending on the evolution process under consideration. For our purpose, consider the evolution of the density matrix {$\rho(t)=\sum_i p_i\ket{\phi_i(t)}\bra{\phi_i(t)}$}, driven by a time-dependent Hamiltonian $H(t)$. The \textit{Bures angular distance} \cite{bengtsson_cambridge_book} defines the distance between two states $\rho_1$ and $\rho_2$ in the density matrix space as,
\begin{equation}\label{eq_bures_angular}
\mathcal{D}(\rho_1,\rho_2)=\arccos\left[\mathcal{F}(\rho_1,\rho_2)\right],
\end{equation}
where $\mathcal{F}(\rho_1,\rho_2)=\Tr\left[\sqrt{\sqrt{\rho_1}\rho_2\sqrt{\rho_1}}\right]$ is the Uhlmann's fidelity \cite{nielsen_cambridge_book}. For two states separated by an infinitesimal evolution, {it can be shown from Eq.~\eqref{eq_bures_angular}} that the distance is given by \cite{braunstein_prl_3439, deffner_jpa_453001},
\begin{equation}
\lim_{\Delta t\to 0}ds^2(\rho(t),\rho(t+\Delta t))=\frac{1}{4}\mathcal{H}_t\Delta t^2,
\end{equation}
where $\mathcal{H}_t$ is the QFI, previously discussed in Sec.~\ref{subsec_qfi}.  Using the general expression for QFI derived in Eq.~\eqref{eq_qfi} and the fact that the eigen values of $\rho$ remain invariant in unitary evolution, we evaluate the QFI with respect to time as follows,
\begin{align}
\mathcal{H}_t&=2\sum_{i\neq j}\frac{\left(p_i-p_j\right)^2}{p_i+p_j}\left|\bra{\phi_j}\frac{\partial}{\partial t}\ket{\phi_i}\right|^2\nn\\
&=2\sum_{i\neq j}\frac{\left(p_i-p_j\right)^2}{p_i+p_j}\Big|\bra{\phi_j}H(t)-\la H(t)\ra\ket{\phi_i}\Big|^2\nn\\
&\leq 2\sum_{i\neq j}\left(p_i+p_j\right)\Big|\bra{\phi_j} H(t)-\la H(t) \ra\ket{\phi_i}\Big|^2\nn\\
&=4\Delta H(t)^2,
\end{align}
where $\Delta H(t)=\sqrt{\la H^2(t)\ra-\la H(t)\ra^2}$ and we have used the inequality $(p_i+p_j)^2\geq(p_i-p_j)^2$ for $p_i,p_j\geq 0$. The equality is satisfied in the above equation for $p_ip_j=\delta_{i,j}$; or equivalently, for pure states.
The speed of evolution is therefore found to be,
\begin{align}
\frac{ds}{dt}=\frac{1}{2}\sqrt{\mathcal{H}_t}\leq\Delta H(t).
\end{align}
Thus, given a pair of initial and final states $\rho(0)=\rho_i$ and $\rho(\tau)=\rho_f$, respectively, the time required to traverse the distance between them is calculated by integrating over the Bures distance between them and the time taken,
\begin{equation}
\int_0^{\mathcal{D}(\rho_i,\rho_f)}ds \leq \int_0^\tau\Delta H(t)dt
\end{equation}
Therefore, we obtain a lower bound on the minimum time required in any unitary evolution as,
\begin{equation}\label{eq_mt_qsl}
\tau\geq\tau_{QSL}\geq\frac{\mathcal{D}(\rho(t),\rho(t+\tau))}{\Delta E_\tau},
\end{equation}
where,
\begin{equation}\label{eq_energy_ta_var}
\Delta E_\tau = \frac{1}{\tau}\int_0^\tau\Delta H(t)dt.
\end{equation}
As already mentioned, the second inequality in Eq.~\eqref{eq_mt_qsl} saturates for pure states $\rho(t)=\ket{\psi(t)}\bra{\psi(t)}$ and we obtain,
\begin{equation}
\tau_{QSL}^{pure}=\frac{\arccos\left|\braket{\psi(t)|\psi(t+\tau)}\right|}{\Delta E_\tau}
\end{equation}
The bound derived in Eq.~\eqref{eq_mt_qsl} is known as the Mandelstam-Tamm \cite{mandelstam_jp_249} bound for arbitrary mixed states. Sometimes, a unified bound \cite{levitin_prl_160502} is used which combines the Mandelstam-Tamm bound  with the Margolus-Levitin bound \cite{margolus_pd_188} and is given as,
\begin{equation}\label{eq_uni_qsl}
\tau\geq\tau_{QSL}\geq\frac{\mathcal{D}(\rho(t),\rho(t+\tau)}{\mathrm{min}\{\Delta E_\tau, E_\tau\}}, 
\end{equation}  
where,
\begin{equation}\label{eq_energy_ta_mean}
E_\tau = \frac{1}{\tau}\int_0^\tau\ H(t)dt.
\end{equation}

Note that the Bures angular distance considered above measures the physical distinguishability of the initial and final states. However, from the perspective of quantum batteries, it is also important to consider how much the states are distinguishable in terms of their average energy. With this motivation, a different speed limit was recently analyzed in Ref.~[\onlinecite{farre_prr_023113}], where the distance between a pair of states is measured in the energy space as follows. Let us consider the probability distribution of a pair of density matrices $\rho_1$ and $\rho_2$ in the energy eigen basis as, $p_k=\Tr\left[\rho_1\Pi_k\right]$ and $q_k=\Tr\left[\rho_2\Pi_k\right]$, respectively, where $\Pi_k=\{\ket{k}\bra{k}\}$ is the projection operator on the energy eigen-state $\ket{k}$. The distance is then defined as the relative entropy distance or the Kullback-Liebler divergence between these two distributions \cite{bengtsson_cambridge_book},
\begin{equation}
\mathcal{D}_{KL}(p,q)=\sum_kp_k\log_2\frac{p_k}{q_k}.
\end{equation}
For states separated by infinitesimal time $\Delta t$, the distance can be expanded upto second order in $\Delta t$ as,
\begin{equation}
\lim_{\Delta t\to0}\mathcal{D}_{KL}(p(t),p(t+\Delta t))=\sum_k\frac{1}{2p_k}\left(\frac{d p_k}{d t}\right)^2\Delta t^2.
\end{equation}	 
The speed limit in the energy space is therefore defined as \cite{farre_prr_023113},
\begin{equation}\label{eq_qsl_energy}
v_E(t)=\lim_{\Delta t\to0}\frac{\sqrt{\mathcal{D}_{KL}(\rho(t),\rho(t+\Delta t))}}{\Delta t}=\sqrt{\frac{\mathcal{I}_E(t)}{2}},
\end{equation} 
where
\begin{equation} \label{eq_fi_ene}
\mathcal{I}_E(t)=\sum_k\frac{1}{p_k}\left(\frac{d p_k}{d t}\right)^2,
\end{equation}
is the classical Fisher information (see Eq.~\eqref{eq_fi}) defined on the energy eigen-space.

\subsection{Quantum advantage}
The power of a quantum battery is defined as the speed at which energy can be deposited on it (charged) or the rate at which useful work can be extracted from it (discharged). Any enhancement in the power of a quantum battery resulting from quantum effects such as entanglement is dubbed as \textit{quantum advantage} \cite{campaioli_prl_150601}. As we shall see below, QSLs enforce certain bounds on any possible quantum advantage in the charging or discharging process of quantum batteries. We discuss these bounds by analyzing QSL restrictions on both the Hilbert space as well as the energy space.

\subsubsection{From speed limit considerations in Hilbert space}
The comparison of the charging power of quantum batteries using local (non-entangling) and non-local (entangling) operations was first considered in Refs.~[\onlinecite{campaioli_prl_150601},\onlinecite{binder_njp_075015}]. It was argued that a quantum advantage can be meaningful only when the thermodynamic resources available to the entangling protocols is same as that of non-entangling protocols. In other words, any trivial enhancement of the charging power simply resulting from availability of more energy in the driving Hamiltonian must be properly accounted for. This necessitates appropriate re-scaling of the driving Hamiltonian in the entangling protocol, which ultimately results in an extensive scaling of the quantum advantage. 

To elaborate, let us consider a quantum battery composed of $n$ cells. We wish to compare two charging protocols as given below:
\begin{enumerate}
	\item \textit{Parallel charging}: Each of the battery cell is driven simultaneously under the action of an identical Hamiltonian,
	\begin{equation}
	H^{\parallel}(t) = h_0+V_{\parallel}(t),
	\end{equation}
	were $h_0$ is the un-driven Hamiltonian of a single cell.
	\item \textit{Collective charging}: The battery is collectively driven with the Hamiltonian,
	\begin{equation}
	H^{\#}(t)=H_0 + V_{\#}(t),
	\end{equation}
	where $H_0=\sum_{i=1}^Nh_{0,i}$, with $h_{0,i}=h_0~\forall i$.
\end{enumerate} 
The charging takes place for a duration $\tau$ and $V^{\parallel}(t)=V^{\#}(t)=0$ for $t<0, t>\tau$. We also assume that the same  final state is achieved using both the protocols starting from a given initial state. The \textit{quantum advantage} for collective charging is then defined as the ratio of the power for collective and parallel charging \cite{campaioli_prl_150601},
\begin{equation}
\Gamma = \frac{P^{\#}}{P^{\parallel}}=\frac{\tau^{\parallel}}{\tau^{\#}}.
\end{equation}

We now make use of the QSL results discussed in Sec.~\ref{subsec_qsl}; specifically in context of the inequality presented in Eq.~\eqref{eq_uni_qsl}. As already mentioned, one must rule out any trivial enhancements in power due to transfer of more energy. For this purpose, let us first put a constraint (C1) on the driving Hamiltonian of the form $E_\tau^{\#}\leq NE_\tau^{\parallel}$, where $E_\tau^{\#,\parallel}$ is the time-averaged mean energy defined in Eq.~\eqref{eq_energy_ta_mean}. From Eq.~\eqref{eq_uni_qsl}, we therefore obtain,
\begin{equation}
 \tau^{\#}\geq\tau^{\#}_{QSL}\geq \mathcal{D}_N/E_\tau^{\#}\geq\frac{\mathcal{D}_N}{NE_\tau^{\parallel}},
\end{equation}
and hence the quantum advantage is upper bounded as,
\begin{equation}\label{eq_qa_c1}
\Gamma_{C1} \leq \frac{NE_\tau^\parallel}{\mathcal{D}_N}\tau^\parallel=N\beta\frac{\mathcal{D}_1}{\mathcal{D}_N},
\end{equation}
where $\beta=\tau^\parallel/\tau_{QSL}^\parallel\geq1$ and $\mathcal{D}_{1}(\mathcal{D}_N)$ is the Bures angular distance between the initial and final density matrices of the cell (battery). Similarly, if the variance is constrained (C2) in stead of the mean energy, one can show that,
\begin{equation}\label{eq_qa_c2}
\Gamma_{C2}\leq \sqrt{N}\beta\frac{\mathcal{D}_1}{\mathcal{D}_N}.
\end{equation} 

The inequalities derived above shows that collective charging protocols provide a quantum advantage over parallel ones  that scales extensively. The exact scaling depends on the particular choice of the constraint applied on the driving Hamiltonian of the collective charging protocol. However, {it is important to note that although non-local `entangling' operations are necessary for a quantum advantage, generation of entanglement between the battery cells is not required. This can be seen as follows. For $N$ systems of $d$ dimensions each, there exist a ball of radius $R(N, d)$ with the maximally mixed state at its centre, within which all the states are separable in the individual systems \cite{gurvits_pra_032322, gurvits_pra_062311}. If the initial state of the battery lies within this \textit{separable ball}, then it continues to remain so as the distance from the maximally mixed state cannot change in a unitary process. Hence, the quantum advantage can be achieved even though the cells remain separable at all instants of time during the charging process \cite{campaioli_prl_150601}.} 

In practical situations, it is difficult to engineer very long-range non-local interactions. This brings into question the scaling of the quantum advantage for arbitrarily large battery size. In fact, it turns out that for a driving Hamiltonian containing $k$-local terms (terms which act over $k$ cells at any given time) and where each individual cell interacts with only $m$ other cells, the quantum advantage scales as \cite{campaioli_prl_150601}
\begin{equation}\label{eq_bound_hil}
\Gamma\leq\gamma\left[k^2(m-1)+k\right].
\end{equation}
The constant $\gamma$ is independent of battery size and hence there is no extensive scaling of the quantum advantage with $n$. We note though that, unlike the constraints $C1$ and $C2$,  this bound is derived with a constraint imposed on the operator norm of the driving Hamiltonian\cite{campaioli_prl_150601}, $\mathcal{E}_\tau^\mathcal{\#}\leq N\mathcal{E}_\tau^\mathcal{\parallel}$, with,
\begin{equation}
\mathcal{E}_\tau^{\mathcal{\parallel}(\mathcal{\#})}=\frac{1}{\tau_{\mathcal{\parallel}(\mathcal{\#})}}\int_0^{\mathcal{\parallel}(\mathcal{\#})}||H^{\mathcal{\parallel}(\mathcal{\#})}(t)||_{\mathrm{op}}dt,
\end{equation} 
where $||H(t)||_{\mathrm{op}}$ denotes the largest singular value of $H(t)$.
\subsubsection{From speed limit considerations in energy space}
An alternate bound on the instantaneous power of a quantum battery can be derived using the QSL restrictions in the energy eigen space (see Eq.~\eqref{eq_qsl_energy}). Considering a driving Hamiltonian of the form $H(t)=H_0+V(t)$, the instantaneous power is defined as,
\begin{equation}
P(t)=\frac{d}{dt}\la H_0\ra=\frac{d}{dt}\Tr\left(\rho(t)H_0\right).
\end{equation}
Expanding in the eigen-basis of $H_0=\sum_j\varepsilon_j\ket{j}\bra{j}$, we can write,
\begin{equation}
P(t)=\sum_j\varepsilon_j\frac{dp_j(t)}{dt}=\sum_j(\varepsilon_j-c(t))\frac{dp_j(t)}{dt},
\end{equation}
where $p_j(t)=\Tr\left(\rho(t)\ket{j}\bra{j}\right)$ and $c(t)\in \mathbb{R}$. Note that in deriving the second equality, we have used $\sum_jdp_j/dt=d(\sum_jp_j)/dt=0$. Next, we square and rearrange the above equation as,
\begin{equation}
P^2(t)=\left(\sum_j(\varepsilon_j-c(t))\sqrt{p_j}\frac{1}{\sqrt{p_j}}\frac{dp_j(t)}{dt}\right)^2.
\end{equation}
Applying the Cauchy-Schwarz inequality, we arrive at,
\begin{equation}
P^2(t)\leq\left(\sum_jp_j(\varepsilon_j-c(t))^2\right)\left(\sum_j\frac{1}{p_j}\left(\frac{dp_j}{dt}\right)^2\right)
\end{equation}
It is easy to check that the r.h.s is minimized for $c(t)=\la H_0\ra$. We therefore arrive at the inequality \cite{farre_prr_023113},
\begin{equation}\label{eq_bound_ene}
P^2(t)\leq\Delta H_0(t)^2\mathcal{I}_E(t),
\end{equation}
where $\Delta H_0(t)^2$ is the variance of the un-driven or bare Hamiltonian of the battery and $\mathcal{I}_E(t)$ is the Fisher information defined on the energy space (see Eq.~\eqref{eq_fi_ene}). The above inequality implies that the power of a quantum battery at a particular instant is crucially dependent on two aspects of the charging/discharging protocol. Firstly, it depends on how much non-local the driving protocol is in the energy space as quantified by the instantaneous variance. In other words, it depends on how many of the energy eigen-states have a significant population at that instant of time. Secondly, it also depends on the instantaneous speed of the driving protocol in the energy space, where the distance between the states is measured in terms of their energetic difference. {For later discussions, it is useful to define a quantity,
\begin{equation}\label{eq_bound_tight}
	\cos\theta_P=\frac{P}{\sqrt{\Delta H_0^2\mathcal{I}_E}},
\end{equation}
which gives an estimate of the tightness of the bound for a given battery setup}.

Let us now have a closer look at the variance $\Delta H_0(t)^2$. Assuming a battery of $N$ identical cells, the bare Hamiltonian reads $H_0=\sum_{i=1}^Nh_i$ and the variance can be expanded as \cite{farre_prr_023113},

\begin{align}\label{eq_batt_scal_entang}
\Delta H_0(t)^2&=\Tr\left[\left(\sum_{i=1}^Nh_i\right)^2\rho(t)\right]-\Tr\left[\left(\sum_{i=1}^Nh_i\right)\rho(t)\right]^2\nn\\
&=\sum_{i=1}^N\left(\Tr\left[h_i^2\rho(t)\right]-\Tr\left[h_i\rho(t)\right]^2\right)+\sum_{i\neq j}^N\Big(\Tr\left[h_ih_j\rho(t)\right]-\Tr\left[h_i\rho(t)\right]\Tr\left[h_j\rho(t)\right]\Big)
\end{align}  

The first term in the second equality above captures the total contribution of the \textit{local variance} from each cell and therefore scales linearly with the battery size $N$. For the parallel charging protocol, the second term vanishes  and hence the maximum power follows the same linear scaling. Thus, a quantum advantage can be achieved for protocols for which the second term remains finite and as well as scales faster than $N$. However, the second term is non-zero only when the cells are entangled and hence the protocol must generate entanglement between the battery cells to achieve any quantum advantage. The variance therefore bridges a connection between the scaling of the quantum advantage and entanglement generating protocols.

It is important to note that the upper bound of the power in Eq.~\eqref{eq_bound_ene} and the resulting extensive scaling of the quantum advantage has been derived without incorporating any constraint or re-scaling of the driving Hamiltonian. This is unlike the bounds (of the quantum advantage) derived in Eqs.~\eqref{eq_qa_c1} and~\eqref{eq_qa_c2} where the constraints imposed on the driving Hamiltonians resulted in the extensive scaling of the quantum advantage.

Once again, it is also necessary to analyze the scaling of the bound with the range of entanglement generated in the system. For a battery consisting of $N$ qubits where at most $k$ qubits are entangled \cite{chen_pra_052302,guhne_njp_229} at a given time $t$, it can be shown that the variance satisfies \cite{toth_pra_022322, hyllus_pra_022321}, $4\Delta H_0(t)^2\leq rk^2+(N-rk)^2$, where $r$ is the integer part of $N/k$. The power is therefore bounded as \cite{farre_prr_023113},
\begin{equation}
P(t)^2\leq\frac{1}{4}\left(rk^2+(N-rk)^2\right)\mathcal{I}_E(t).
\end{equation}
The inequality derived above thus provides a deeper insight into how the instantaneous entanglement generated during the charging or discharging protocol affects the power of a quantum battery. 

{Before concluding this subsection, we would like to note that very recently, a rigorous bound was derived on the maximum scaling power of quantum batteries \cite{kim_arxiv_02491}. It was shown that the quantum advantage can at most be extensive with the size of the battery and such an advantage can be obtained only with global charging protocols in which all the battery cells are charged collectively.}  

\subsection{Role of inherent entanglement}

So far, we have only considered entangling protocols in charging batteries composed of independent cells. How does the situation change if some amount of entanglement pre-exists among the battery cells and is not generated by the charging Hamiltonian? This question was addressed by Le \textit{et.al.} \cite{le_pra_022106}, who considered a many-body spin chain with two-body interactions as a model of quantum battery and analyzed the power while charging with a local external driving field. In particular, they considered the XXZ Hamiltonian as the bare or un-driven Hamiltonian of the battery, which is given by,
\begin{subequations}
\begin{equation}
H_0 = H_B + H_g,
\end{equation} 
where,
\begin{equation}
H_B = B\sum_{i=1}^n\sigma_i^z,
\end{equation}
\begin{equation}
H_g = -\sum_{i<j}g_{i,j}\left[\sigma_i^z\sigma_j^z+\alpha\left(\sigma_i^x\sigma_j^x+\sigma_i^y\sigma_j^y\right)\right].
\end{equation}    
\end{subequations}
In the above set of equations, $B$ corresponds to an external field in the transverse direction and the coupling $g_{i,j}>0$ may either be short ranged ($g_{i,j}=g\delta_{i,j-1}$) or long-ranged ($g_{i,j}=g|i-j|^{-\nu}$ with $\nu>0$). To charge the battery, the transverse field is switched off and a perpendicular field is applied, so that the battery evolves under the action of the Hamiltonian,
\begin{subequations}
\begin{equation}
H_c=H_g+V,
\end{equation}
where 
\begin{equation}
V = \omega\sum_{i=1}^N\sigma_i^x.
\end{equation}
\end{subequations}

The initial state is assumed to be the ferromagnetic ground state $\rho(0)=\ket{\downarrow}\bra{\downarrow}^{\otimes N}$ where $\{\ket{\uparrow},\ket{\downarrow}\}$ is the eigen-basis of $\sigma^z$.
For the isotropic XXX chain ($\alpha=1$), it turns out that the power scales linearly with system size and no quantum advantage is obtained. This is because the interacting part of the Hamiltonian $H_g$ commutes with $H_c$ for the isotropic model and hence plays no role during the evolution process. Hence, the chain effectively behaves as a collection of $N$ spins which results in the linear scaling of the power with battery size. 

In the strong coupling regime $g\gg\omega$ of the anisotropic ($\alpha\neq1$) model, the power is worse than that of the parallel charging case. However, in the weak coupling regime $\sum_{i<j}g_{i,j}\ll n\omega$, the power scales as $\sim\mathcal{O}(\log N)$ when $\nu=1$ and $\sim\mathcal{O}(N)$ when $\nu\to 0$. For $\nu>1$ and nearest-neighbor models, the power is enhanced only by a constant factor independent of $N$. Note that these results are in agreement with the bounds in quantum advantage derived for $k=2$ order interactions with $m$ number of participating cells derived in Eq.~\eqref{eq_bound_hil}. An extensive scaling of the quantum advantage is possible only in the limit of $\nu\to0$ when $m\sim N-1$, while it scales as $\sim\mathcal{O}(2m)=\mathcal{O}(const)$ for $\nu>1$ when $m$ is finite. However, the crucial difference is that the advantage is facilitated by the entanglement generated by the internal Hamiltonian $H_g$ and not by the charging field $V$.
 
\subsection{Usability of stored energy}\label{subsec_usable_batter}

The quality of a quantum battery is not only determined by its energy storage capacity and rate of charging or discharging, but also on the `usability' of the stored energy, which may depend on a number of factors. As for example, a battery having a high uncertainty of the stored energy, characterized by the  ratio $\Delta H_0(\tau)/E(\tau)$ where $E(\tau)$ is the mean energy stored, can be potentially hazardous to use for work extraction \cite{friis_q_61, farre_prr_023113}. This is because a high variance typically means that the energy which the battery can deliver has a large fluctuation about its mean energy. Similarly, large temporal fluctuations during the charging process implies that the amount of stored energy is highly sensitive to the charging duration, thereby making it difficult to estimate the stored energy. 

Another important figure of merit of the quality of the stored energy is the fraction of the mean energy available for extraction after the charging process. Mathematically, it is defined as \cite{andolina_prl_047702, rossini_prb_115142},
\begin{equation}
f(\tau) = \frac{\mathcal{E}(\tau)}{E(\tau)},
\end{equation}  
where $\mathcal{E}(\tau)$ and $E(\tau)=\la H_0\ra$ are the ergotropy (see Eq.~\eqref{eq_ergo}) and mean energy of the battery, respectively, after charging for a duration of time $\tau$. This fraction is unity if the ground state energy is zero and the battery is in a pure state. Therefore, the quantity $f$ becomes relevant when the battery exist in a mixed state. This may happen for \textit{charger-battery} systems (see Sec.~\ref{subsubsec_charger_assist} below) where energy is transferred from another system which acts as a 'charger' to the battery. While the composite system evolves unitarily and hence remains pure if initialized in the same, the reduced state of the battery may be mixed in case of entanglement generation between the charger and the battery. Similarly, in some cases, one may need to consider energy extraction from only a subset of $M$ number of cells from the battery as the full system may not be accessible \cite{andolina_prl_047702, rossini_prb_115142}. The reduced system of the $M$ cells $\rho_M$ is likely to exist in a mixed state and hence one needs to consider the fraction of usable energy from this subset.

It has been argued (and numerically demonstrated for some models) \cite{friis_q_61, andolina_prl_047702, rossini_prb_115142} that, whereas entanglement provides a boost in the optimal charging power of batteries, the same may be responsible for lesser availability of usable energy from the battery. This is because of the fact that the amount of entanglement within the battery cells directly influences the mixed nature of $\rho_M$. Similarly, highly entangled states of the battery and charger also renders the battery state mixed, thereby leading to $f(\tau)\ll 1$. However, for systems having certain integrals of motion in which the dynamical evolution is constrained to a small part of the full Hilbert space, it is expected that in the thermodynamic limit \cite{andolina_prl_047702},
\begin{equation}
\lim_{N\to\infty}f(\tau)=1.
\end{equation}
This can be understood from the fact that the entanglement entropy of subsystems in integrable systems (a measure of bipartite entanglement), does not scale extensively with the system size. On the other hand, the amount of energy stored in the battery scales linearly with the size of the battery. Hence, in the thermodynamic limit, the energy locked away due to entanglement is expected to be negligible in comparison with the energy stored. The presence of entanglement thus serves as a double-edged sword. {As we shall see in the specific models below, while it may provide quantum advantage in terms of charging power in special settings, the usability of the energy stored in the process may turn out to be far from optimum.}  

\begin{figure}
	\includegraphics[width=0.5\textwidth]{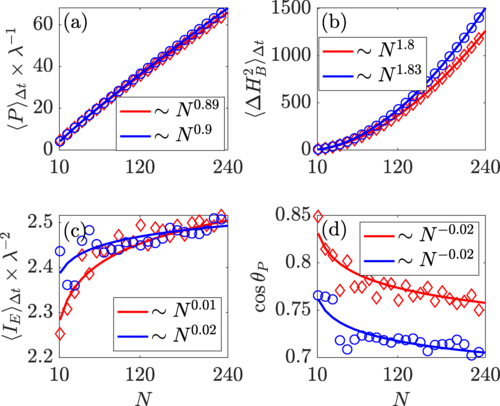}
	\caption{{Scaling of the (a) power, (b) variance of the bare Hamiltonian (here $H_B$ is the same as $H_0$) (c) Fisher information in energy space and (d) $\cos\theta_P$ (see Eq.~\eqref{eq_bound_tight}) averaged over a charging time $\Delta t$ with battery size for the LMG battery with charging Hamiltonian as defined in Eq.~\eqref{eq_batt_lmg}. The final time has been chosen as the one for which the energy capacity is maximized. The blue colored plots correspond to $\lambda = 20$,
	whereas the red colored plots correspond to $\lambda = 5$. (see Ref.~[\onlinecite{farre_prr_023113}] for details).}} \label{fig_farre_lmg}
\end{figure}

\subsection{Models}
\subsubsection{Spin models}

A plethora of physically realizable models of quantum batteries have been proposed over the past few years. A common starting point in a large chunk of these models is to consider an array of non-interacting spins or qubits, placed in a local external field. The Hamiltonian of the battery thus reads,
\begin{equation}\label{eq_hamil_spin_batt}
H_0=B\sum_i^N{\sigma_i^z},
\end{equation}
while a charging Hamiltonian $V(t)$ is used for the charging process so that the battery evolves under the action of the Hamiltonian,
\begin{equation}\label{eq_hamil_spin_charge}
H(t)=H_0+V(t)=B\sum_{i}^N\sigma_i^z+V(t)
\end{equation}
The exact form of charging Hamiltonian $V(t)$ characterizes the model under consideration and the capacity and power can then be analyzed. We discuss some of these models below.

Firstly, one can rule out all integrable models of spin chains, such as the Ising, XY or extended Ising models, as potential candidates for attaining any quantum advantage \cite{farre_prr_023113}. This is because the Hilbert space of such models have a local decoupled structure in the quasi-momentum basis. As the size of the quasi-momentum basis increases linearly with the system size, the power can also be shown to scale linearly, at best. Also, the different quasi-momentum modes do not maximize their local expectation energy simultaneously, which severely limits the maximum storage capacity of these models.

{Secondly, it is important to keep in mind that the presence of interactions that can generate long-range entanglements may not guarantee a quantum advabtage. This is best illustrated in the case of the} Lipkin-Meshkov-Glick (LMG) Hamiltonian \cite{lipkin_npa_188},
\begin{equation}\label{eq_batt_lmg}
H_{LMG} = \frac{\lambda}{N}\sum_{i<j}\left(\sigma_i^x\sigma_j^y+\gamma\sigma_i^y\sigma_j^y\right),
\end{equation}
which appears to be a good candidate for $V(t)$ that could attain a quantum advantage as the long range uniform two-body interactions are capable of generating long-range entanglements. However, as shown in Ref.~[\onlinecite{farre_prr_023113}], this is not the case as shown in Fig.~\ref{fig_farre_lmg} and can be understood by inspecting Eq.~\eqref{eq_bound_ene}. Although the variance indeed shows a quadratic scaling due to presence of long-range eantanglements , the Fisher information defined on the energy space is independent of the system size $N$ (see Figs.~\ref{fig_farre_lmg}~(b) and~\ref{fig_farre_lmg}~(c)). As a result, the power scales linearly (Fig.~\ref{fig_farre_lmg}~(a)) and no quantum advantage emerges for the LMG battery. {Indeed, it remains an open question whether it is possible to identify or construct charging Hamiltonians based on spin models that can truly achieve a quantum advantage in charging of quantum batteries.} 

\begin{figure}
	\includegraphics[width=0.5\textwidth]{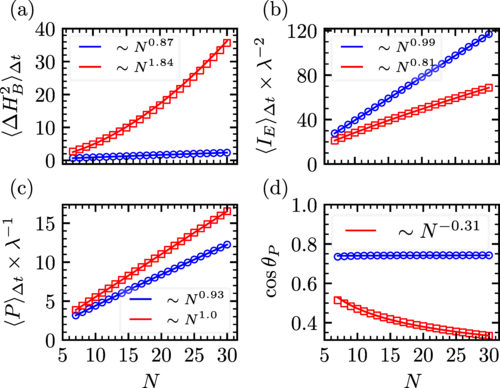}
	\caption{{Scaling of the (a) power, (b) variance of the bare Hamiltonian (here $H_B$ is the same as $H_0$) (c) Fisher information in energy space and (d) $\cos\theta_P$ (see Eq.~\eqref{eq_bound_tight}), averaged over a charging time $\Delta t$ for the Dicke battery (see Eq.~\eqref{eq_batt_dicke}) with the coupling scaled as $\lambda\to\lambda/\sqrt{N}$ and $\omega=\omega_c=1$. The final time has been chosen as the one for which the energy capacity is maximized. The blue colored plots correspond to $\lambda = 0.01$ (weak-coupling),
	whereas the red colored plots correspond to $\lambda = 0.5$ (strong-coupling). (see Ref.~[\onlinecite{farre_prr_023113}] for details).}}\label{fig_farre_dicke}
\end{figure}

\subsubsection{Cavity assisted charging}\label{subsubsec_charger_assist}

A collection of TLSs charged collectively through cavity induced excitations also forms a promising model of quantum battery, as the collective charging may naturally lead to a quantum advantage. The Dicke model, described by the Hamiltonian \cite{dicke_pr_99},
\begin{equation}\label{eq_batt_dicke}
H_{DK}=\omega J_z+ \omega_ca^\dagger a+2\omega_c\lambda J_x\left(a+a^\dagger\right),
\end{equation}
was explored in Ref.~[\onlinecite{ferraro_prl_117702}] as the charging Hamiltonian of a quantum battery composed of $N$ TLSs. In the above equation, $J_{\alpha}\sum_{i=1}^N\sigma_i^\alpha$ and the operator $a$ ($a^\dagger$)  annihilates (creates) a cavity photon. The model is initialized in a direct product state of the ground state of the TLSs and a Fock state, $\ket{\psi(0)}=\ket{G}\otimes\ket{n}$, where $\ket{G}=\otimes^N\ket{g}$ is the ground state of $N$ TLSs and $\ket{n}$ is the Fock state in the Hilbert space of the cavity with $n$ number of photons. The charging occurs when a finite $\lambda$ is switched on. Note that this setup is different from the ones previously discussed in the way that the state of both the TLSs which form the battery as well as the cavity which acts like a \textit{charger}, is tracked during the evolution. However, only the energy deposited on the TLSs is considered relevant in the context of energy storage and hence the battery Hamiltonian is considered to be $H_0=\omega J_z$. Such battery-charger composites have also been explored in a number of other works \cite{zhang_pre_052106, andolina_prl_047702, andolina_prb_205423, farina_prb_035421, andolina_prb_205437,crescente_arxiv_09791}. 

Under resonant conditions ($\omega=\omega_c$), the charging power of the above model was compared  to that of a \textit{parallel} model where the $N$ TLSs interact with $N$ distinct cavities. Through numerical analysis, it was found that the model indeed exhibits a quantum advantage which scales as $\sim \sqrt{N}$. However, if one re-scales the coupling between the TLSs and the cavity as $\lambda\to\lambda/\sqrt{N}$ so as to have a well defined thermodynamic limit, the quantum advantage has been shown to disappear with the power scaling only linearly with $N$ as shown in Fig.~\ref{fig_farre_dicke}~(c). \cite{farre_prr_023113}. This is despite the fact that the variance of the bare Hamiltonian $\Delta H_0^2$ with $H_0=\omega J_z$ scales quadratically with $N$ in the strong-coupling regime (Fig.~\ref{fig_farre_dicke}~(a)) and the Fisher information $\mathcal{I}_E$ scales linearly in both the strong and weak coupling regimes(Fig.~\ref{fig_farre_dicke}~(b)). The absence of super extensive scaling of the quantum advantage has been attributed to a poor saturation of the bound in Eq.~\eqref{eq_bound_ene} as shown in Fig.~\ref{fig_farre_dicke}~(d).

In addition, the super-extensive scaling of the variance in the strong-coupling regime is also found to be present at the end of the charging protocol, which means the ratio $\delta H_0(\tau)/E(\tau)$ does not vanish even in the thermodynamic limit, thereby undermining the quality of the energy stored \cite{farre_prr_023113} (see Sec.~\ref{subsec_usable_batter}). In addition, a significant amount of entanglement is also found to be present between the TLSs and the cavity in the final state which implies that a large fraction of the energy stored is likely to be unavailable for extraction.

\begin{figure}
	\includegraphics[width=0.8\textwidth]{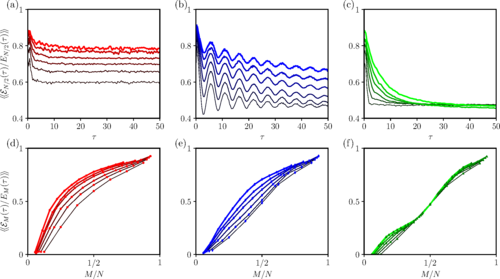}
	\caption{{(a-c)The fraction of energy available for extraction from half  of the battery  $\la\la f_{N/2}(\tau)\ra \ra=\la\la\mathcal{E}_{N/2}(\tau)/E_{N/2}(\tau)\ra\ra$, as a function of charging time $\tau$. (d-f) The extractable energy $f_M(\tau)$ at a fixed time $\tau=32$ as a function of $M/N$. (a) and (d) correspond to the AL phase; (b) and (e) correspond to the MBL phase and (c) and (f)
			correspond to the ergodic phase. The solid lines thicken with increasing $N = 8, 10, 12, 14, 16, 18$. The averaging is performed over 103 disorder realizations for $N$ up to $14$, and over 102 realizations in the cases $N = 16$ and $18$. (see Ref.~\onlinecite{rossini_prb_115142})}}\label{fig_rossini_mbl}
\end{figure}

\subsubsection{Disordered chains}

Disordered quantum many-body systems are known to host certain `localized' phases which may have potential applications in constructing quantum batteries as hinted in recent results. Consider once again, a battery composed of an array of spins with the bare and the driving Hamiltonians given as in Eqs.~\eqref{eq_hamil_spin_batt} and \eqref{eq_hamil_spin_charge}, respectively.  The charging Hamiltonian is of the form \cite{rossini_prb_115142} $V(t)=\lambda(t)H_1$, where

\begin{equation}
H_1=-\lambda(t)\sum_{i=1}^N\left(J_i\sigma_i^x\sigma_{i+1}^x-J_2\sigma_i^x\sigma_{i+2}^x\right).
\end{equation}  
As before, we have $\lambda(t)=1$ during the charging time $0\leq t\leq\tau$ and is zero otherwise. The nearest-neighbor coupling is of the form $J_j=J+\delta J_j$, where $\delta J_j$ is random chosen from a uniform distribution satisfying $-\delta J \leq\delta J_j\leq\delta J$. For $J_2=0$ the model described by the Hamiltonian $H_0+H_1$ exist in the Anderson-localized (AL) \cite{anderson_pr_1492} phase. Similarly, for $J_2\neq 0$ and $\delta J> \delta J^c>0$, the model exhibits a many-body localized (MBL) phase \cite{nandkishore_arcmp_15, alet_crp_498, abanin_rmp_021001}. On the contrary for $J_2\neq 0$ and $\delta J< \delta J^c$, the spectrum of the model is characterized by a \textit{mobility edge}, i.e.  all energy eigen-states below an energy threshold are localized while states with energy above the threshold are thermal and show ergodic behavior.

\begin{figure}
	\includegraphics[width=0.8\textwidth]{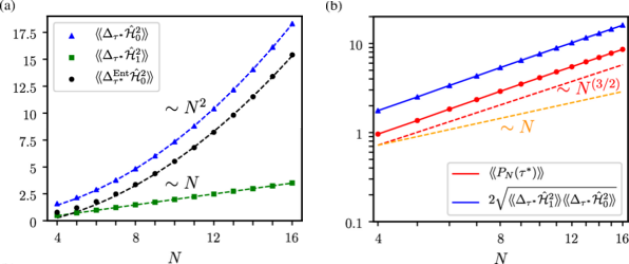}
	\caption{{(a) Scaling (in units of $J^2$) of the variance of the bare Hamiltonian, charging Hamiltonian and the entanglement contribution (see Eq,~\eqref{eq_entang_contr}) to the variance of the bare Hamiltonian for the SYK battery (see Eq.~\eqref{eq_syk_battery} and the following discussion). The dashed lines are linear (green) and quadratic (blue and black) fits to the numerically calculated  data. (b) Scaling of the power (in log-log scale) at optimal time and comparison to a weaker form of the bound in Eq.~\eqref{eq_bound_ene} (see Ref.~[\onlinecite{rossini_arxiv_07234}] for details). The dashed lines are provided for visual reference to deduce a rough estimate of the scaling exponent. All the quantities in the figure are disorder averaged over time-averaged variances and calculated for the optimal charging time $\tau^*$ for which the power is found to be maximum. the disorder average is performed over 		$10^3$ (for $N = 4; …; 10$), $5 \times 10^2$ (for $N = 11, 12$) and $10^2$ (for $N = 13; …; 16$) instances of disorder in the couplings $\{J_{i;j;k;l}\}$.}} \label{fig_rossini_syk}
\end{figure}

Numerical analysis have shown that the fraction of the stored energy in the battery which can be subsequently extracted as work, i.e. the quantity $f(\tau)$ discussed in Sec.~\ref{subsec_usable_batter}, drastically improves in the localized phases of the AL and the MBL as compared to ergodic phases \cite{rossini_prb_115142}. {This is illustrated in  the top panel of Fig.~\ref{fig_rossini_mbl}, where the extractable energy from half the system size $N/2$ is plotted as a function of time. The bottom panle of the same figure also shows the extractable energy at a fixed time as a function of the fraction of $M/N$, where $M$ is the number of cells accessible}. In the AL phase, this enhancement arises due to the integrability of the model as {the entanglement of a part of the system (entanglement-entropy) is known to scale in an area-law fashion in one dimensional integrable models \cite{vidal_prl_227902, calabrese_jsmte_p06002}.} Similarly, the MBL phase is also characterized by the presence of an extensive number of localized integrals of motion, thus restricting the dynamics in the Hilbert space. Further, the presence of interactions in the MBL phase suppresses temporal fluctuations in the stored energy which are otherwise prominent in the AL phase.

Similarly, it has also been demonstrated that for a battery prepared in the ground state of the quantum XYZ chain and charged through a local magnetic field (in $x -$ direction), the power is enhanced when the nearest-neighbor couplings are chosen randomly from a Gaussian distribution \cite{ghosh_pra_032115}. Note that in this case, the disorder is introduced in the battery Hamiltonian $H_0$; unlike the previous cases where disorder was present in the couplings of the charging Hamiltonian.

{We would like to mention that the exploration of disordered or random interactions in models of quantum battery has only just begun to pick up pace and is still in its nascent stages. In this regard, it has been recently shown that using the fermionic Sachdev-Ye-Kitaev (SYK) Hamiltonian for charging a battery of spins, it is possible to achieve explicit quantum advantage in the power of quantum batteries as well as improve the quality of the energy stored by suppressing unwanted fluctuations \cite{rossini_arxiv_07234, rosa_arxiv_07247,carrega_arxiv_03551}. The authors of Ref.~[\onlinecite{rossini_arxiv_07234}] consider a battery Hamiltonian similar to Eq.~\eqref{eq_hamil_spin_batt}, $\mathcal{H}_0=\sum_{j}h_j=(\omega_0/2)\sum_j\sigma_y^j$ which is charged using the complex SYK Hamiltonian,
\begin{equation}\label{eq_syk_battery}
	\mathcal{H}_1^{c-SYK}=\sum_{i,j,k,l=1}^NJ_{i,j,k,l}c_i^\dagger c_j^\dagger c_kc_l,
\end{equation}
where $c_i (c_i^\dagger)$ are spinless fermionic annihilation (creation) operators and $J_{i,j,k,l}$ are zero-mean Gaussian-distributed complex random variables, with
variance $\la\la J_{i,j,k,l}^2\ra\ra =J^2/N^3$, satisfying $J_{i,j,k,l}=J^*_{k,l,i,j}$ and $J_{i,j,k,l}=-J_{j,i,k,l}=-{J_{i,j,l,k}}$. Note that under appropriate Jordan-Wigner representation, the fermionic operators correspond to string operators in the spin representation, $c_j^\dagger=\sigma_j^\dagger(\prod_{m=1}^{j-1}\sigma_m^z)$. In  Fig.~\ref{fig_rossini_syk} (a), it can be seen that the disorder averaged variance of the battery Hamiltonian (calculated at optimum time $\tau^*$ at which the power is maximum) scales quadratically with $N$. The figure also shows that the disorder averaged quantity $\la\la\Delta_{\tau^*}^{Ent}\mathcal{H_0}^2\ra\ra$, where,
\begin{equation}\label{eq_entang_contr}
	\Delta_{\tau^*}^{Ent}\mathcal{H}_0^2 = \frac{1}{\tau}\int_0^\tau dt\sum_{i,j}\left[\la h_ih_j\ra-\la h_i\ra\la h_j\ra\right],
\end{equation} 
also scales quadratically which confirms that the super linear scaling of the variance of the battery Hamiltonian arises from the presence of long-range entanglement (see Eq.~\eqref{eq_batt_scal_entang} and the following discussion). The resulting quantum advantaged achieved is evident in Fig.~\ref{fig_rossini_syk} (b) where the power is found to scale super linearly with the system size.}

Finally, we also note that a general approach to analyze battery models based on disordered systems has been recently introduced in Ref.~[\onlinecite{caravelli_prr_023095}]. 

\subsection{Quantum batteries as open systems}

Finally, we would like to mention that the working of quantum batteries has also been explored in the context of open system dynamics. The charger-battery setting, discussed in Sec.~\ref{subsubsec_charger_assist}, has been extended to the case where the charger extends as a mediator between an external energy supply and the battery with the energy transfer facilitated by thermalization with the bath or coherent pumping by classical fields coupled to the charger \cite{farina_prb_035421}. The evolution of the charger-battery composite is thus dissipative in nature and is consequently analyzed using the GKSL master equation. Using this approach, different implementations of  charger-battery setup has been considered using harmonic oscillator and qubit systems to analyze the interplay between coherent pumping and thermalization mechanisms in the battery operation. We refer the reader to Ref.~[\onlinecite{farina_prb_035421}] for detail.

An interesting protocol of charging a quantum battery was considered in Ref.~[\onlinecite{barra_prl_210601}], where the system to be charged is coupled sequentially to a series of auxiliary systems prepared in Gibbs state. The coupling with each of the auxiliary system lasts for a time $\tau$ during which the systems undergoes dissipative dynamics. Using this method, it is possible to drive the system to an \text{active equilibrium state}, in which no external work is required to sustain the state once it is reached. Importantly, the equilibrium state is not passive thus allowing work extraction and thus provides advantage over regular thermalization processes in which the steady state is generally thermal and passive in nature. Once again, we refer to Ref.~[\onlinecite{barra_prl_210601}] for detail. {Similarly, stable charged states of quantum batteries have been shown to be realizable using adibatic protocols that are also robust to moderate environment induced dissipative effects \cite{santos_pre_032107,santos_pre_062114}.} Recently, other aspects of dissipative charging of quantum batteries are also being explored \cite{hovhannisyan_njp_052001, liu_jpc_18303, quach_arxiv_10044, carrega_arxiv_14034, zakavati_arxiv_09814, ghosh_arxiv_12859, hovhannisyan_prr_033413,bai_arxiv_06982, kamin_njp_083007,gherardini_prr_013095,caravelli_quantum_505}. 

\section{Outlook}\label{sec_outlook}
It is fair to say that quantum thermal machines have come a long way since the inception of the early prototypes. A rigorous analysis of their operating mechanism over the years has provided a much deeper understanding of how thermodynamic signatures are manifested at the scale of a few-level systems. The nature of the heat reservoirs has been found to significantly impact the performance of the QTMs; which at times have been shown to outperform their classical counterparts. Nevertheless, all QTMs proposed thus far have been found to operate within the premises of the well established classical thermodynamic laws. While {the basics of ideal QTMs} employing simple systems such as qubits or harmonic oscillators as working fluids and operating within the framework of Markovian dynamics are well understood by now, {investigations are still on with respect to their finite-time operations as well as fluctuations in their performance. In addition}, the trend is also shifting more towards incorporating various sources and aspects of non-Markovian dynamics into the operation of QTMs. Likewise, the pros and cons of using quantum many-body systems as working fluids are being actively explored. Newer variants of QTMs, aimed at utilizing entanglement as a resource \cite{tavakoli_pra_012315, khandelwal_njp_073039, josefsson_prb_081408, bresque_arxiv_03239} {and measurement driven QTMs which require only a single bath to function have also been recently explored \cite{yi_pre_022108, ding_pre_042122}.} It is also to be noted that there has also been a spurt in the number of experimental implementations of the QTMs and the verification of their predicted performance. 

On the other hand, there remains much to be understood as of how the \textit{quantumness} of batteries exactly influences their utility as energy storage devices. In particular, the stability of the stored energy, particularly with respect to environmental exposure, also needs to be carefully examined. Nevertheless, given the current pace of development and the amount of research being devoted, these questions are expected to be answered in the near future.  

\begin{acknowledgments}
We are grateful to Utso Bhattacharya, Victor Mukherjee, Wolfgang Niedenzu and Saikat Mondal for collaboration and useful discussions on related works. We also thank Arnab Ghosh, Souvik Bandyopadhyay and Somnath Maity for critical comments and suggestions. S.B. acknowledges CSIR, India for financial support. A.D. acknowledges financial support from SPARC program, MHRD, India and SERB, DST, India.
\end{acknowledgments}

\end{document}